\documentclass{aastex}   
\usepackage{spr-astr-addons}  
\usepackage{graphicx}
\usepackage{txfonts}
\usepackage{amsmath,amssymb}

\catcode`\@=11
\def\gapprox{\mathrel{\mathpalette\@versim>}}
\def\lapprox{\mathrel{\mathpalette\@versim<}}
\def\@versim#1#2{\lower2.45pt\vbox{\baselineskip0pt\lineskip0.9pt
     \ialign{$\m@th#1\hfil##\hfil$\crcr#2\crcr\sim\crcr}}}

\catcode`\@=12

\newcommand{\hii}{H\,{\sc ii}}

\newcommand{\Msun}{\ensuremath{\rm{M}_\odot}}

\newcommand{\OmM}{\ifmmode {\Omega_{\rm M}}\else $\Omega_{\rm M}$\fi}
\newcommand{\OmL}{\ifmmode {\Omega_{\Lambda}}\else $\Omega_{\Lambda}$\fi}
\newcommand{\kmps}{\ifmmode {\rm\,km~s^{-1}} \else ${\rm\,km\,s^{-1}}$\fi}

\newcommand{\mum}{\ensuremath{\mu\mbox{m}}}
\newcommand{\apjvec}[1]{\mbox{\boldmath{$#1$}}}

\newcommand{\vx}{\apjvec{x}}
\newcommand{\vy}{\apjvec{y}}

\newcommand{\vk}{\apjvec{\xi}}

\newcommand{\leqsim}{\mathrel{\lower2.3pt\hbox{$\mathpalette\oversim<$}}}

\begin{document}
\title{Fitting the integrated Spectral Energy Distributions of Galaxies}

\shorttitle{Fitting Integrated SEDs}
\shortauthors{Walcher, Groves, Budavari, Dale}

\author{Jakob Walcher} 
\affil{Research and Scientific Support Department, European Space Agency, Keplerlaan 1, 2200AG Noordwijk, The Netherlands}
\email{jwalcher@rssd.esa.int} 
\and 
\author{Brent Groves}
\affil{Sterrewacht Leiden, Leiden University, P.O. Box 9513, 2300 RA Leiden, The Netherlands}
\and
\author{Tam\'as Budav\'ari}
\affil{Dept. of Physics and Astronomy, The Johns Hopkins University, 3400 N. Charles Street, Baltimore, MD 21218, USA}
\and
\author{Daniel Dale}
\affil{Department of Physics and Astronomy, University of Wyoming, Laramie, WY 82071, USA}

\tableofcontents

\pagebreak

\begin{abstract}
Fitting the spectral energy distributions (SEDs) of galaxies is an almost universally 
used technique that has matured significantly in the last decade. Model predictions 
and fitting procedures have improved significantly over this time, attempting to keep 
up with the vastly increased volume and quality of available data. 
We review here the field of SED fitting, 
describing the modelling of ultraviolet to infrared galaxy SEDs, the creation of 
multiwavelength data sets, and the methods used to fit model SEDs to observed galaxy 
data sets. We touch upon the achievements and challenges in the major ingredients of 
SED fitting, with a special emphasis on describing the interplay between the quality of 
the available data, the quality of the available models, and the best fitting technique to 
use in order to obtain a realistic measurement as well as realistic uncertainties. 
We conclude that SED fitting can be used effectively to derive a range of physical 
properties of galaxies, such as redshift, stellar masses, star formation rates, dust masses, 
and metallicities, with care taken not to over-interpret the available data. Yet there still 
exist many issues such as estimating the age of the oldest stars in a galaxy, finer details of
dust properties and dust-star geometry, and the influences of poorly understood, luminous 
stellar types and phases. The challenge for the coming years will be to improve both the 
models and the observational data sets to resolve these uncertainties. 
The present review will be made available on an interactive, moderated web page (sedfitting.org), 
where the community can access and change the text. The intention is to expand the 
text and keep it up to date over the coming years. 

\end{abstract}


\section{Introduction}
\label{s:introduction}

Integrated spectral energy distributions (SEDs) are our primary source of 
information about the properties of unresolved galaxies. Indeed, the different 
physical processes occurring 
in galaxies all leave their imprint on the global and detailed shape of the spectrum, 
each dominating at different wavelengths. Detailed analysis of the SED of a galaxy 
should therefore, in principle, allow us to fully understand the
properties of that galaxy. SED fitting is thus 
the attempt to analyze a galaxy SED and to derive one or several physical properties 
simultaneously from fitting models to an observed SED. This is in contrast to searching 
a single feature that could constrain a single parameter (a prominent example would be the 
use of the H$\alpha$ line to derive the star formation rate, SFR). 

The aim of this review is to present the state of the art in the area of fitting 
the ultraviolet (UV) to far infrared (FIR) SEDs of galaxies. 
It grew from and presents a summary of a workshop held 
at the Lorentz center in Leiden in November 2008. As the field is 
extremely large we attempt here a somewhat novel approach to 
the process of writing a review. We consider this text as a starting point. 
The text will then be made available at www.sedfitting.org and we invite the 
community to sent us corrections and additions. We particularly hope that 
this will help in covering the work that we might have missed. We also 
made no attempt (yet) to cover the new results of the last year, in 
particular from Herschel. 

Progress in many areas that affect SED fitting has been made recently. 
A major development in the last decade has been the advent of new 
observing facilities and large surveys at all wavelengths of the 
spectrum, enabling astronomers for the first time to observe the full
SEDs of galaxies at wavelengths from 
the X-rays to the radio. The same surveys have also pushed the 
distance of the farthest galaxy whose SED is amenable to study
to redshifts higher than 6. At the same time, tools and models have 
been created that aim to extract the complex information imprinted 
in the SEDs. Also, useful semi-analytic models of galaxy formation 
have appeared that provide realistic predictions for the 
properties of galaxies as they 
would be formed in the current cosmological standard model. 
Not only are astronomers exploiting the available data, 
but the next generation of surveys is now in the planning phase. 

For this review we concentrate on observations from the ultraviolet to the far infrared, including 
both multi-band photometric and spectral data. We thus treat the light emitted 
by stars, either directly or processed by the gas and dust of the surrounding interstellar 
medium. At wavelengths outside the regime considered here, such as the X-ray
and radio wavelengths, non-stellar processes (or at least those not
directly associated with stellar light) such as shocks, accretion onto
compact objects, etc.~dominate. While many of these can be associated
with the star formation history of a galaxy (e.g.~supernova rate and
recent star formation), these processes require a higher order of
complexity generally not considered  in most current models of galaxy
spectra, and hence we do not discuss these wavelengths in the rest of
this review. We also do not treat the contribution of active galactic nuclei 
to the SEDs of galaxies. 

Our initial aim of both the workshop and this review was to set up a
basic framework to answer the main question relating to SED fitting:  
considering the difficulties with the models, considering the limitations of the 
data and considering the fitting technique, what is the true
uncertainty and limitations on the properties that can be determined
from galaxy SED fitting?  

This review is structured as follows: in Section \ref{s:models} we review 
the basics of galaxy ultraviolet
to infrared SED modelling, from galaxy formation to the production of 
and processing of the radiation from these galaxies. We especially mention 
some of the current issues and main uncertainties of the modelling of
galaxy SEDs. In Section \ref{s:obs} we provide a short overview of the 
intricacies of assembling multi-wavelength SEDs. 
In Section \ref{s:method} we present techniques and algorithms for SED 
fitting, and -- most importantly -- efforts at 
validating the results from the SED fitting procedure with independent data. 
Section \ref{s:photoz} presents a review of photometric redshift determinations, 
a special case 
of a physical property derived from SED fitting, as it can be compared to and 
calibrated on independently determined data, spectroscopic redshifts. 
Finally, Section \ref{s:results2} showcases some recent results from 
application of the SED fitting procedures, where we hope to emphasize the 
variety of problems to which SED fitting can significantly contribute.

\section{Modelling galaxy SEDs}
\label{s:models}

Galaxies emit across the
electromagnetic spectrum. Excluding those galaxies dominated by an
accreting supermassive black hole at their nucleus (AGN), the
ultraviolet to infrared spectra of all galaxies arises from stellar
light, either directly or reprocessed by the gas and dust of the
surrounding interstellar medium (ISM). Thus the UV-to-IR spectral energy
distribution or SED contains a large amount of information about the
stars of a galaxy, such as the stellar mass to light ratio, and
the surrounding ISM, such as the total dust mass. However, to extract such
information, models are necessary in order to connect physical
properties of the galaxy with the observed SED. 
In this section we
discuss such models, beginning with the stellar spectrophotometric models,
moving on to the transfer of the radiation of these stars in a galaxy through
the ISM, and finally how to connect these with the larger picture of
galaxy formation and evolution. We use the following abbreviations 
for designing wavelength ranges, though the exact boundaries between
wavelength regimes are not sharp: ultraviolet (UV) for $\lambda$$<$3500
{\AA}, optical for 3500$<$$\lambda$$<$8000 {\AA}, near infrared (NIR) 
0.8$<$$\lambda$$<$3 $\mu$m, mid-infrared (MIR) 3$<$$\lambda$$<$25 $\mu$m, 
far-infrared (FIR) 25$<$$\lambda$$<$250 $\mu$m, sub-mm 
0.25$<$$\lambda$$<$1 mm, and radio $\lambda$$>$1 mm.

\subsection{Stars} 
\label{s:stars}

In its simplest sense, a galaxy is a population of stars ranging from
numerous, low-luminosity, low-mass stars, to the bright, short-lived,
massive OB stars. On closer examination, these stars are distributed
in both metallicity content and age ranging from when the galaxy first
formed to those newly born. The method of creating a galactic 
spectrum through the sum of the spectra of its stars is called stellar 
population synthesis and was pioneered in works by \citet{tinsley72}, 
\citet{searle73} and \citet{larson78}. A simplification for the modelling 
of galactic SEDs is that 
the emitted light can be represented through a sum of spectra of simple 
stellar populations (SSPs) with different age and element abundances. 
Here a SSP is an idealized single-age, single-abundance ensemble of 
stars whose distribution in mass depends on both the initial
distribution and the assumed age of the ensemble. 

There are two main methods used by current stellar spectrophotometric
models to compute the SEDs of SSPs: The first is called 'isochrone 
synthesis'. It uses the locus of stars with the same age, called an isochrone, 
in the Hertzsprung-Russel diagram and then integrates the spectra of all 
stars along the isochrone to compute the total flux. This method was established 
by \citet{chiosi88,maeder88} and in particular \citet{charlot91} and is currently 
used by the majority 
of stellar population models. The second uses the `fuel consumption' approach. 
One of the problems of the isochrone synthesis method was that 
isochrones are calculated in discrete steps in time and therefore phases 
where stellar evolution is more rapid than theses timesteps were not well 
represented (the most famous example of the last years being the
thermally pulsing asymptotic giant branch
stars). Models using the fuel consumption theorem circumvent this 
problem by changing the integration variable above the main sequence 
turnoff to the stellar fuel, i.e.~the amount of hydrogen and helium used in 
nuclear burning. The fuel is integrated along the evolutionary track. 
The main idea is that the luminosity of the post-main sequence
stars, which are the most luminous, is directly linked to the fuel
available to stars at the turnoff mass \citep[for full details, see
e.g.][]{buzzoni89,maraston98,maraston05}. 
While these methods are fundamentally different in their integration
methods, most of the issues discussed here in terms of stellar
evolution and stellar libraries apply to both.

\subsubsection{Simple stellar populations}
\label{s:ssp}

The spectrum (flux emitted per unit frequency per unit mass), $L_{\nu}$, of a
SSP of mass M, age $t$, and metallicity $Z$ is given by the 
sum of the individual stars: 
\begin{equation}\label{e:f_ssp}
L_{\nu}(t,Z)=\int_{\rm{M}}\phi({\rm{M}})_{t,Z}L_{\nu}({\rm{M}},t,Z).
\end{equation}

In practice the emitted light is dominated by the most massive, luminous 
stars. 

The stellar mass function, $\phi(\rm{M})_{t,Z}$, is computed from an
initial mass function (IMF, $\phi_{0}(\rm{M})$) and stellar evolution,
which describes when and which stars will stop contributing to the 
SSP spectra because they end their lives either as Supernovae or as 
white dwarfs. 
The IMF describes the distribution in mass of a putative zero-age main sequence
stellar population and is an input parameter of stellar population
synthesis models. The IMF is usually limited between a minimum
and maximum stellar mass (generally M$_{\rm{min}}\sim 0.05-1.0$\Msun;
M$_{\rm{max}}\sim 100-150$\Msun). Three empirical forms are most 
commonly used: a simple power-law model \citep{salpeter55,massey98}, 
a broken power-law \citep{kroupa01}, or a lognormal form
\citep{chabrier01}. However whether these forms hold in all conditions
and for all redshifts is still an open question \citep[A good coverage
of this field can be found in the ``IMF@50''
proceedings,][]{corbelli05}. 

The true difficulty of calculating equation \ref{e:f_ssp} lies in the second
part, determining the SED ($L_{\nu}$) of a star of initial mass, M, age, 
$t$, and metallicity, $Z$. This requires;
1) the computation of stellar evolutionary tracks that determine where a star of 
given stellar parameters (e.g.~mass M, age t and abundance $Z$) 
lies on the Hertzsprung-Russel diagram or log $g$ - T$_{\rm{eff}}$ 
diagram, to build up the stellar 'isochrone', and 2) the computation
or empirical building of a stellar library of $L_{\nu}$ 
with full coverage of $\log g,\rm{T}_{\rm{eff}},$ and $Z$ to determine 
what the resulting spectrum of such a star is. 

\begin{figure}
\includegraphics[width=\hsize]{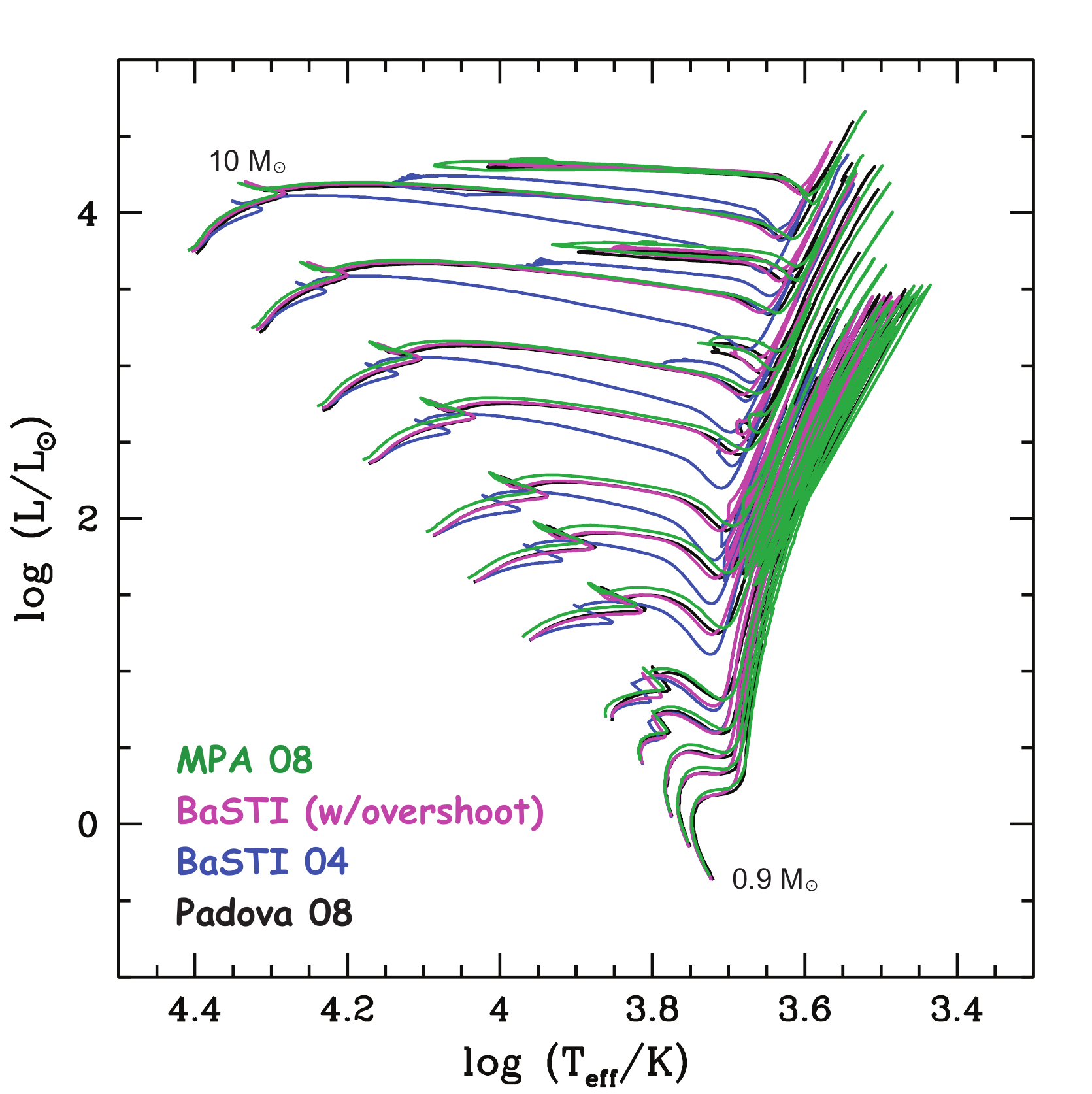}
\caption{Evolutionary tracks of solar composition low mass stars 
(0.9 - 10 \Msun) demonstrating the differences between four different 
models (as labelled): MPA08 \citep{{weiss08}}, BaSTI04 
\citep[with/without overshoot;][]{pietrinferni09}, and Padova08 
\citep{marigo07,marigo08} [Courtesy S. Charlot]. }\label{f:model_tracks}
\end{figure}

The creation of stellar isochrones requires a large grid of
evolutionary tracks, created by modelling the evolution of stars of a
given initial mass and metal content. Over the past few decades, 
much work has gone into providing homogeneous sets of stellar tracks 
from different groups, e.g.~Padova \citep{marigo07,marigo08}, Geneva 
\citep{lejeune01}, Yale \citep{demarque04}, MPA\citep{weiss08}, BaSTI 
\citep{pietrinferni09}. For SSP modelling, the models generally run
from the start of the 
main sequence (Zero Age Main Sequence, ZAMS) to some end point of the
star, such as a supernova or the asymptotic giant branch (AGB) phase.  
Originally computed only for a solar metallicity composition and a few
stellar masses, sets of homogeneous stellar evolutionary tracks now
exist for a wide range of initial masses \citep[from $\sim0.1 $\Msun\
to $\sim120$ \Msun; see e.g.][]{girardi00,meynet05} and
metallicities ($\sim0.01$ to $\sim4$Z$_{\odot}$). 
In most stellar evolutionary modelling it has been assumed that for
all stellar masses the elemental composition is the same for a given
metallicity, however more recently  the evolutionary effects of
elemental variations such as $\alpha$-enhancement
\citep[e.g.][]{salasnich00} or individual element variations
\citep[e.g.][]{dotter07} have been investigated.  
However problems still remain in the field, with the different treatments
by the different groups still giving distinct evolutionary tracks
even with the same inputs, as shown in figure \ref{f:model_tracks}.

\begin{figure}
\includegraphics[width=\hsize]{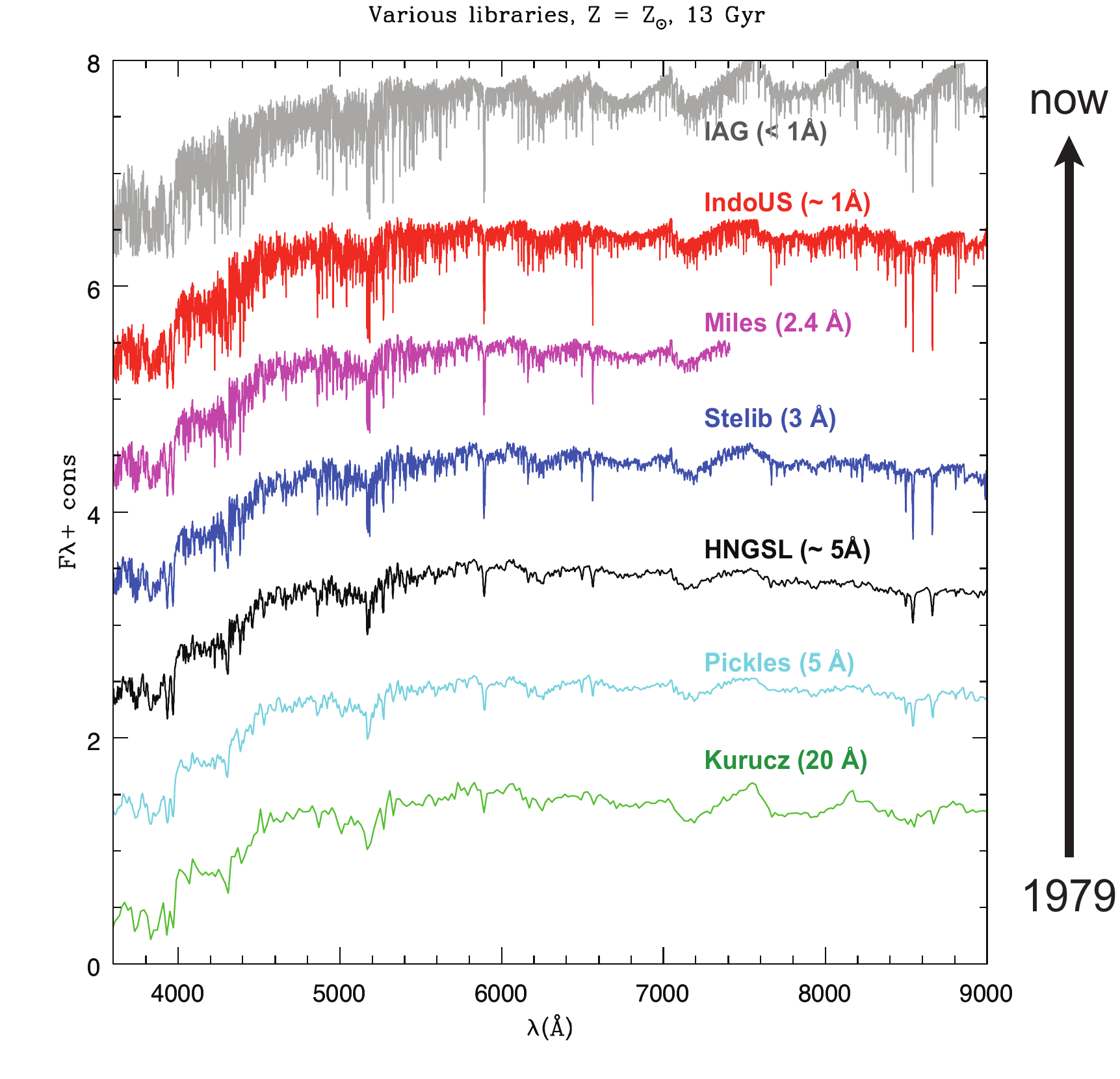}
\caption{Optical spectra from both theoretical and empirical stellar
  libraries (as labelled) demonstrating the improvement of spectral
  resolution over time with the associated improvement in library size
[Courtesy S. Charlot]. }\label{f:stellib}
\end{figure}

While the evolutionary tracks lead to the generation of an isochrone,
to determine a SSP spectrum, a library of stellar spectra is needed, 
covering the necessary parameter space in log T$_{\rm{eff}}$, log $g$, 
metallicity etc. As with the evolutionary tracks, stellar libraries have improved 
significantly in recent years, with both fully theoretical
\citep[e.g.][]{kurucz92, westera02, smith02, coelho05, martins05, lancon07} 
and empirical, e.g.~STELIB \citep{leborgne03}, 
MILES \citep{sanchezblazquez06,cenarro07}, Indo-US \citep{valdes04}, 
ELODIE \citep{prugniel01,prugniel07}, HST/NGSL \citep{gregg04} libraries 
covering much greater parameter spaces and increasing in both spectral 
and parameter resolution (see figure \ref{f:stellib}). Unlike the optical, the UV 
still suffers from incomplete libraries which is specially important for fully 
exploiting data on high redshift galaxies \citep[see e.g.][]{pellerin09}. 
The question about which 
of empirical or theoretical libraries is preferable is a matter of the specific 
application \citep[for a short review on both sets of libraries and their respective 
issues, see][]{coelho09}. The main benefit of the empirical  libraries is that 
they are based on real stars and thus avoid uncertainties in stellar atmosphere 
structure or in opacities. On the other hand, due to the observational limits they 
cover a restricted parameter space biased towards Milky way compositions
\citep[see e.g.][]{cenarro07}. Additionally, the determination of their 
fundamental parameters can be difficult for some types of stars and is
itself based on stellar models. Conversely theoretical libraries can
cover a much larger parameter space and at any chosen resolution
\citep[see e.g.][]{martins05}. There are still known problems in the 
comparison between the observed and theoretical stellar spectra 
\citep[][]{martins07}. Two specific examples for problems of theoretical models 
are incomplete line lists \citep{kurucz05}, problematic particularly at high spectral 
resolution, and the modelling of the IR emission \citep{lancon07}, which is 
particularly difficult for stars in the luminosity classes I and II (A. Lan\c{c}on, 
talk at workshop). The way forward may be a synthesizing approach, as 
suggest by \citet{walcher09}, aimed at using the strengths of both kinds of 
libraries. As with the evolutionary tracks, most libraries 
are limited to single compositions for a given metallicity. However
recently this also has been changing, with stellar libraries exploring
abundance changes such as $\alpha$-enhancement as well
\citep[e.g.][]{coelho07} .

It is not much of an overstatement to say that the magic of stellar population 
evolutionary synthesis spectrophotometric codes lies in interpolation. Indeed, 
to go from evolutionary tracks to isochrones \citep[quoting][]{maeder88} 
``the interpolation between evolutionary tracks must be properly 
based on point of corresponding evolutionary status'' \citep[see
also][]{prather76}, and similarly to go from an isochrone to SSP SED,
the correct, often interpolated, spectrum must be found for each
mass bin. These stellar population synthesis (SPS) codes, using
stellar evolutionary tracks and stellar  
libraries, then calculate Equation \ref{e:f_ssp}. Besides interpolation, the 
challenge is to create the most homogeneous and most accurate set 
of input ingredients, interpolating in an appropriate manner when necessary 
in dependence on the coverage and strengths of these sets. Stellar population 
models predicting full spectra include; 
\citet[][PEGASE]{fioc97}, 
\citet[][used in GRASIL]{bressan98} and 
\citet{leitherer99} and \citet[][]{vazquez07} (both Starburst99),
\citet{vazdekis99}, 
\citet{schulz02}, 
\citet{cervino02}, 
\citet{robert03},  
\citet[][GALAXEV, also commonly referred to as BC03]{bruzual03}, 
\citet[][PEGASE-HR]{leborgne04},
\citet[][M05, based on fuel consumption theorem]{maraston05},  
\citet{lancon08},
\citet{molla09}.
Currently less frequently used are fully theoretical stellar population 
models such as \citet{gonzalezdelgado05, coelho07}. However, 
as recently shown by \citet{walcher09}, a combination of semi-empirical 
and fully theoretical models holds great promise for the future. 
As a result of the improvements in stellar evolutionary tracks and
stellar libraries, as well as the codes themselves, stellar population
synthesis models today can recreate broad-band UV to NIR SEDs and
high-resolution spectra in the optical remarkably well.  

\subsubsection{Validation of SSP predictions}
\label{s:validssp}

As SSP spectra form the basis of all fitting of galaxy SEDs and as 
the complexities of real galaxies may introduce degeneracies and 
further uncertainties in the resulting interpretations, it is of primary importance 
to validate directly the predictions of the stellar population synthesis models 
for SSPs \citep[see e.g.][]{bruzual01}. For the impact of uncertainties in the stellar 
parameters effective temperature, surface gravity, and iron abundance on 
the final SSPs see \citet{percival09}.

The ideal testbed for such validations are simple stellar populations 
occurring in nature, i.e.~co-eval stellar populations such as 
globular clusters (GCs), open clusters, and young star clusters. 
These objects have been used as such
for some time \citep[e.g.][]{renzini88,gonzalezdelgado10}, but unfortunately, the exact
equivalence between star clusters and SSPs breaks down for a number of reasons: \\
1) In star clusters, the stellar populations are affected by the dynamical evolution of the cluster, 
which leads to mass segregation, and evaporation of low-mass stars in
GCs and, in young star clusters, the expulsion of gas early in their
life time may lead to  dissolution (open clusters) or
to the loss of a significant number of stars \citep[though see][for recent work on
dealing with this in SSP models]{anders09}. \\
2) Clusters also contain exotic stars (e.g.~blue straggler stars,
discussed in the following section) which influence the integrated
light of the cluster but are not accounted for by
most population synthesis models \citep[though see][for a discussion
on how to account for these]{xin07}.\\
3) Finally, star clusters only have a finite number of stars. In a less than $10^5$ M$_{\odot}$ 
star cluster, the number of bright stars is so small, that stochastic 
fluctuations in the photometric properties of the cluster are common 
\citep{barbaro77, lancon00a, cervino04, cervino06,piskunov09}.

In the study of individual clusters, most of these problems might be alleviated 
by concentrating on the most massive specimens (W3, $\omega$ Cen, starburst 
clusters), but these have the tendency to exhibit multiple rather than
single (simple) stellar 
populations \citep[e.g.][]{lee99}. Thus, one needs to study star cluster populations 
for comparison with SSP models and to account for the influence 
of the stochastic fluctuations on the color-luminosity distribution. Looking the other 
way around, within a given error box for the observed colors, a 
complex distribution of possible ages is possible (Fouesneau et al., talk at workshop). 
While multi-wavelength observations help, they do not completely eliminate 
the problem.

\subsubsection{Current Issues with SSPs}
\label{s:sspissues}

Even though significant improvements in the evolutionary
tracks and stellar libraries have been made in the last decade, 
significant challenges remain, as some parts of stellar evolution 
are only weakly understood, and hence poorly treated. The most
important of these tend to be short lived but bright phases:
massive stars, thermally pulsing asymptotic giant branch (TP-AGB) stars, extreme
horizontal branch stars (EHB) and blue stragglers.

Massive stars, due to their rapid evolution and short lifetime, prove
to be difficult to model and observe at all phases. Additional difficulties 
arise in that they tend to be buried by the interstellar
material that they formed from for a large fraction of their lifetime,
and experience high stellar winds (and hence strong mass evolution
over their lifetime). 
Yet massive stars are a vital component of SSP modelling because they are
so luminous and can thus dominate a SSP spectrum, and because they 
give rise to most of the 
ionizing flux and resulting nebular emission-line contribution.
Previous modelling of massive star evolution paid particular 
attention to the size of the convective core and stellar mass loss,
yet recent theory has indicated the significant, if not dominant, role that stellar
rotation has on the evolution of these stars \citep{meynet05}.  \citet{vazquez07} 
show in their recent stellar population model that, as rotating stars 
tend to be bluer and more luminous than in earlier models, even the ionizing 
spectrum can be significantly altered. These differences have
consequences when interpreting the SEDs of young galaxies, such as decreasing the
determined mass or star formation rates.

\begin{figure}
\includegraphics[width=\hsize]{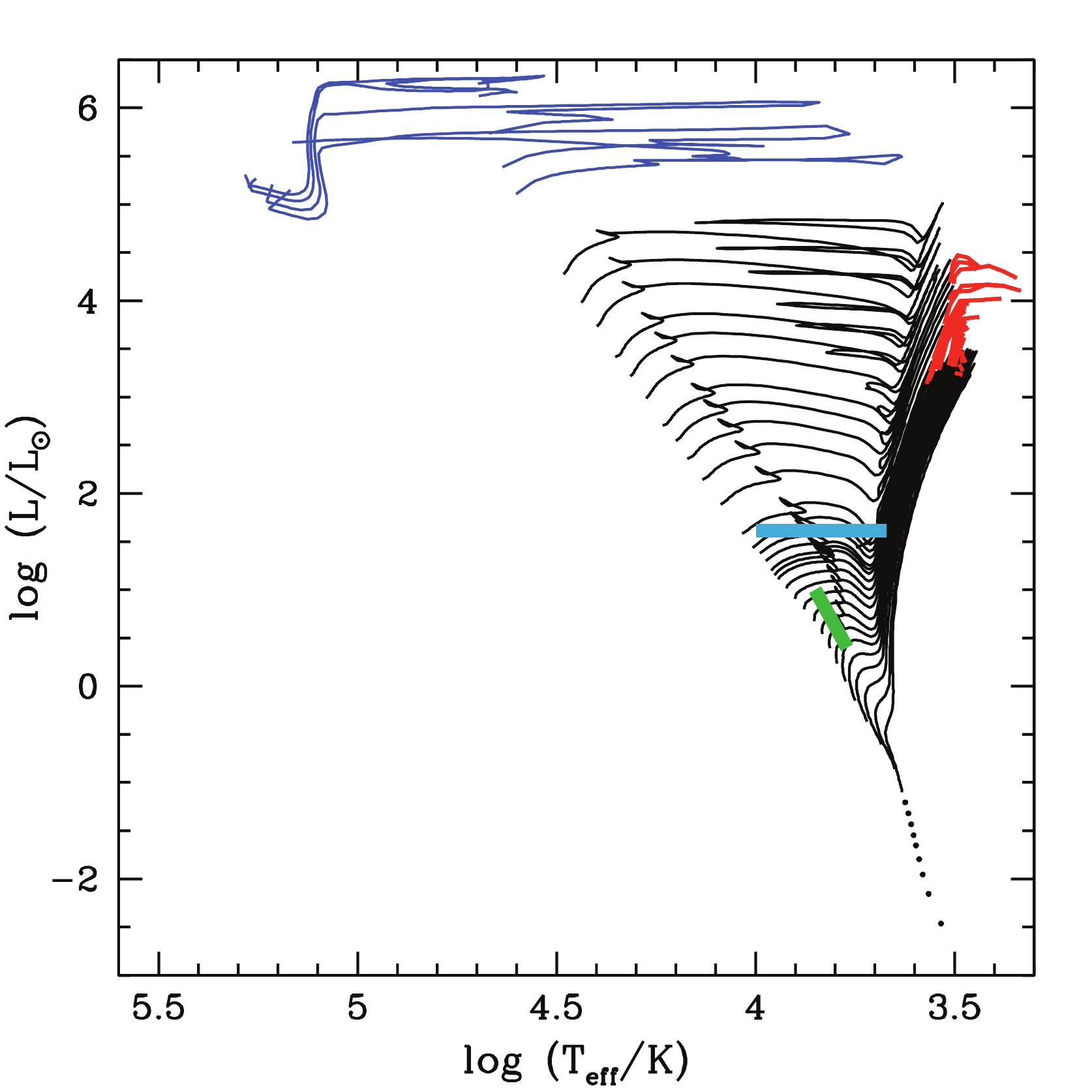}
\caption{Effective stellar temperature versus luminosity diagram
  showing stellar evolutionary tracks, with the  problem/uncertain
  areas marked. As a basis, the Padova \citep{marigo08} tracks for
  0.1 to 15 \Msun\ are shown in black. At higher luminosities (blue
  curves) are
  evolutionary tracks for the rapidly evolving high mass ($\sim100$
  \Msun) stars. The red curves show the new evolutionary tracks for
  the thermally pulsing asymptotic giant branch (TP-AGB) phase, 
  while the blue and green rectangles show the regions
  dominated by extreme horizontal branch (EHB) and blue straggler
  stars respectively. [Courtesy S. Charlot]  }\label{f:SSP_probs}
\end{figure}

TP-AGB stars are short-lived, cool but luminous components of evolved
stellar populations that tend to be more prominent at low metallicities. Due
to the short lifetime of this stellar phase, as well as the inherent
instability of the pulsations, such stars are difficult to model. Yet
due to their relatively high luminosities (see figure \ref{f:SSP_probs})
they can significantly alter the mass-to-light (M$_{*}$/L) ratio of intermediate-age
populations and it is thus important to properly include them. 
Previously, several different theoretical and semi-empirical recipes
had been used in stellar population synthesis models leading to large
discrepancies between SSP spectra \citep{vassiliadis93, maraston98}. 
Attention has been focused on these stars since \citet{maraston05}
raised this issue,
leading to rapid progress in the modelling \citep[][]{marigo07}, largely 
reducing these differences in the broad-band photometry. 

The emission of SSPs near 10 Myr is dominated by luminous red supergiants, 
showing that more problems exist in the NIR than only TP-AGB stars. At 
higher spectral resolution in the NIR, comparisons of SSPs based 
on different libraries of synthetic stellar spectra and different isochrones 
show large residuals in the whole range from 1 to 2.5 $\mu$m and in 
particular for young to intermediate ages (A. Lan\c{c}on, talk at workshop). 
Additionally, ages derived from NIR and optical spectroscopy 
are discrepant by factors of two in this age regime. 
Resolving these problems will surely lead to improved 
predictions of stellar population models at all wavelengths. 

EHB stars represent the most luminous hot component in old 
stellar populations. Understanding these stars and implementing 
them in SSP codes is important because they could be mistaken 
for low-level star formation in more evolved, early-type galaxies.
Unfortunately, the evolution of EHB stars is not fully understood. While HB
morphology may be dependent upon metallicity, some metal-rich stellar
populations show HB stars bluer than expected \citep{heber08}. 
The second-parameter problem with the morphology of the horizontal 
branch \citep[for a review see][]{catelan09} will need to be solved 
before significant progress can be expected in this field. Meanwhile, 
the comparison between model SSP spectra and the data at old 
ages is affected by these uncertainties \citep{ocvirk10}.

Blue straggler stars are, as the name suggests, stars that extend
beyond the main sequence turn-off. Their origin is still unknown,
though it is believed to be associated with binary star evolution,
either through mass transfer or merging \citep{tian06, ferraro06}. 
As with EHB stars, blue stragglers can affect the interpretation of 
early-type spectra giving younger average ages. These stars point 
towards a fundamental limitation of current SSP modelling, 
in that effects of binary
evolution are not included. While this causes little difference in
most cases, in some situations (such as where blue stragglers may
dominate) recent binary stellar population models may be more suitable
\citep{han07}. 

The reader is also referred to the series of papers 
\citet{conroy09, conroy10b} for a recent systematic 
study of some of the uncertainties affecting SSP models. 

Yet, even with all these remaining issues, SSP modelling has advanced
significantly in recent years, with simple Charlot \& Bruzual (2010,
in preparation) exponentially-declining SFH + burst models able to reproduce 
1000's of optical SDSS spectra (of $\sim5$\AA\
resolution) to a few percent (see Section \ref{s:invervalid}).

\subsection{The ISM around the stars}
\label{s:ism}

SSP models are state of the art for producing the spectra of
stellar populations, yet they are not sufficient alone for reproducing the
spectra of galaxies. Stars are the dominant power sources within
galaxies (excluding AGN). However, the radiation from stars is 
absorbed and processed by the gas and dust that lies between the stars, the
interstellar medium (ISM). This absorption must be accounted for when
comparing SSP models with optical/UV observations and a treatment of
the radiative transfer of the stellar light through the ISM and
subsequent ISM emission is necessary if the full UV--IR SED is to be understood.

While the gas and dust are in reality intermingled within the ISM, in practice 
they are often treated as separate components because their absorption 
properties have a different wavelength dependence. 

\subsubsection{Interstellar gas}
\label{s:isgas}

Interstellar gas is predominantly treated as atomic in the
modelling of galaxies. While molecular gas is clearly present in many 
galaxies, it has generally a low volume filling factor, meaning that rarely 
contributes significantly to the overall opacity in a galaxy. 
It is only a noticeable opacity
source in specific spectral features or in galaxies dominated by
nuclear/heavily obscured sources, such as AGN and ultra-luminous IR
galaxies (ULIRGs). Molecular
gas emission in galaxies is predominantly seen at longer wavelengths
(NIR and longer) and is generally treated to arise mostly from
``Photodissociation Regions'', where the gas is heated by 
the diffuse interstellar radiation field of the galaxy. Although 
it provides insight into the molecular phase of the ISM, 
molecular emission is not considered to be a significant
contributor to the overall SED of a galaxy \citep[for further 
details see reviews by e.g.][]{young91,hollenbach97}.

Atomic gas however is the dominant opacity source in the
extreme-UV ($\gapprox13.6$ eV). It reprocesses this light into strong
emission lines in the UV, optical and IR. It is thus especially important 
for  young, actively star-forming galaxies. 
Usually, it is assumed that all hydrogen ionizing photons
($h\nu >13.6$ eV) are absorbed locally, within a small volume around the
ionizing sources
(approximately the Str\"omgren
sphere\footnote{The Str\"omgren sphere is defined as;
  V$_{s}=Q({\rm H^0})/(n_{\rm H}^2\alpha_{B})$, where $Q({\rm H^0})$ is the
total number of ionizing photons, $n_{\rm H}$ is the hydrogen number
density, and $\alpha_{B}$ is the case $B$ hydrogen recombination rate.}),
which is appropriate due to the high opacity in this wavelength regime. 
This energy is then re-emitted in the hydrogen recombination lines, which correlate
directly with the total number of ionizing photons 
\citep[see e.g.][]{storey95, leitherer99}. However to determine the emission from
other elements, or to take account of both gas and dust absorption, 
full radiative transfer is needed using
photoionization codes such as CLOUDY \citep{ferland98} or Mappings III
\citep{groves08}. For further details, see reviews by
\citet{ferland03} and \citet{stasinska07}.

As both the number density and absorption cross-section of dust is low
relative to hydrogen in the EUV, dust is often ignored as an opacity
source. However, as hydrogen absorption is limited by the
recombination rate, dust absorption becomes relatively more important
as the strength of the ionizing radiation field increases, becoming
the dominant EUV-opacity source when
$q({\rm H^0})>\alpha_{B}/\kappa\sim5\times10^8 {\rm cm~s^{-1}}$
assuming typical values for the dust opacity, $\kappa$
\citep{dopita02}. 
This value of the ionization parameter $q$ is well  above the average
value measured for star-forming galaxies \citep[see e.g.][]{kewley01},
meaning negligible EUV absorption by dust in typical \hii\ regions ($<5$\%),
but such high values may be reached within compact \hii\ regions and
AGN meaning dust will absorb a significant fraction of EUV photons
\citep{dopita02,draine10}.

\subsubsection{Interstellar dust}
\label{s:isdust}

Interstellar dust has been a field of constant inquiry since it was
first realized that an obscuring material existed between the stars
and a large body of research exists on the composition, shape and
distribution of dust exists \citep[see][for a detailed review of the
field, and some remaining questions about dust]{draine03}.  

Most of our understanding of interstellar dust has come locally, from
observations within our own Galaxy and the Magellanic clouds, and also
through theoretical and experimental laboratory work. It is generally
accepted that the grains can be considered to be composed of three
different compositions; graphitic/amorphous carbon grains, amorphous
silicate grains, and polycyclic aromatic hydrocarbons (PAHs), which
may  or may not be an extension of the carbonaceous grains. The
former two were found to reproduce the observed extinction along
different lines-of-sight within our galaxy \citep{mathis77}, while the
latter were added to explain unidentified emission bands in the mid-IR
\citep{leger84}. Other forms of dust have been suggested, such as SiC
\citep{treffers74}, and ice is expected to form  on grains in the
coldest environments such as deep in molecular clouds, but generally
only these forms are considered in the SED modelling of galaxies.

The size distribution of interstellar dust grains is thought to be
power-law in nature, with a distribution N$(a)\propto a^{-3.5}$ or
similar, with the average cross-section dominated by small grains, but mass
dominated by large. This slope arises from both theory \citep{jones96}
and matching observations \citep{mathis77,draine84,weingartner01}. 

To obtain the optical data used for dust calculations in SED
modelling, the size distribution and types are then convolved with
absorption/emission cross-sections and scattering cross-sections and
phase functions which are determined by both laboratory observations
and Mie theory \citep[see][and references
within]{draine07a,zubko04}. PAHs are treated slightly differently as
their composition is not fully understood, and their properties can change significantly
with the charge of the grains, and thus have more empirical based
treatments \citep[see e.g.]{weingartner01}. 
Altogether these form the dust models which are used most often in SED
modelling, such as \citet{draine07a} or \citet{zubko04}, that have
been successfully compared with determined depletion patterns within
the ISM and observations in the UV, optical, and IR.
These models are either used as an
ensemble of individual grain sizes, or integrated to give the opacity
data of dust as a whole. Of course empirically based laws and
templates are also often used in SED modelling such as the Milky way
extinction law and Calzetti law (see below).

For the purposes of 
SED modelling and fitting, dust absorption and scattering, and dust emission
are often treated as distinct components. As the hottest dust is 
constrained by sublimation to $\lapprox2000$K (corresponding 
to $\sim 3-4$\mum\ peak emission),  in practice only the scattering
and absorption of light needs 
to be considered for modelling the optical-UV emission of galaxies.
Conversely, as dust opacity strongly decreases with increasing 
wavelength, in the far-infrared (FIR) only dust emission needs to be considered. 

\paragraph{Attenuation by dust}

The effects of dust on the optical-UV light are often described
by two parameters - the reddening and total obscuration.  
Reddening is the wavelength dependence of dust effects, including
features, and takes account of the fact that shorter wavelength
photons are more readily scattered and absorbed by dust. This is often
parametrized by the color excess $E(B-V)$ or the Balmer decrement
${\rm H}\alpha/{\rm H}\beta$.
The total obscuration is a measure of the total light absorbed or
scattered out of our -line-of-sight by dust either bolometrically or
in a single band and can be considered the normalization of the
reddening. This is generally parametrized as $A(V)$. For relative
measures correcting only for reddening is sufficient,  however for
absolute quantities the total obscuration must also be taken into
account. This is especially important when the reddening is close to flat,
i.e.~ only small visible effects by dust on the spectrum.

For individual stars in the Milky Way, the Large and Small Magellanic clouds, 
extinction laws have been measured
\citep[e.g.][]{cardelli89}. However, when considering a galaxy as a whole,
it must be taken into account that stars reside at different optical depths, depending on whether they 
lie on the side of the galaxy facing the observer or averted from the
observer, and that the stellar light can be scattered into the observer's
line-of-sight as well as out of it.
Additionally, stellar populations of different age will have different extinction 
optical depths, and this extinction might have a different wavelength 
dependence. These issues lead to the concept of `attenuation', where the 
complexity of the actual star-gas geometry is wrapped into a single attenuation 
law, now not applied individually to each star in the galaxy, but applied 
to the full spectrum of the galaxy. 

Using an attenuation law, the dust obscuration of stellar light is expressed 
through a screen approximation (see Equation \ref{e:dust}),
as if the dust was lying between us and the stellar population of the galaxy, with a
wavelength-dependent reddening law ($a_{\lambda}$). 
The total amount of attenuation then depends only upon the thickness of the screen ($\Delta\tau$),
\begin{equation}\label{e:dust}
I(\lambda)_{\rm{obs}}=I_{\rm{star}}(\lambda)e^{-a_{\lambda}\Delta\tau}.
\end{equation}
The attenuation law was derived empirically for starburst
galaxies by \citet{calzetti94,calzetti97} who fit the law with a
simple polynomial as a function of $1/\lambda$. They found a law much
greyer than the extinction laws of the Milky Way and LMC demonstrating
the effects of geometry and mixing compared to simple
extinction. Generally an simple power-law ,
$a_{\lambda}\propto \lambda^{-0.7}$, is able to reproduce the observed
effective attenuation in galaxies \citep{charlot00}.

However, a simple attenuation law cannot account for differential geometries 
and star formation histories within and between galaxies. This can be
seen with the higher optical depths observed for nebular emission
lines relative to the underlying stellar continuum, indicating that
the stars and gas that give rise to the lines and to the continuum see
different amounts of dust \citep{calzetti94,calzetti97}.   
These observations led to the improvement over a simple attenuation
law in the approaches of \citet{silva98} and \citet{charlot00}, who
created a more physical two-step model  in which young stars which
emit ionizing photons are likely to be still surrounded by the clouds
of gas and dust from which they formed.  In this model all stars are
attenuated by `diffuse' 
dust in the same manner as equation \ref{e:dust}. However young ($<
10$Myr) stars undergo an additional `birth cloud' attenuation. In
practice this means that the UV light and nebular emission lines
associated with the short-lived massive stars are more obscured than
the optical light dominated by the longer-lived stars, as observed in
real galaxies. 

While the empirically calibrated \citet{charlot00} model is an
improvement over a simple attenuation law, it still does not take
account of the differential dust and star geometries that are clearly
visible in resolved galaxies, such as bulges, disks, and dust
lanes. The clumpiness of the ISM, both within the diffuse phase
\citep[see e.g.][]{kuchinski98,witt00} and within the birth clouds \citep[see
e.g.][]{popescu00,dopita05}, will also affect the resulting attenuation of
galaxies. However the greatest difficulty that simple,
empirically-based attenuation laws face is the anisotropic scattering
of light by dust, as photons are not only scattered out of the
line-of-sight, but can also be scattered \emph{into} it. This can cause bluer
integrated spectra than can be accounted for by simple attenuation
laws, especially for face on galaxies \citep[see
e.g][]{baes01b,fischera03, pierini04,inoue06}. 

However, to take account of all these issues, proper radiative
transfer (RT)
calculations must be done, which require intensive computations. To
limit these calculations several treatments exist, which can be
broadly grouped into iterative methods and Monte Carlo methods
\citep[for a more detailed description for several of the methods used
in RT calculations, see][]{baes01a}. In the iterative approach, the
light is broken up into emitted and scattered components, with the RT
equation solved separately for each component, and the solution from
the previous component being used for the subsequent (i.e.~directly emitted photons
by stars, then photons scattered once by dust, photons scattered twice
etc.) and these equations iterated to convergence \citep[see
e.g.][]{kylafis87,xilouris98,xilouris99,tuffs04}. Monte Carlo methods use a method closer to
reality, where the paths of individual 'photons' are followed through
their interactions (absorption and scattering) through the galaxy. The
photons are emitted in a random direction from the sources, such as
stars, and interact randomly with the surrounding ISM with a certain
probability based on the mean free path length, and are followed
through these scattering events until the photons escape or are
absorbed. To build up an integrated SED of a galaxy, many photons must
then be followed, though many treatments now exist to limit this
number, such as only following photons which end up in the observer's
line of sight \citep[see e.g.][for some early work on
Monte Carlo RT in galaxies]{witt92,bianchi96,witt96}. Both of these approaches are currently used,
with the iterative quicker for given geometries, while Monte Carlo is
more able to handle complex distributions of stars and dust (several
existing codes are discussed in the following section). 

While obviously the most realistic approach, the limitation of the
radiative transfer is that it 
requires complex calculations and thus it is not directly applicable to large
sample of galaxies. RT codes have been used to provide template libraries of
attenuation for a range of galaxies \citep{bruzual88,ferrara99,pierini04}, and also
analytic functions for the attenuation of the components of galaxies
\citep[i.e.~bulge, disk, clumps etc.,][]{tuffs04}, to deal with this
issue, yet these introduce several free parameters which may be
difficult to determine for unresolved galaxies for which only
broad-band SED is available. It is for these reasons that a simple
attenuation law is still the most commonly used way to account for the
effects of dust on the UV-optical SED.

One final note about the attenuation by dust is the silicate dust features
that can appear in absorption at 9.7 and 18 \mum. These features
require large optical depths to be observed, and thus are generally
only seen in galaxies with strong nuclear sources (i.e. nuclear
starburst/AGN). As this absorption occurs againstmodeled dust emission, it is
usually only modeled with a simple absorbing screen, otherwise it
requires self-consistent radiative transfer (discussed in
section \ref{s:fullsed}). 

\paragraph{Emission by dust}

Dust emission in the FIR and sub-mm is most commonly modeled by a single
black body ($F_{\rm FIR} \propto B_{\lambda}({\rm T}_{\rm dust})$)
or emissivity-modified black body ($\propto B_{\lambda}({\rm T}_{\rm
dust})\lambda^{-\beta}$, also called grey body), or a simple sum
over a limited (2--3) number of these. The first form assumes that all dust is
in thermal equilibrium at one temperature ${\rm T}_{\rm dust}$. 
The emissivity of dust grains is generally taken to be a
power-law at these long wavelengths, with models and 
laboratory data suggesting indices ranging from
$\beta=1.0$--$2.0$. Actually the $\beta$ index is expected to be a
function of both grain size, composition and temperature \citep[see
e.g.][with a nice discussion on the constraints on $\beta$ in the
latter]{andriesse74,draine84,agladze96,mennella98}.  
When introducing more than one black body, one is generally limited by the
number of wavelengths observed and the details of the model
\citep[see e.g.][]{dunne01}. In general, two modified black-bodies are 
sufficient to model these wavelengths, encompassing the idea
of warm and cold components of the ISM (see
e.g.~\citealp{popescu02,hippelein03}, and the review by
\citealp{sauvage05}).  

In the MIR range simple black bodies are not sufficient and more detailed modelling is 
necessary. This is due to strong dust (PAH) emission features  and the
stochastic heating processes 
that become important for smaller dust grains. 
As the size of a dust
grain decreases, the impingement of photons onto the dust grain surface
becomes less frequent and more random, thus less statistically representative of the
interstellar radiation field, allowing significant cooling between
photon impacts \citep[Figure 13 of][]{draine03}. Thus, rather than
having a single temperature, the dust has a range of temperatures and
is parametrized rather by the strength of the radiation field heating
it. To model this one can use either Monte Carlo calculations simulating the
arrival of photons and subsequent emission, or more simply one assumes 
and solves for a steady-state distribution of temperatures 
given the strength and shape of the impinging radiation field and
dust size and composition\citep[see e.g.][]{guhathakurta89, desert90,
draine07a}. Once this temperature distribution is known, it can be
convolved with black bodies modified by the dust emissivity in the
MIR, including any features.

Polycyclic aromatic hydrocarbons (PAHs) could be either 
called the largest molecular species or the tiniest dust -- emit strong features in the
MIR \citep[see e.g.][]{smith07}. These features arise from specific bending and stretching modes
of the large aromatic molecules \citep{bauschlicher09}. As PAH emission bands
are so complex they are generally incorporated into 
the models by either assuming 
a template form for the MIR emission features \citep[see e.g.][]{desert90} or by modelling 
the physical processes in a way similar to the small dust grains 
\citep[e.g.][]{weingartner01,draine07a}. On the whole, while
aromatic molecules within galaxies are accepted to be the source of the MIR features, 
the typical shapes, sizes, and ionization-charges of these molecules are an active 
field of research.

More realistic FIR dust emission models must take into account that 
the dust within the ISM of galaxies will exhibit a range of
temperatures, from the hot dust around young stars and in outflows to
the coldest dust in cold molecular cores, driven by the range of
radiation fields and dust sizes. Such complex emission models
calculate, for a given radiation field, the emission from each grain
size and composition and then integrate over these for a given dust
distribution to obtain the total dust emission. The largest grains
are generally considered to have a single temperature, as they will be
in thermal equilibrium, leading to a simple distribution of
temperatures dependent upon grain size and composition. 
In more accurate models, the smallest grains are considered to be
stochastically heated and the temperature
distribution of the individual grains is calculated \citep[using, e.g., the 
treatment of][]{guhathakurta89}.  
To finally calculate the IR emission from a galaxy, the distribution
of dust masses over heating radiation field are also needed. Simpler IR
emission models assume a functional form of dust mass over heating
intensity; $dM_{d}=f(U)dU$, with $f(U)$ most often assumed to be a
power law \citep[see e.g.][]{dale01,dale02,draine07b}. The most
complex IR emission models use radiative transfer to calculate the
radiation field distribution over a galaxy, where the distribution of
dust and stars are assumed (i.e. parameters of the model), and thus
these models directly link the dust absorption and dust
emission. These are discussed in Section \ref{s:fullsed}.  

However, as the temperature distributions of the dust in the galactic
ISM are dependent upon dust--gas geometry and cannot be determined
from optical-UV data alone,
empirically-based templates are often used for representing the IR
SED of galaxies, especially when IR data is limited due to sensitivity
or confusion. These templates take dust models as described above
(i.e.~multiple modified black bodies, or dust heated by a range of radiation
fields)  and match these to observed IR SEDs (or IR colors) of groups
of galaxies. These templates then tend to have galaxy-wide properties
such as IR luminosity or galaxy type as parameters, though intrinsic
properties such as average interstellar radiation field intensity are
also used. Well known examples of templates include those of 
\citet{chary01}, \citet{dale02}, \citet{lagache04}, and, more
recently, \citet{rieke09}. Though these templates tend to be limited
by the samples that define them, they provide a good alternative to
models when no or very little information is available about the 
actual IR emission of a galaxy.

\subsubsection{Combining stellar and dust emission}
\label{s:fullsed}

The full UV to IR SED of a theoretical galaxy can be created through the
combination of the techniques and modelling discussed in the previous
sections (\ref{s:stars}--\ref{s:isdust}). However, the different  
wavelength regimes need to be consistently connected. The 
simplest method is to take the energy absorbed in the optical-UV (see 
Equation \ref{e:dust}) and to distribute it across the MIR and FIR, assuming 
simple emission properties for the dust, such as black bodies. This
is the method used by \citet{devriendt99} and \citet{dacunha08}. These
authors attempt to strike  
the balance between the capability to model large datasets and the 
minimum sophistication necessary for a realistic model. 

To associate full UV--submm SEDs with their semi-analytic models
(discussed in the following section) \citet{devriendt99} created
``STARDUST''. This model assumes that stars and dust are
homogeneously-mixed in the galaxy. The light from the stars,
i.e.~summed from SSPs, is then passed through an ISM with the amount of
dust determined from a simple galaxy chemical evolution model. The
dust-absorbed radiation is then re-emitted via a series of templates
generated from the \citet{desert90} model and fitted to observed IRAS
points, parametrized by the total IR luminosity. 

\begin{figure}
\includegraphics[width=\hsize]{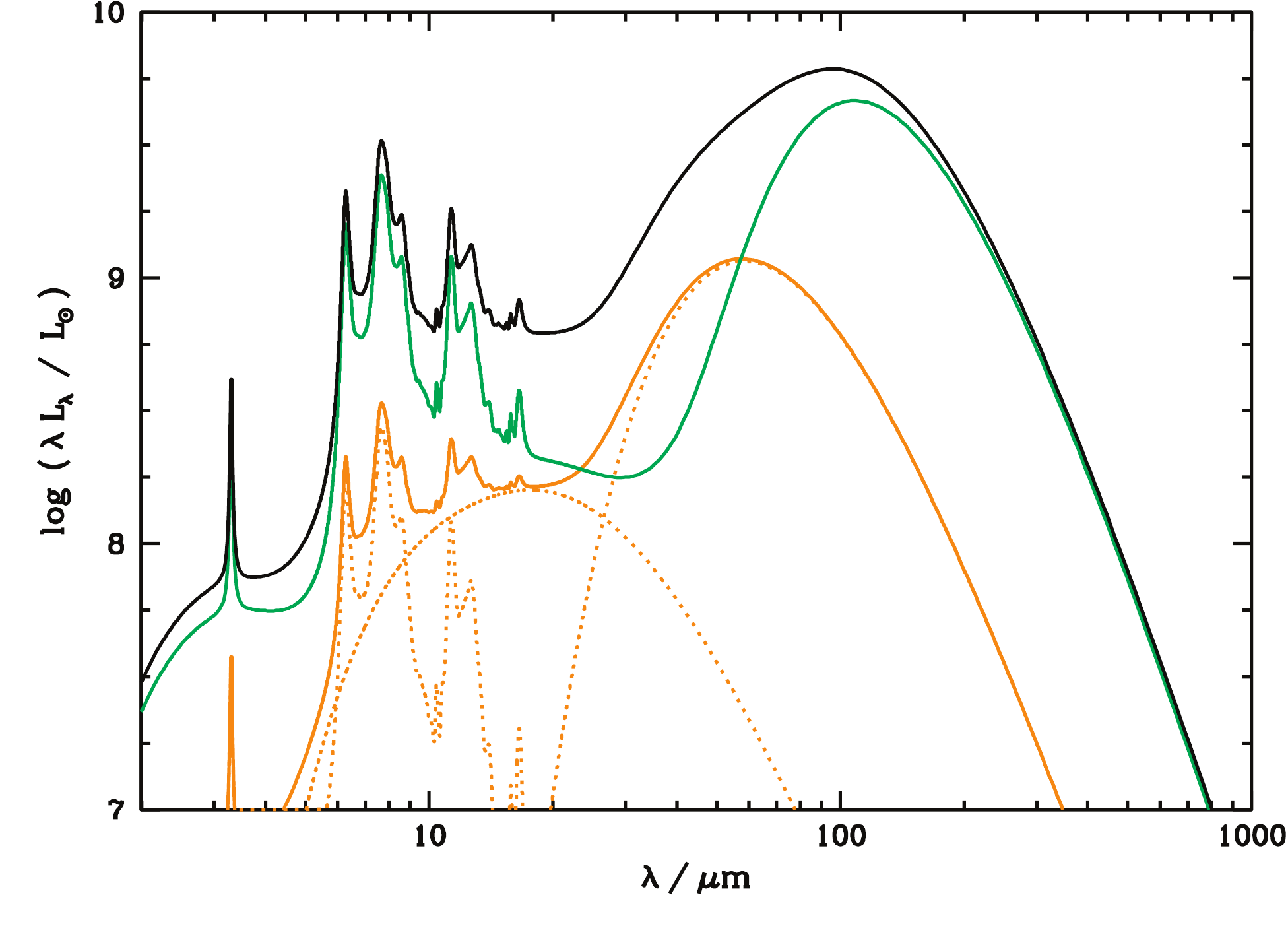}
\caption{IR emission of a simulated galaxy from the
\citet{dacunha08} model (black curve) 
demonstrating the individual contributions from the `birth cloud' dust (orange)
and `diffuse ISM' dust (green) [Courtesy E. da Cunha]. }\label{f:dacunha}
\end{figure}

\citet{dacunha08} follow a similar idea, but improve upon this by
using the \citet{charlot00} recipe for the attenuation. They thus  
obtain naturally corresponding `birth cloud'  and `diffuse ISM' dust emission
components over which the absorbed energy is
distributed (see Figure \ref{f:dacunha}). The two emission
components are both made up of a PAH template and variable grey body
contributions, with the birth cloud emission consisting of shorter
wavelength (hotter dust) emission. Such a model 
can simultaneously determine quantities such as stellar mass and dust mass of a
galaxy, and provide quantitative uncertainties for all parameters 
(see section \ref{s:physprop}). While this
method is quick, and hence suitable for comparison against large datasets, it
is self-consistent across the two emission components only in terms of
the total amount of radiation absorbed and re-emitted; physical 
properties, such as the dust temperature or the shape
of the emission within the components, are based on educated assumptions 
and are not constrained directly by the optical-UV absorption in the model. 

A very similar method was followed by \citep{noll09} with the CIGALE code,
which uses either the \citet{maraston05} or PEGASE codes for the stellar
populations and only a Calzetti attenuation law to attenuate the
stellar light. The major differences lies in the use of existing
empirically calibrated templates, such as from \citep{dale02}. rather
than a free IR emission made up of several parameter-controlled
components. 

To be properly self-consistent, the absorption and emission must occur more 
`simultaneously', such that the exact temperatures (including stochastic 
effects) of the dust causing the absorbing can be directly calculated. Such 
models require radiative transfer calculations to be
performed, such that the exact radiation field, or at least the
heating intensity, is
known at each point in the dusty ISM. This, along with assumptions
about the stellar ages and distribution, and the dust distribution and
properties can then give the full UV-IR SED of a model galaxy.

The models of \citet{efstathiou00} and \citet{siebenmorgen07} do this 
radiative-transfer calculation using the ray-tracing method for
starburst galaxies, which, being dominated  
by young stars and their birth clouds, are well represented by simple spherical
approximations. These models build upon a strong history of dust
radiative transfer and emission modelling and star-formation region modelling 
work to create simple models for the understanding of the UV--submm SEDs 
of starburst galaxies \citep{rowan80, rowan89, rowan92, siebenmorgen92a,
siebenmorgen92b, rowan93, siebenmorgen93, krugel94}. These works are 
based on the observation that young stars are both relatively more luminous 
and more obscured (thanks to the birth clouds) than older stars, and that in
strongly star-forming galaxies these young stars will be the dominant
IR (and significant bolometric) sources. In particular,
\citet{siebenmorgen07}, reduce the results of complex modelling to
a series of templates, based upon the physical properties of starbursting
galaxies, such as the total luminosity, size and extinction of the
star-forming regions and the contribution of the young stars to the
total luminosity of the galaxy.

\begin{figure}
\includegraphics[width=\hsize]{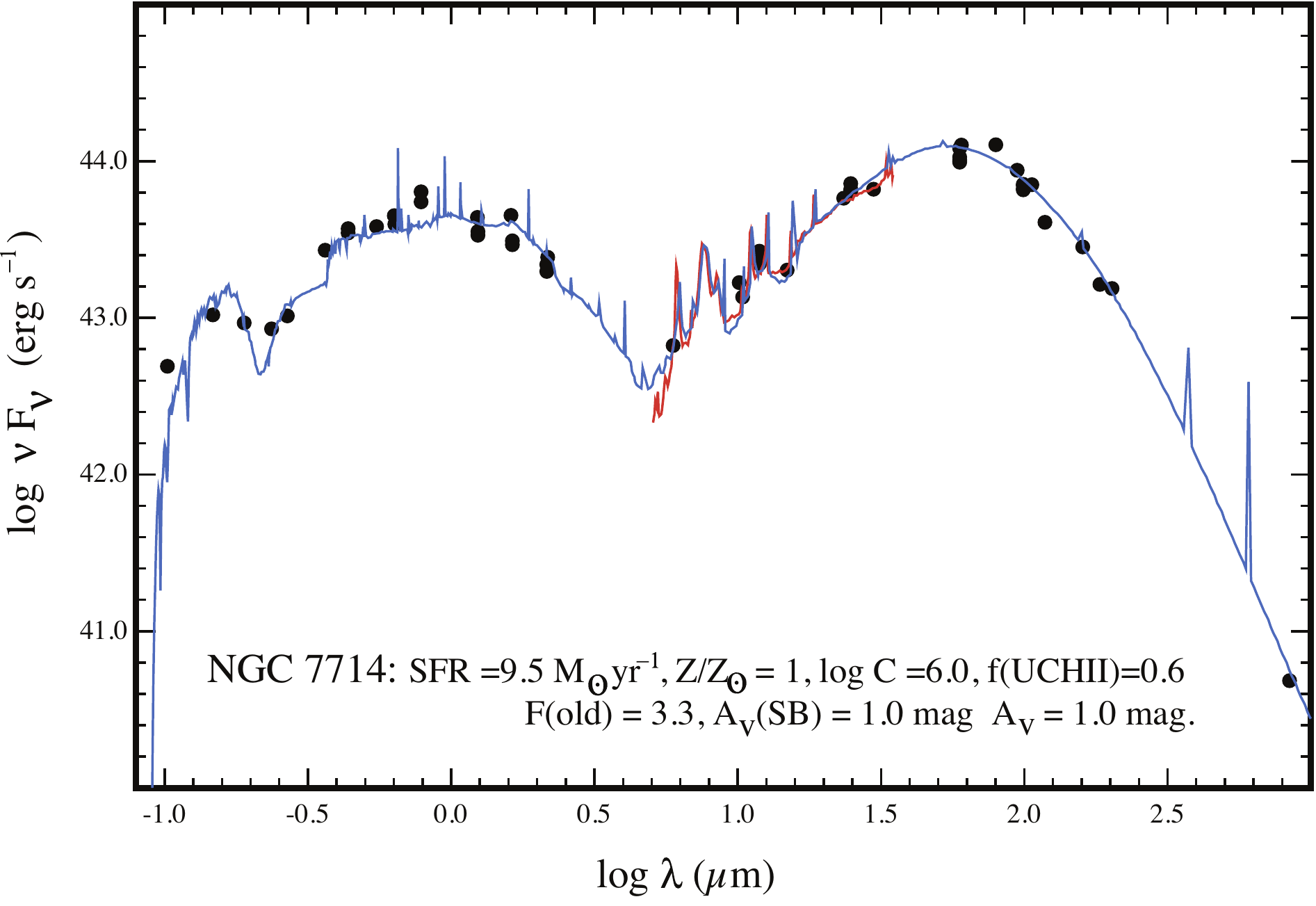}
\caption{\citet{groves08} model fit (blue curve) of the starburst
  galaxy NGC 7714 SED (black points and red curve mid-IR spectra),
  demonstrating the determination of physical galaxy properties such
  as star-formation rate (SFR) and metallicity \citep[as labelled,
  see][for full description of parameters]{groves08} 
  [Courtesy M. Dopita].}\label{f:dopita}
\end{figure}

\citet{groves08} \citep[building on previous
works;][]{dopita05,dopita06a,dopita06b}, take this work a step further
by self-consistently calculating the emission of a
star-forming region, including the radiative
transfer through the surrounding  gas and dust simultaneously. Like
\citet{efstathiou00} they allow for the \hii\ regions to evolve over
time, using empirically calibrated models. This
model is well suited for modelling starburst
(star-formation dominated) galaxies, where young stars and their
'birth clouds' dominate the emission, determining conditions such as
star-formation rate and compactness of the gas and stars. Like 
\citet{dacunha08} and \citet{siebenmorgen07} it
provides physical templates with as few parameters as
possible. Yet, while it fits well the SEDs of star-formation dominated 
galaxies (see Figure \ref{f:dopita}), this model is not suited for
non-starbursting galaxies, where the distribution of the diffuse dust
and stars must be accounted for. 
  
\begin{figure}
\includegraphics[width=\hsize]{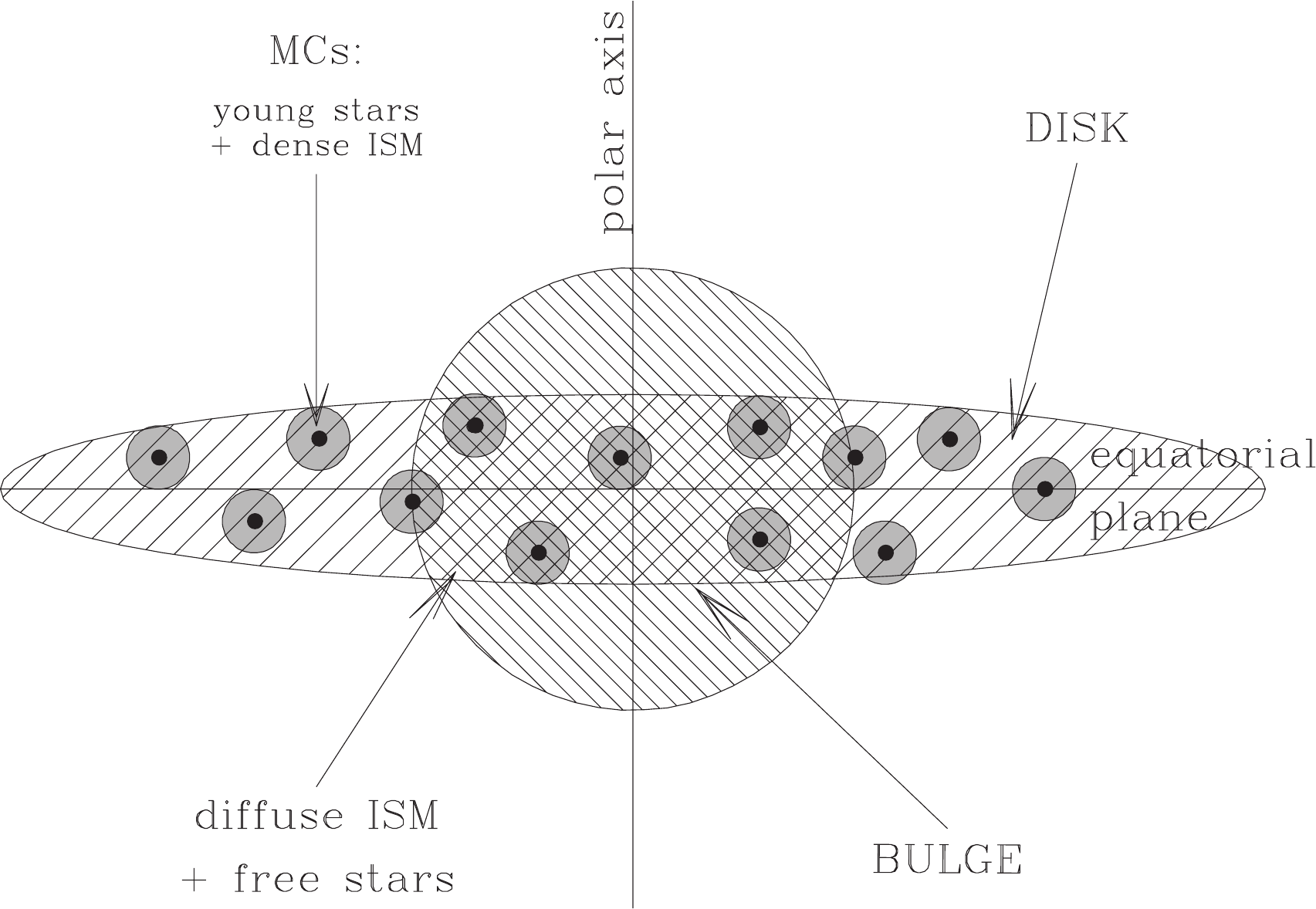}
\caption{Sketch of the geometry assumed within the GRASIL model
\citep[Figure 1 from ][]{granato00}.}
\label{f:GRASIL}
\end{figure}

By assuming a simple molecular cloud-disk-bulge
geometry (as shown in figure \ref{f:GRASIL}), the GRASIL model
\citep{silva98, granato00} is able to
account for the differential extinction suffered by the stars of
different ages associated with each of these components in a
galaxy. In addition, by varying the contribution of each component,
galaxies from spirals to ellipticals can be modeled. Unfortunately,
the more general geometry means that some parts \citep[such as the gas-dust
connection calculated in][]{groves08} cannot be calculated, and
also means more parameters are needed to define the model. As
with the \citet{groves08} model, the more accurate dust calculations mean
a longer calculation time, as compared with simpler models such as
\citet{dacunha08}. The GRASIL team is currently working on speeding 
up their calculations  for semi-analytic models (see following
section) by the use of neural networks \citep{silva10}. 

The main issue with all models discussed above is that, while they
take account of absorption (and emission) reasonably well,
they do not accurately take account of dust
scattering, which, as discussed above, can make some galaxies appear
bluer or redder depending 
upon inclination. This can be even more obvious in spatially resolved SEDs
of galaxies, where light from stars which are obscured along our line
of sight can be seen in reflection. However, as scattering is an
inherently stochastic process, it is difficult to model simply in a
galaxy, especially when multiple scatterings can occur. 

\citet{tuffs04}, following on from \citet{popescu00} and
\citet{misiriotis01}, use the iterative ray-tracing radiative transfer
method of \citet{kylafis87} to efficiently calculate the radiation field
throughout model galaxies consisting of a stellar bulge, stellar and
dusty disks and dusty clumps. Their resulting SEDs are then
self-consistent across the UV-IR range. In addition, one of the strong
benefits of radiative transfer is that the resulting SEDs can also be
spatially resolved, and be compared to multi-wavelength studies of
resolved galaxies, which they have done with edge-on galaxies such as
NGC 891 \citep{popescu00} and NGC 5097 \citep{misiriotis01}.

The other common approaches is to use
the Monte-Carlo radiative transfer method to model the UV-IR SED of galaxies.
Existing Monte-Carlo codes that have been applied to galaxies include
SUNRISE \citep{jonsson06, jonsson10},  DIRTY\citep{gordon01,misselt01}, TRADING
\citep{bianchi96,bianchi00,bianchi08}, SKIRT \citep{baes03}, and
RADISHE \citep{chakrabarti09}. These 
are able to model arbitrary and complex geometries of dust and gas, including
spiral arms, dust lanes, bulges and clumpy ISM. However unlike the ray
tracing method, the radiation field within the galaxy 
is not directly calculated (as only
individual photons or photon packets are followed). Thus dust heating
and emission must be treated through approximations
(discussed in detail within the papers listed above).  One treatment is to 
integrate within set volumes (i.e.~a grid) the amount of energy
absorbed by dust, and to redistribute this energy 
over large equilibrium grains. In some cases \citep{bianchi00}, 
small stochastic grains are also considered (using template assumptions). 
This approach can suffer from stochastic noise if the number of photons 
used is not sufficient. A similar treatment is to convert the absorbed energy into a
radiation field using the dust cross-sections, and thus with the
radiation field known the methods described in the Section \ref{s:isdust}
can be used \citep[see e.g.][]{misselt01}, though this still suffers
from issues of stochastic noise. Another treatment is called 
the ``dust temperature update method''. Here, the
temperature of the grains is updated with the absorption and emission of
each photon \citep[described in detail in][]{bjorkman01,baes05}.
All these methods must iterate in the case of self-absorption of dust. 
A more efficient method iterates on the calculation of the radiation
field density by using the previous estimates as a base and only
calculating for the difference at each iteration. This method  will always
converge as each iteration only adds a small amount of dust emission
which will provide an even smaller amount of dust emission. The
new radiation field is 
converted to IR emission using models such 
as \citet{dale02} or \citet{draine07a} \citep[see e.g.][]{juvela05, jonsson10}. 
While definitely more accurate in the treatment of dust, Monte Carlo 
codes require some representation of the ISM as
input and are much more
expensive computationally, especially in the cases where the dust is
optically thick to its own (IR) emission and many iterations may be
required. Current models are also, due to resolution effects both
within the RT and galaxy models, unable to
calculate the absorption on both diffuse (kpc) and local (pc) scales,
and thus currently use approximations or sub-resolution models
\citep[see e.g.][]{jonsson10}. Hence, while reproducing ``real'' galaxies, 
they cannot be directly used to fit observations of individual galaxies.

In summary, the modelling of the transfer of stellar light through the
ISM is well advanced, yet two significant challenges still exist. The first is
simply the computational effort needed to represent the radiative transfer
accurately. Many of the above models are limited in their resolution
to trace the ISM accurately, and thus need sub-resolution
approximations to treat some of the coldest or hottest dust
\citep[e.g. SUNRISE uses the starburst templates of][]{groves08} .
The second is our general lack of 
understanding of the dust composition in the ISM. Generally, dust is 
assumed to consist mainly of carbonaceous and silicate-like grains 
(such as olivine), in some power-law size distribution, \citep[see e.g.][]{mathis77}. 
This form is reasonably well constrained by observations of extinction in the 
optical-UV and emission features in the IR \citep[see][]{draine03}. 
Yet there are still open questions on shape \citep[how ordered or ``fluffy''
are the grains, e.g.~][]{zubko04}, on whether there are other kinds of dust, 
and on what formation and destruction processes lead to this power-law
distribution of sizes \citep[e.g.~][]{jones96}. Conversely there are
still spectral features associated with dust that are yet to be
properly explained, such as the 2175\AA\ absorption feature, the
diffuse interstellar bands in the optical, and the ``Extended Red
Emission'' band observed around 7000\AA\ \citep[see ][for a discussion
on these features and other remaining issues]{draine03}.

\subsection{Evolution of Galaxies}
\label{s:sam}

Together, the SSP and ISM radiative transfer models of the previous
sections are able to reproduce  the full UV--sub-mm SED of galaxies
with a reasonably high degree of accuracy \citep[see
e.g.][]{dacunha08, groves08}. Yet, by themselves, these models
are inherently static. Only limited model assumptions about the
past evolution of the galaxy can be introduced through the star 
formation history. In particular, the ISM is rarely
evolved along with the stars, and is presumed to be the same
metallicity as the latest generation of stars in the SSP models. 
It is common form to assume that the dust in the ISM is a constant 
fraction of the metals within the gas, distributed in a form similar 
to that found in our Milky Way.

These assumptions are sufficient to reproduce the observed SEDs of
real galaxies using empirically based priors
(e.g. Section \ref{s:bayesian}), or multiple components
(e.g. Section \ref{s:inversion}, and see Section \ref{s:method} for full
discussion).
However to produce fully-theoretical SED models that are at least conceptually similar to local 
galaxies, one needs to
fall back on galaxy evolution codes. There are three 
levels of these. At the innermost level are 
galactic chemical evolution codes, which, given some star-formation
history and/or some ``pristine'' inter-galactic medium (IGM) infall rate, trace
the evolution of the ISM metallicity, allowing for outflows, infalls,
and pollution by stars \citep[see reviews
by][]{hensler08,matteucci08}. The more recent of these codes also
evolve the dust along with the gas, taking into account the different
pollution rates of different elements, and the evolving
temperature/phases of the ISM \citep[e.g.][]{calura08}.  Once these
codes have given the corresponding ISM evolution with the
star-formation history (input or calculated), these can be associated
with SSP and ISM codes to give a more self-consistent instantaneous
spectrum of a galaxy \citep[e.g.][]{schurer09, conroy10a}. Some of the main issues with
these are the limited knowledge of the external gas losses and infalls,
meaning that exact evolution cannot be obtained, and the computational time
needed to calculate this evolution and associate it with a spectrum,
meaning that only specific sets of SFH or infall can be calculated at a time.

The next scale above the chemical evolution models are models that 
evolve the whole galaxy. These models are based upon hydrodynamic 
and N-body codes that follow the evolution of the ISM and stars within 
a dark matter halo representing a galaxy \citep[e.g.][]{springel05}. These 
codes use empirically based relations to 
follow the detailed evolution, such as the formation of stars from
gas, and the feedback from stars to the gas \citep[see
e.g.][]{tormen96,Cox06}. Containing both the stars (or ``stellar
particles'') and the ISM (with known metallicity), these galaxy
simulation/evolution codes are perfectly suited for linking with the
Monte-Carlo radiative transfer codes such as SUNRISE
\citep{jonsson06} or RADISHE \citep{chakrabarti09} which have
been purposely built to create spectra and broad-band images of 
these simulated galaxies.

The outermost layer are the cosmological models. These trace the
formation of structure in the Universe from the original perturbations
in the cosmic microwave background to redshift zero, using N-Body codes to simulate dark
matter and its gravitational interaction \citep[see][for a
review]{dolag08}.  While some of these models trace baryonic matter
as well as the dark matter, most trace only the dark matter due to the
more complex interactions of baryonic matter.  Thus to trace the
formation of galaxies within the forming dark matter halos
semi-analytic models (SAMs) are used \citep[e.g.][]{cole00, kauffmann00, 
hatton03,delucia04, somerville08}. These models use the outputs
from the dark matter simulations and approximate the physics of galaxy formation 
within the dark matter halos by empirical relations (e.g.~for gas cooling, star 
formation, AGN fueling, feedback). 

The SAMs return (and trace) the star formation history of each galaxy
that is created, including the effects of mergers, as well as the gas
content and metallicity of the gas (and stars). These results can be
used in association with SSP models (as discussed in section
\ref{s:stars}) to determine the stellar spectra of each galaxy. As
little geometrical information is returned by the SAMs, associating the
ISM effects on the stellar spectra is more difficult, especially so
for the IR emission. 
For the gas, most tend to use the associated emission lines added to
the SSP models \citep[see e.g.][]{leitherer99, charlot01}.
For dust attenuation, a simple treatment taken by many is to determine
the extinction assuming a uniform mixing of the stars and gas in a
galaxy, a fixed `template' attenuation curve, and basing the optical
depth on either empirical relations between galaxy
luminosity \citep[e.g.][]{kauffmann99, delucia04}, or amount of
dust in the galaxy \citep[e.g.][]{guiderdoni87,devriendt00}. More
advanced treatments include the use of the \citet{charlot00} model
\citep[e.g.][]{delucia07} or attenuation libraries like that of
\citet{ferrara99}, made for such purposes \citep[e.g.][]{bell03b}.

For the dust emission, the situation is more challenging. 
The simplest treatments assume that all of the radiation attenuated in
the optical (by the above treatments) are re-emitted in the IR. This
radiation is either distributed through modified Planck functions with
empirically-calibrated temperatures \citep[e.g.][]{kaviani03} or
empirically-based templates
\citep[e.g.][]{guiderdoni98,devriendt00}. Yet such models do not
take into account the strong geometrical dependence of dust heating or
the strong variations in the spectral shape and they are clearly not
self-consistent with the extinction in the optical--UV (see Section \ref{s:fullsed}).

For self-consistent SED models, the SAMs need to be coupled with radiative 
transfer (RT) calculations such as GRASIL, which has been done only for 
a few models \citep[e.g.][]{granato00,lacey08}. 
However one of the main strengths of SAMs is their computational
efficiency and speed which allows the calculation of the physical parameters of the
many galaxies in large cosmological volumes and over large redshift intervals
for many different implementations of the galaxy formation
physics. Yet RT is computationally intensive, and
severely slows the SAMs, meaning only relatively small volumes were investigated
in the SAM-RT models. In addition some of the details necessary for
the RT calculations are generally poorly modeled within the SAMs. 
Thus currently there is a  choice between poorly
representative but fast, or better modelling and slow \citep[see][for
an overview]{fontanot09}. The currently most advanced models
choose to compromise by using a RT-based library, empirically linked
with the SAMs \citep{fontanot09} or even linked through artificial
neural networks to account for the large and complex variations in
galactic UV--IR SEDs (Silva et al., in prep). 

Models of galaxy SEDs thus exist of varying resolution and complexity, adapted 
to model everything from individual galaxies or to large catalogs of galaxies on 
cosmological scales. While at each level of the SED models our knowledge of the 
important physical processes could be improved, SED modelling today is
much more accurate across the wavelength range than it was even a decade ago.

\section{Constructing observed galaxy SEDs}
\label{s:obs}

A major development in the last decade has been the advent of new 
observing facilities and large surveys at all wavelengths of the 
spectrum, enabling astronomers for the first time to observe the full
SEDs of galaxies from the UV to the FIR, from the local universe out to redshifts 
beyond 6. While databases such as those referenced in Table \ref{t:surveys} 
make it tempting to simply go ahead and fit full galaxy SEDs, it is important to 
pause for a moment and review which are the difficulties associated with the 
construction of a single SED.
Indeed, the ``true SED'' of a galaxy as defined in the models considers the sum of all 
photons emitted from inside the volume defining the galaxy. To make the observed 
SED of a galaxy, however, this "true SED" is then filtered through the spectral response 
curve of the instruments and is redistributed spatially over the point spread function 
(PSF). Additionally, the measurement process not only adds noise, but also makes it 
necessary to join data from different instruments.

The construction of multi-wavelength SEDS is a complex and rich subject and a dedicated 
review would be a welcome addition to the literature. In 
keeping with the scope of the present text we can here only give a very cursory 
treatment of the issue. As a starting point for further reading we suggest consulting some of
the major multi-wavelength surveys and their overview articles provided
in Table \ref{t:surveys} below.

\subsection{Spectral response curve and resolution}
\label{s:response}

Be it in spectroscopy or in photometry, one identifies the SED as a series of wavelengths 
and associated fluxes. In both cases, this is only a simplification of 
the fact that the measurement process convolves the true SED with a spectral response 
curve, yielding a transmitted flux at an effective wavelength. 
In spectroscopy, the response curve is almost invariably assumed to be Gaussian, 
with a $\sigma$ determined by the slit width and the dispersing device. 
Therefore, in practice the distinction between the instrumental broadening 
and the broadening due to the intrinsic velocity dispersion of the astronomical object 
is not very sharp. To mimic the instrumental broadening, one should first convolve with 
the appropriate Gaussian and then resample onto the spectral bins. Care needs to 
be taken that the wavelength calibration of both models and data are better than 
a tenth of a pixel over the full wavelength range 
\citep[this is not always the case, see e.g.][]{koleva08}.

In photometry, the response curve is much broader and therefore needs to be 
represented with more care, i.e.~tabulated as a response function. The response 
function in turn depends on the detector quantum efficiency, the instrument transmission 
and the filter in use. Photometric calibration and response characterization is a vital 
task \citep[see][for just two prominent examples]{koornneef86,landolt92}.

The signatures available for determination of the physical properties of galaxies of course depend 
on wavelength and on the achieved resolution. For example, in the optical many of the strongest features 
of galaxies can be adequately resolved at a resolution of R=$\lambda/\Delta\lambda$$\sim$2000, 
while the low-resolution part of the Spitzer IRS can easily resolve PAH features
at R$\sim$100. However, spectroscopy is more expensive 
in terms of telescope time, making photometry very attractive for obtaining large samples. 
In the last decade, successful use of narrow-band filters have blurred the distinction 
between spectroscopy and photometry, see for example COMBO-17 \citep{wolf03}, 
COSMOS \citep{scoville07} and NEWFIRM \citep{dokkum09}. Narrow-band filters have 
even been used to directly measure emission line equivalent widths \citep[e.g.][]{kakazu07}.

\subsection{Spatial resolution, aperture bias and matching}

One of the main tasks when assembling a multi-wavelength SED, indeed any 
catalog that contains more than one measurement, is to control whether 
what is measured in each band is actually physically the same. Due to either 
the seeing of the atmosphere or the diffraction of the telescope, the flux 
from a point-like source is re-distributed over the point spread function 
(PSF) of a width that typically depends on the time of observation and on 
the wavelength used. Moreover, galaxies are intrinsically extended and their 
morphology may depend on the wavelength in which they are observed. 

One of the main problems in the process of matching is the size of the PSF. 
Typically, the PSF is narrowest at optical wavelengths, while UV and IR PSFs 
are broader. This can lead to situations in which there is more than one optical 
counterpart to the UV, IR or sub-mm source. The agnostic way to deal with 
this is to simply exclude such objects from the sample, however, this may introduce 
a bias if the multiple optical counterparts are actually physically associated. 
A more intricate, but also more uncertain, way is to redistribute the flux according 
to optical priors \citep{guillaume06}. Finally, the use of all available information, 
spatial as well as spectral, seems to provide a promising way forward for 
multiwavelength datasets \citep{roseboom09}.

Another problem related to the resolution of the telescope are aperture biases. 
A rather simple manifestation of this is that objects that are further away will be seen 
as smaller on the sky. Therefore, in order to construct comparable samples at different 
redshifts, one has to adapt the size of the extracted aperture to the same physical 
size. A more complicated problem is the definition of the ``total light'' from an object. 
Indeed, the surface brightness profiles (SBP) of galaxies usually extend much beyond the 
threshold observational surface brightness. In the case of specific objects, such 
as cD galaxies, these extended wings might contain a significant part of the total 
light from a galaxy \citep{oemler76,carter77}. 
Different strategies have been developed to avoid these biases, 
such as either integrating over a full model for the SBP or simply using specific 
apertures to integrate the light only inside some physical radius, but each method 
has its own problems. One of these is also that galaxies have different intrinsic 
morphology in different bands, thus complicating the application of consistent 
procedures, even when using data with comparable angular resolution. 

A particular concern with fibre spectroscopic surveys such as the SDSS
is that the fibre apertures (3'' in SDSS) only sample part of the object, 
with this fraction different at each redshift. The effect of this can be tested by 
comparing result of a fit to the photometry of the whole galaxy against the 
photometric fit corresponding to the area of the fibre only \citep[e.g~][]{gomez03,brinchmann04}. 


\subsection{Examples of multi-wavelength datasets }

We now describe three "real-life" examples of assembled multi-wavelength datasets. 
It would here be impossible to describe all existing datasets, we therefore chose just three 
that highlight different goals and methods. 

\subsubsection{The Spitzer Local Volume Legacy - spatially resolved SEDs}
\label{s:LVL}

The Local Volume Legacy (LVL) is a Spitzer legacy program built upon a foundation 
of GALEX ultraviolet and ground-based H$\alpha$ imaging of 258 galaxies within 
11~Mpc.  The goal of the LVL survey is to fill a vital niche in existing multi-wavelength 
surveys of present-day galaxies with a statistically robust, approximately volume-complete 
study of our nearest star-forming neighbors.  Although star formation rates based on 
optical spectroscopy as well as GALEX ultraviolet and Spitzer infrared imaging have 
been measured for many thousands of galaxies, most currently available datasets 
are derived from flux-limited samples, and thus suffer from well-known biases against 
low-mass, low surface brightness systems.  Multi-wavelength datasets that do include 
such systems often only provide representative samples of this galaxy population 
\citep[e.g., SINGS;][]{kennicutt03}, and are thus not suitable for studies that both seek 
to probe the low metallicity dwarf galaxy regime and require datasets which are true to 
the statistics rendered by volume-limited sampling.  LVL consolidates and builds upon 
recent Local Volume galaxy surveys which have acquired ground-based narrowband 
H$\alpha$ \citep{kennicutt08}, GALEX ultraviolet (Lee et al., in prep.) and HST resolved 
stellar population imaging \citep{dalcanton09}, by collecting Spitzer IRAC and MIPS 
infrared imaging for a sample of 258 galaxies derived from these programs.  The collection 
of these observations enable a wealth of spatially-resolved and spatially-integrated 
studies probing present-day star formation, chemical abundance, stellar structure, and 
dust properties as well as galaxy evolution, particularly for metal-poor, low-mass galaxies 
which dominate the LVL sample by number.  The coupling of the infrared and ultraviolet 
data in this survey are explored in Section \ref{p:IRXbeta}.

\subsubsection{The Herschel ATLAS - unresolved SEDs}
\label{s:ATLAS}

\begin{figure}
\includegraphics[width=6cm,angle=-90]{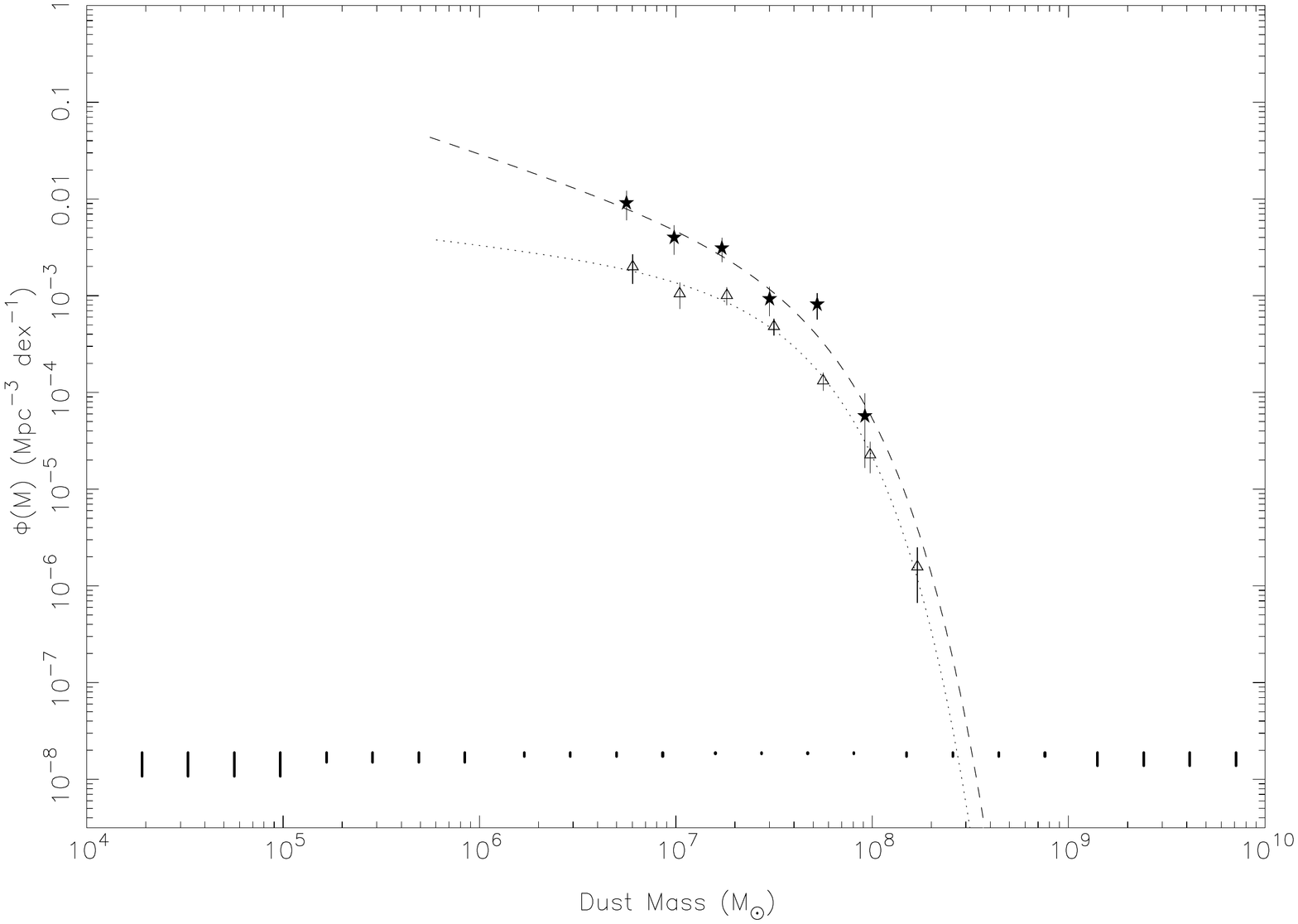}
\includegraphics[width=6cm,angle=-90]{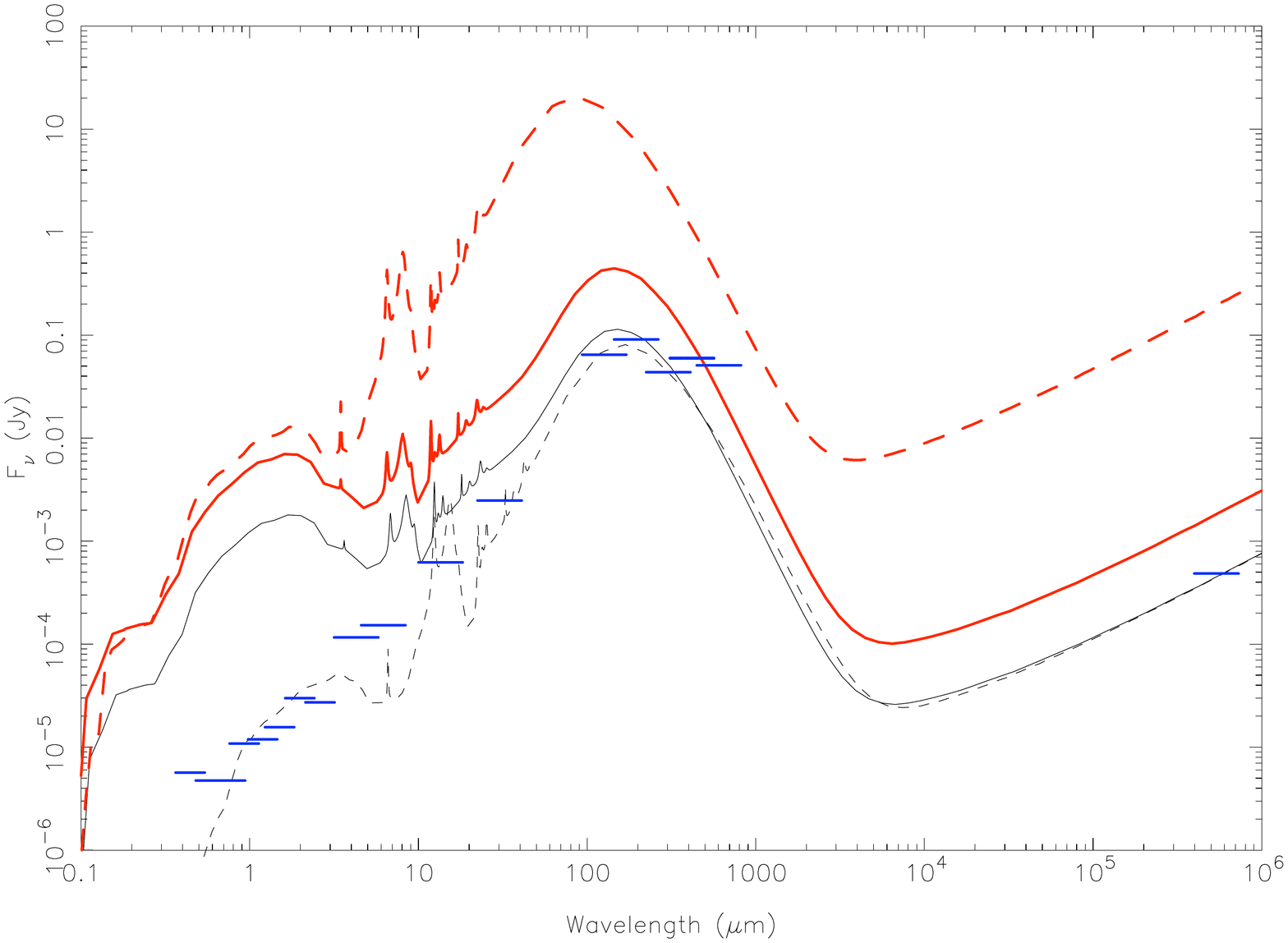}
\caption {Top: Recent estimates of the dust mass function from targeted observations 
of an IRAS sample \citep[triangles;][]{dunne00}) and an optical sample 
\citep[stars;][]{vlahakis05}.  The error bars at the bottom show the accuracy and range 
for the dust mass function that will be measured with ATLAS.  Bottom: Optical-radio 
SEDs from GRASIL \citep{silva98} for M~51 (solid)---a typical spiral, and Arp~220 
(dashed)---the archetypal ULIRG.  Thick red lines show the SEDs at $z=0.05$ 
where they have a similar optical/ultraviolet flux but very different infrared-radio 
SED. Optical / ultraviolet fluxes alone are therefore unreliable as a measure of a 
galaxy's star formation rate in dusty galaxies.  Thin black lines show each SED at the redshift where 
it is just detected at 250~$\mu$m.  This is $z\sim0.1$ for M51 and $z\sim1$ for 
Arp~220.  The far-infrared-radio SEDs are now identical, while the optical fluxes 
are very different.  Photometric redshifts from far-infrared/radio alone are totally 
unreliable without templates and optical / infrared fluxes to break the degeneracy 
between dust temperature and redshift.  Horizontal dashes show the 5$\sigma$ 
flux limits of SDSS/UKIDSS-LAS/WISE/ATLAS and LOFAR all-sky surveys.
[Courtesy L.~Dunne]}
\label{f:LDunne}
\end{figure}

ATLAS (Astrophysical Terahertz Large Area Survey) intends to provide the first 
unbiased survey of cool dust and 
obscured star formation in the local Universe.  ATLAS will detect $\sim 250,000$ 
sources with a median redshift of $\sim 1$ over $\sim 550$ square degrees of 
sky in five bands covering the wavelength range 
110-500~$\mu$m.  
The five photometric bands cover the peak of the dust SED for local galaxies and, 
crucially, can probe the cold dust component ($T\leq 20$~K) which makes up 
$\simeq 90$\% of the dust mass in most galaxies \citep{dunne01}. Herschel 
can simultaneously provide an accurate measure of the bolometric far-infrared 
luminosity (related to the current star formation rate) and the total mass of dust 
(related to the gas mass). Figure~\ref{f:LDunne} shows the improvements 
expected to be made to the 
measurements of the dust mass function and the need for a multi-wavelength 
approach to understanding the properties of galaxies.


The ATLAS will be {\em unbiased} as it does not rely 
on prior detection in other wavebands. Nevertheless, 
the ATLAS fields contain $>10^5$ redshifts and are the best studied 
fields of this size; they are the targets of surveys being carried out with GALEX, 
VST, VISTA, UKIRT and the South Pole Telescope and will be the natural targets of 
many future surveys, including ones carried out by DES, WISE, LOFAR and the two 
SKA precursor telescopes in the south.  



The greatest challenge for the Herschel ATLAS is likely to be the matching of 
the submillimeter sources to the correct optical / infrared counterparts.  At very 
low redshift this is fairly straightforward as the density on the sky of bright 
optical galaxies is low enough that associations are unlikely to be random.  
At intermediate--high redshifts the number of potential matches increases 
dramatically and the large Herschel beam (18--36\arcsec) means that correct 
identification cannot simply be a matter of probability and distance.  
This will require a multi-wavelength SED modeling code 
which can `join together' the ultraviolet / optical / near-infrared portion of the 
spectrum with the far-infrared/submillimeter `bump' in a self-consistent way.

\subsubsection{The SWIRE templates}

\citet{polletta07} published a set templates that combine the SEDs of 
galaxies with those of AGN and thus fill an important hole in many SED 
fitting template sets\footnote{The full library can be downloaded from:
http://www.iasf-milano.inaf.it/~polletta/templates/swire\_templates.html.}. 
The library contains 20 templates including 1 elliptical, 7 spirals, 3
starbursts, 6 AGNs, and 3 composite (starburst+AGN) templates covering the
wavelength range between 1000{\AA} and 1000 $\mu$m. The elliptical, spiral
and starburst templates were generated with the GRASIL
code~\citep{silva98}.
Templates of moderately luminous AGN, representing Seyfert 1.8 and Seyfert 2
galaxies, were obtained by combining models, broad-band photometric data, and 
ISO-PHT-S spectra of a random sample of 28 Seyfert galaxies. Four additional 
AGN templates represent optically-selected QSOs with different values of
infrared/optical flux ratios and one type 2 QSO. The QSO1 templates are 
derived by combining the SDSS quasar composite spectrum and rest-frame 
infrared data of a sample of SDSS/SWIRE quasars divided in three groups, all, 
and the 25\% brightest and 25\% weakest measurements per rest-frame  
wavelength bin. The type 2 QSO template (QSO2) represents the SED of the red 
quasar FIRST J013435.7$-$093102~\citep{gregg02}.
The composite (AGN+SB) templates are empirical templates that well reproduce 
the SEDs of  the following sources: the heavily obscured BAL QSO Mrk
231~\citep{berta05}, the Seyfert 2 galaxy IRAS 19254$-$7245
South~\citep{berta03}, and the Seyfert 2 galaxy IRAS
22491$-$1808~\citep{berta05}.

\begin{figure}
\plotone{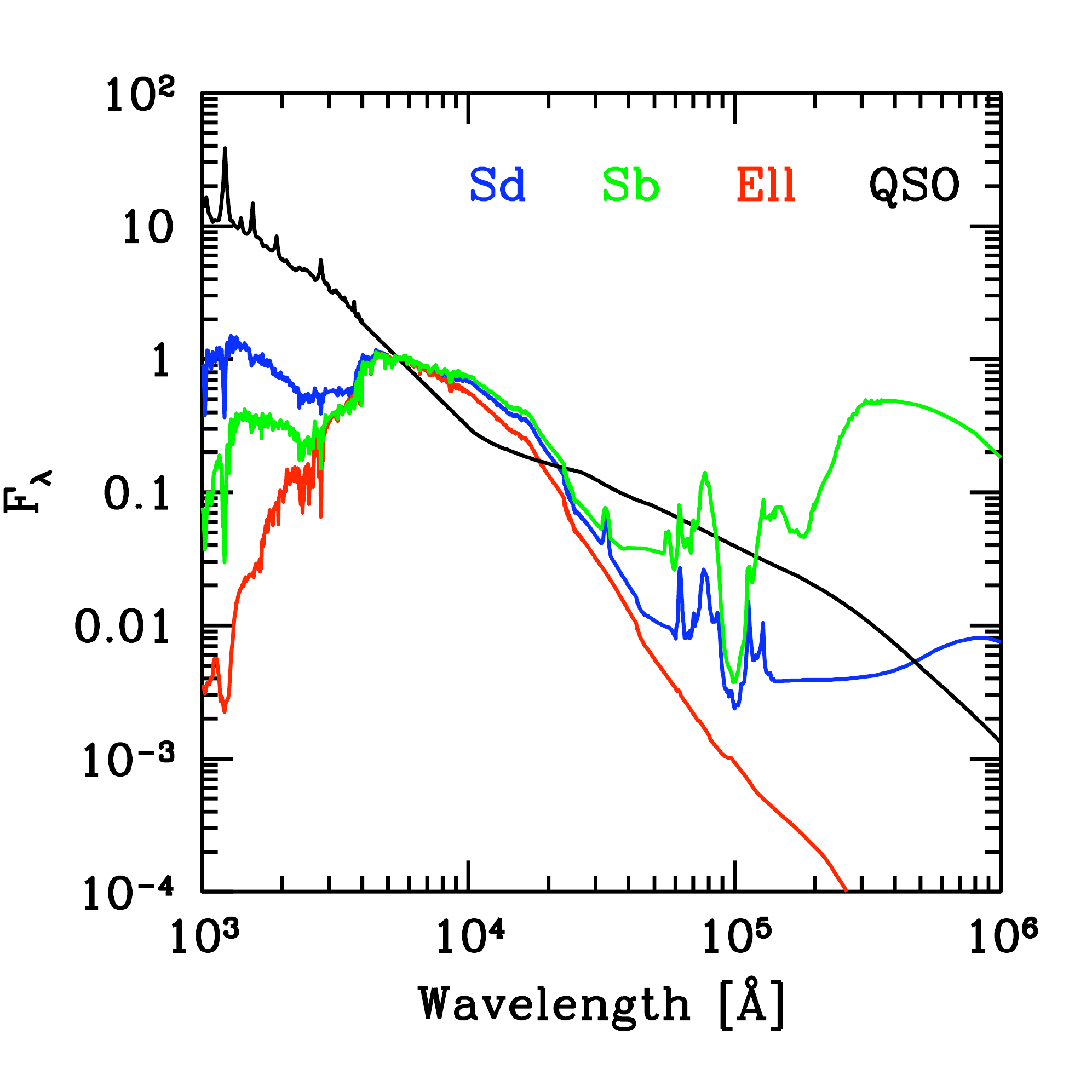} 
\caption {Four of the 20 templates from \citet{polletta07}, as an example of a set 
of empirically calibrated, multi-wavelength templates [Courtesy M. Polletta]. }
\label{f:polletta}
\end{figure}

\subsubsection{Further examples}

While we presented above only three examples, the number of public 
datasets (potentially) useful for SED fitting is truly staggering. Many science projects 
can be carried out without ever writing an observing proposal, the prime example being 
use of the SDSS database. This is possible thanks to the work of a countless number of 
individuals on survey planning, data reduction and quality insurance. Below we provide 
a table of some of the major datasets, for exploration and further reading. We will keep 
a similar table online on sedfitting.org and we hope to expand it in the future.

\begin{table*}
\begin{center}
\caption{Some major multi-wavelength databases}
\begin{tabular}{rcccccccc}
Acronym & Full name & Reference & Website \\
\hline
AEGIS & All-wavelength Extended Groth Strip International Survey & \citet{davis07} & aegis.ucolick.org \\
COMBO-17 & Classifying Objects by Medium-Band Observations & \citet{wolf03} & www.mpia.de/COMBO \\
COSMOS  & Cosmic Evolution Survey & \citet{scoville07}  & cosmos.astro.caltech.edu  \\
GAMA  & Galaxy And Mass Assembly  &  \citet{driver09} & www.eso.org/$\sim$jliske/gama \\
GOLDMINE & Galaxy On Line Database Milano Network & \citet{gavazzi03} & goldmine.mib.infn.it  \\
GOODS  & Great Observatories Origins Deep Survey & \citet{giavalisco04}  & www.stsci.edu/science/goods \\
LVL  & Local Volume Legacy & \citet{dale09} & www.ast.cam.ac.uk/research/lvls/ \\
MUSYC  & Multiwavelength Survey by Yale-Chile &  \citet{taylor09} & www.astro.yale.edu/MUSYC \\
SDSS  & Sloan Digital Sky Survey & \citet{abazajian09} & www.sdss.org \\
SINGS  & Spitzer Infrared Nearby Galaxies Survey & \citet{kennicutt03} & sings.stsci.edu \\
SSGSS  & Spitzer SDSS GALEX Spectroscopic Survey & \citet{treyer10} & www.astro.columbia.edu/ssgss \\
SWIRE  & Spitzer Wide-area InfraRed Extragalactic survey & \citet{lonsdale03} & swire.ipac.caltech.edu  \\
VVDS  & Vimos VLT Deep Survey  &  \citet{lefevre05} & cencosw.oamp.fr \\ 
\end{tabular}
\label{t:surveys}
\end{center}
\end{table*}

\section{Methods and validation of SED fitting}
\label{s:method}

In Section \ref{s:models} we have described how models \emph{predict} 
the SEDs of galaxies. Yet the ultimate goal of these models is to allow 
the reverse process, that is to derive the properties of galaxies from their observed 
SEDs through fitting procedures. Therefore we describe in this section
several of the main methods used to fit model or observed template
SEDs to multiwavelength observations of galaxies. Exactly which of
these procedures should 
be used is contingent on which physical parameters are required and 
with what precision. In turn, this, and the accuracy with which the models reproduce 
specific observations, determines what observational data are necessary.

The following sections deal almost exclusively with stellar emission. This reflects 
a real dearth of specialized codes for carrying out the fitting process in the dust emission 
range \citep[besides simple $\chi^2$ minimization over a parameter grid, e.g.~][]{klaas01}. 
This state of affair is understandable, however, in light of the rapid 
development and large uncertainties still inherent to our modelling of the dust emission 
of galaxies. 

\subsection{Parametrizing SED models }
\label{s:physprop}

In the context of a SED model, one must distinguish between independent, input parameters 
and derived parameters. Input parameters are those properties that are needed 
to define the SED model. Some of them may, first, not be measurable or, second, may 
not have a proper physical meaning by themselves. A good example for the first case is the 
shown in Figures 10 and 11 of \citet{dacunha08}, where the width of the probability 
distribution functions of different input parameters indicates which ones are well 
constrained, such as the global contribution of PAHs to the total luminosity, and which 
are not constrained, such as the temperature of warm dust in stellar birth clouds. 
\citet[][their Section 3]{iglesias07} state another example, namely that only a few of the many 
input parameters of the GRASIL code need to be varied in order to produce a library of model 
spectra that cover a realistic range in observational properties. Further, more systematic 
study of the relative importance of the input parameters to dust emission models would 
seem to be a important addition to the literature. 
A good example of the second case is the ``decay 
timescale'', $\tau_{*}$: in modelling the SEDs of early- and even
late-type galaxies,  the star formation histories (SFHs) are often
represented as falling exponential functions modulated by this timescale,
i.e.~SFR$(t)\propto e^{(-t/\tau_{*})}$. However, in a hierarchical
universe the SFH is likely to be more stochastic in form, modulated by discrete accretion 
events. Thus the width of the falling exponential that one determines from an 
SED fit is a good measure of the mean age of the galaxy. It should 
by no means be assumed that this represents the true SFH of the galaxy. The reason 
that falling exponentials provide reasonable fits in practice is simply that the spectral features of SSPs 
vary smoothly with time and there is thus considerable degeneracy between the mass 
contributions of SSPs of different age, effectively smoothing the SFHs of galaxies. 
This smoothing makes it also impossible in practice to robustly disentangle the epoch 
of the formation of the most luminous stellar population in a galaxy from the timescale 
over which this star formation took place. 

Input parameters constitute a discrete set with a maximum number of members 
determined by the quality of the data and the models. On the other hand the set 
of derived parameters can be extended at will and not all derived parameters 
can or need to be independent. For example, the commonly used specific 
star formation rate (sSFR$=SFR/M_{*}$) and the birthrate parameter
($b=SFR/ \langle SFR \rangle _{t}$) are derived parameters that are 
very similar. It is clearly relevant to properly define and understand the 
derived physical properties that one intends to measure. Let us again consider the 
SFR. While the SFR derived from the UV flux traces an average SFR over the 
last 100 Myr, the SFR determined from Balmer emission lines measures
the much shorter timescale of the ionizing stellar flux, $\lapprox10$
Myr. These two measures, while correlated, thus 
need to be properly corrected before being able to compare them 
\citep[see e.g.][and references therein]{kennicutt98,boquien07, kennicutt09,lee09a}. 

As an integrated model of the optical SED of a galaxy is based more or less explicitly 
on an equation similar to Equation \ref{e:f_ssp}, the independent and derived 
parameters should be defined in the same terms 
\citep[see e.g.][Table 1]{walcher08}. Actually writing down the definition helps 
avoid common confusions (such as those between metallicity Z and [Fe/H], or 
concerning the time scale over which the SFR is measured) and should be 
considered good practice.

\subsection{Spectral indices}
\label{s:indices}

It has been said in the introduction that SED fitting can only yield useful results if 
the models are as or more precise than the effect on the data of the property to 
be measured. Historically this was the case only for very limited wavelength 
ranges in the optical. The solution to this problem has then been to 
not fit the entire SED, but to define indices, i.e.~measure the equivalent widths, for 
certain absorption features. 

In the standard side-band method 
\citep[][]{burstein84,faber85} a careful analysis 
leads to the definition of one central and two side bands (a blue and a red). The 
continuum compared to which the equivalent width is measured is a linear 
interpolation between the average fluxes found in these two sidebands. 
Much work has gone into optimizing the set of available indices, as well in terms 
of coverage, as well as in terms of model precision 
\citep[e.g.][and references therein]{rose84,worthey94,trager98,tripicco95,cenarro02,thomas04,lee09b}.

We are still waiting for standard indices in other wavelengths than the optical, 
though see \citet{rix04}, \citet{keel06}, \citet{maraston09}, \& \citet{chavez09}, 
and \citet{lancon08} \& \citet{cesetti09} for headway into the UV and NIR 
respectively. These would make the index approach a true multi-wavelength 
approach. 

\begin{figure}
\plotone{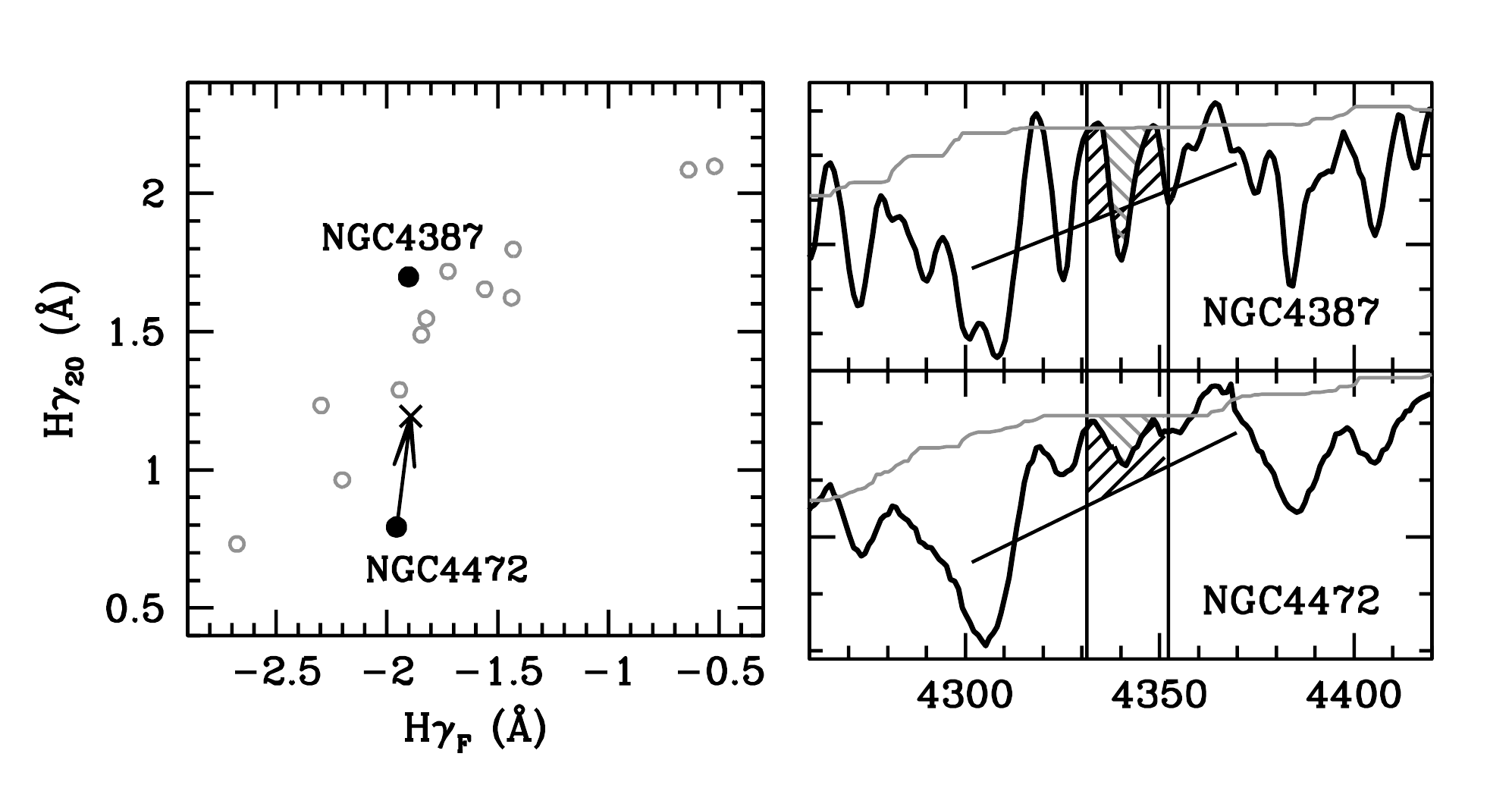} 
\caption {Comparison of the H$\gamma_F$ equivalent width measured with 
the side-band (SB) and the proposed Boosted Median method 
(H$\gamma_{20}$; BMC).  The black dots mark two galaxies with a significant 
difference between both methods. On the right-hand panels, the SED of these 
two galaxies is shown, with the slanted line representing the continuum defined 
by the SB method, and the gray line showing the BMC pseudo-continuum.  
The hatched black (gray) regions correspond to the equivalent width according 
to the side-band (BMC) method.  The arrow and cross in the left panel shows 
the change in equivalent width caused by the difference in velocity dispersion 
between NGC~4387 (112~km/s) and NGC~4472 (303~km/s). 
[Courtesy I. Ferreras]}
\label{f:IFerreras1}
\end{figure}

\citet{rogers10a} have attempted to improve on the classical sideband 
definition by introducing a 
``boosted median'' method. The main feature of this method is that for each 
side-band only the $L$th percentile of the distribution of fluxes is used to 
determine the flux within the side-band. As this procedure 
automatically chooses the largest fluxes in the side-band, it will define the 
pseudo-continuum from those points that are least affected by ``secondary'' 
absorption features, i.e.~absorption features that are not part of the central band. 
It thus improves the robustness of the pseudo-continuum if the 
spectral resolution is high enough to avoid blending of all features. 
Figure~\ref{f:IFerreras1} illustrates the difference between the boosted median 
equivalent width and a standard side-band method for a measurement of 
H$\gamma$.  The sample corresponds to a set of 14 elliptical galaxies in 
the Virgo cluster observed with a 2--2.4\AA\ resolution (FWHM) and with high S/N 
\citep{yamada06}. 

Index fitting can be considered a special case of SED fitting \citep[for a fitting code 
see e.g.][]{graves08}. It has the advantage 
of compressing the information available in galaxy spectra into a set of discrete 
numbers. Much of what is considered secure knowledge in stellar element 
abundances and age of integrated stellar populations is still largely based 
on fitting indices 
\citep[e.g.][to cite only a few]{trager00,kauffmann03a,thomas05, gallazzi08, graves09}.

The clear disadvantage of absorption line indices is that some of the 
information is lost. While indices 
have been defined with great care to approach the ideal of representing 
one element species, in practice many small lines interfere in particular in 
the side-bands. Spectral fitting (Section \ref{s:inversion}) 
can in principle use more information but it 
places much higher requirements on model accuracy.

\subsection{Principal Component Analysis }
\label{s:pca}

\begin{figure}
\includegraphics[width=0.99\hsize]{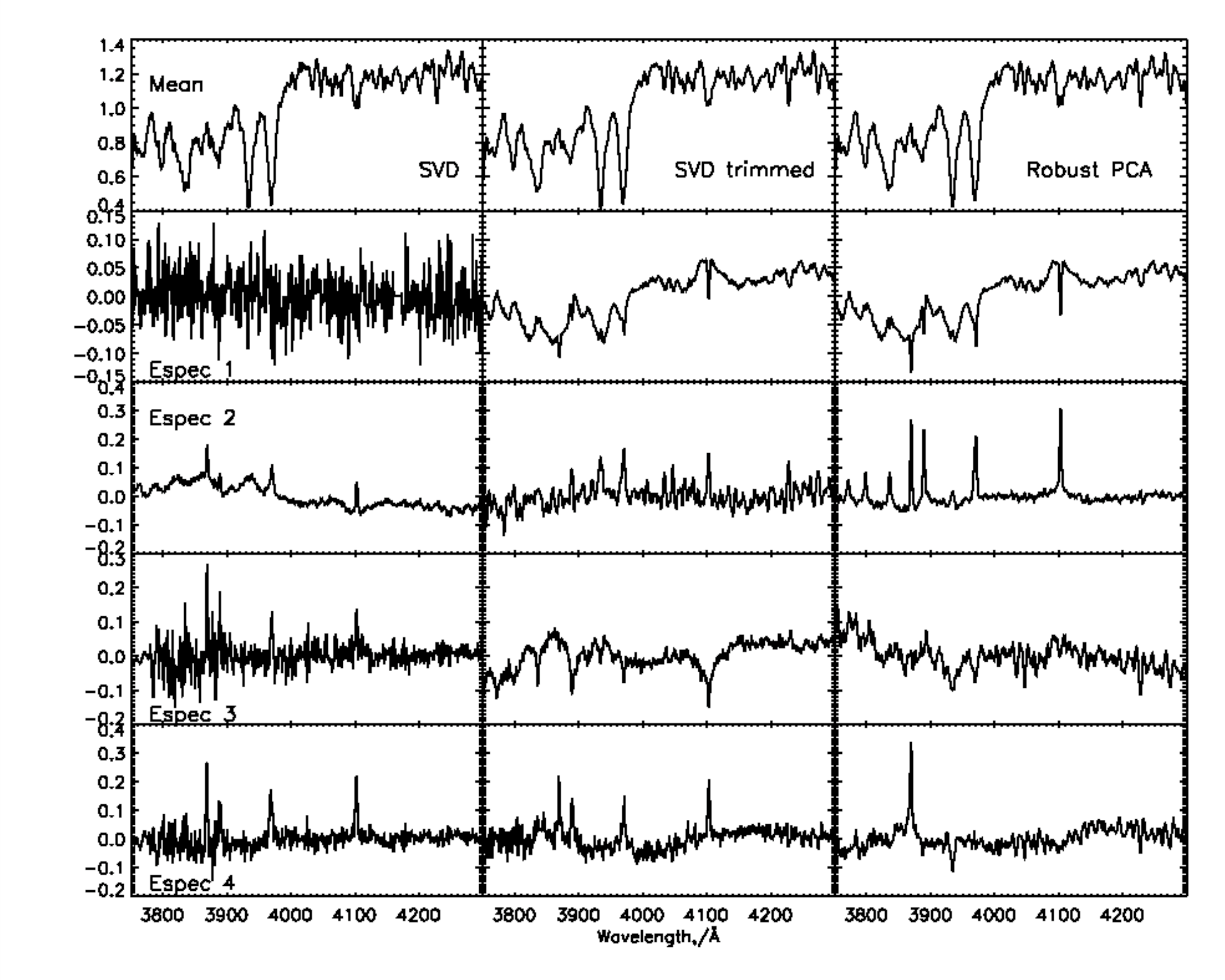}
\caption{Comparison between the eigenspectra from a PCA of 2000
randomly selected SDSS galaxies: {\it left,} using a simple SVD
algorithm, the first eigenspectrum is dominated by a single noisy galaxy
spectrum; {\it center,} using sigma clipping to remove outliers
iteratively from the dataset, the eigenspectra are visibly less noisy;
{\it right, } a robust PCA of the dataset, now the distinct patterns in
the dataset are strongly visible, in particular the separation of narrow
nebular emission lines from broader photospheric absorption features.
[Courtesy V.~Wild]}
\label{f:wild}
\end{figure}

Ideally we would like to represent a galaxy spectrum by a small set of 
continuous parameters that uniquely determine the best-fit spectrum.
Principal Component Analysis (PCA) is one algorithm commonly used to
derive an optimal set of linear components, diagonalising the
covariance matrix of the data points to find the directions of
greatest variation. Its representation of data through a linear combination 
of independent (orthogonal) components, or eigenvectors is thus an 
alternative method to using a set of discrete SSP templates 
(Section \ref{s:inversion}). Since the convolution with transmission curves is a
linear operation, these methods are as simple as solving a linear
equation, even for photometric datasets \citep{connolly99,budavari00}.

PCA has been successfully applied to astronomical
spectral datasets, although not yet to photometric datasets which
suffer the additional complication of observed-to-rest frame
translation \citep{connolly95b}. The main difficulty with PCA is that the  
interpretation of the empirically determined PC components in 
terms of physical properties is complex at best 
\citep[though see][]{wild07, wild09, rogers07, rogers10b}. 
This is exacerbated by its sensitivity to
outliers and hence out-of-the-box algorithms are of limited use for
astronomical datasets.

Recent work by \citet{budavari09b} solves the problem of reliable
eigenspectra determination by an iterative procedure that
is efficient to compute and robust in the statistical sense.
Figure~\ref{f:wild} illustrates the comparison of three PCA methods
on the blue optical region of 2000 randomly selected SDSS galaxy
spectra.  Each column contains (from left to right) the results of the
classic PCA, classic PCA using iterative sigma clipping of the
dataset, and the new robust algorithm.
What is immediately striking in the robust case is the following: (a)
The very clean appearance of the nebular emission lines in the second
eigenspectrum.  (b) These are correlated with the weaker Balmer
absorption (seen as broad wings on the narrow emission lines) and rise
in the blue from the continua of O and B stars which are the
dominant source of ionisation of the nebular emission lines.  (c) The
emission line in the 4th eigenspectrum is the only one in this
wavelength range which is not a Balmer line, and with a higher
ionisation state is sometimes attributed to the presence of an AGN.
Without any prior physical knowledge, the robust PCA has separated out
a line which is physically distinct from the others, and tied together
HII emission lines with the O and B stars that excite them. The
results are clearly of more use for characterising the galaxy
population than traditional PCA algorithms. The robust PCA algorithm
provides a new, fast and easy to use method for the investigation of
real astronomical datasets in a model independent manner. 


\subsection{Spectral fitting by inversion}
\label{s:inversion}

As has been mentioned in Section \ref{s:ssp}, the stellar SED of a galaxy can 
be represented by a sum over the SEDs of individual SSPs with appropriate 
weights which reflect the SFH. As long as any complications, such as dust, can 
be neglected this is a linear problem, i.e.~a matrix inversion. More generally, 
inversion is the attempt to invert the observed galaxy SED onto 
a basis of independent components (SSPs, dust components) drawn from a SED model. 
Inversion is typically used for spectral data and a big success 
of modern stellar population models and inversion codes is that we can 
now fit the models to data to better than 5\% in the optical wavelength range 
(see Section \ref{s:invervalid}).

\subsubsection{Method}

Nearly all inversion methods start with assumptions that reduce the complex physics of 
SEDs (see Eq \ref{e:f_ssp}) to a problem that can be written as a linear function 
of its parameters. Such assumptions are typically that one deals 
with a stellar system in which all generations of stars have the same metallicity, 
i.e.~$\zeta(t) = Z^0$ and the same attenuation, i.e.~$T(t)=T^0$.
The problem of solving for the star formation history of a stellar 
system is then equivalent to defining and minimizing the merit function 
\begin{equation}
 \chi^2 = \sum_{i=0}^{n} \left[ \frac{F_i - \sum_{i=1}^{M} a_k S_{i}[t_k,Z^0,T^0]}{\sigma_i} \right]^2, 
\label{eq:merit}
\end{equation}
over all non-negative $a_k$. Here $F_i$ is the observed spectrum in 
each of $n$ wavelength bins $i$, $\sigma_i$ is the standard deviation 
and $a_k$ are weights attributed to each of $M$ SSP 
models $S_{i}[t_k,Z]$ of age $t_k$ and metallicity $Z$. 
This merit function is linear in the $a_k$ and can thus be solved by standard 
mathematical methods involving singular value decomposition (e.g.~non-negative 
least squares, bounded least squares)\footnote{It needs to be emphasized here 
that a correct solution involves a simultaneous fit for the velocity dispersion of 
a galaxy. This is beyond the scope of the present review though.}. 
A particular advantage of the inversion method is that, besides the 
assumptions necessary to linearize the problem, no parametrization of the 
solution is necessary, in particular not of the star formation history. 
Codes that have been used for scientific analysis are now common and 
include MOPED \citep{heavens00}, PLATEFIT \citep{tremonti04}, VESPA \citep{tojeiro07}, 
STECKMAP \citep{ocvirk06}, STARLIGHT \citep{cid-fernandes05}, 
{\tt sedfit} \citep{walcher06}, NBURSTS \citep{chilingarian07}, 
ULySS \citep{koleva09}. 
Different codes give very comparable results \citep[e.g.][]{koleva08}.

There is one problem that has to be addressed when assessing the 
unparameterized information content 
of galaxy spectra. All methods relying on singular value decomposition 
and derived algorithms inherit one of the features, which is that the method 
tends to search for the smallest number of templates it can use to fit the data. 
For galaxies, which presumably have a smooth SFH, this feature is a grave caveat 
concerning the significance of the recovered weights of each population. In particular,
the recovery of realistic error bars from a simple bootstrap algorithm is not possible, 
as the method will tend to always choose some templates over others, inside a range 
of random errors. One way to address this issue is regularization, detailed in 
\citet[][the STECKMAP program]{ocvirk06}. 

Another robust exploration of this issue is provided in the code {\tt
VESPA} \citep[described in detail in][]{tojeiro07}. As independent 
parameters (see \ref{s:physprop}) Tojeiro et al.~chose 2 values 
of metallicity and a logarithmic binning in age that can be varied between coarse 
(3 age bins) and fine (16 age bins). Dust attenuation and varying metallicity are 
explored by repeating the fit with VESPA over a grid of parameters. Error bars 
are derived by creating noised representations of the input spectra and repeating 
the fit $n$ times. The codes main feature 
in the present context is that it explores the SFHs in an iterative process that goes 
from coarse to fine resolution. It uses the method described in \citet{ocvirk06} 
to estimate at each step, how many parameters can be recovered for a linear 
problem perturbed by noise. The best fit will thus only 
use as many independent stellar populations as required by the data. Nevertheless, 
one caveat remains, which is that the SFH inside each bin is fixed to be either a 
constant SFR or such that the light contribution of each age is more or less constant. 
When recovering parameters such as M/L and in those cases where the age bins are 
coarse, this will lead to an underestimate of the true uncertainty in this parameter, 
as compared to a truly non-parametric method. 

The main lesson learned from this careful analysis is that from spectra  typical of the 
SDSS (S/N$\approx$20, optical wavelength range), one can recover between 2 and 4 stellar 
populations described by an age and a metallicity. These values can be improved 
by a factor of 2 when going either from S/N=20 to S/N=50 or when enlarging the 
wavelength range to include not only the optical, but also the UV range. Such 
data will be available in the near future from the XSHOOTER instrument at the VLT. 
It should be kept in mind, however, that while the formal constraints may be better, 
the accuracy of the results will then depend critically on the data calibration and model 
accuracy over a very large wavelength range. 

\subsubsection{Non-linear inversion codes}

For the optical wavelength range, an exception to the pre-condition of linearity exists 
in the form of the code STARLIGHT \citep{cid-fernandes05}, which is based on simulated 
annealing and thus does not need linearity. Consistency between STARLIGHT and other 
codes based on linearity assumptions has been found. Another code avoiding linearity 
conditions was presented in \citet{richards09}.

For decomposing MIR spectra into their stellar, PAH, dust continuum, and line emission 
constituents, Levenberg-Marquardt algorithms have been used \citep[][see Section 
\ref{s:pahs}] {smith07, marshall07}. This allows among other things to compute 
uncertainties on the derived parameters. With the improvement of dust emission models 
and in particular the more widespread availability of spectral data, more development 
in the technicalities of fitting dust spectra can be expected. 

Finally, a possible extension of the inversion onto SSP models has been 
presented in \citet{nolan06} and would merit further exploration. 
Also \citet{dye08} has presented a new method, which integrates 
inversion into a Bayesian framework and applies it to photometric 
data.

\subsubsection{Non-linear physics}
\label{s:nonlinear}

Dust attenuation is important in the optical and UV wavelength ranges. In the context 
of inversion, the importance of dust is most easily assessed by an iteration 
over a grid of attenuation values. More sophisticated schemes proceed in an 
iterative way, i.e.~alternate linear inversion and a non-linear minimization scheme 
\citep{koleva09}, or use non-linear minimization schemes for all parameters 
\citep{cid-fernandes05}. Any estimate of the dust attenuation at optical wavelengths 
from the SED alone (i.e.~using neither Balmer decrement, nor UV photometry, nor 
dust emission) is bound to be uncertain, in particular because no inversion code 
at present uses a physical model for different attenuations likely experienced 
by young and old stars. 

A very useful way is therefore to normalize the continuum of the observed and model 
spectra before the fit \citep[e.g.][]{wolf07,walcher09} or even simultaneously with the fit 
\cite[][]{koleva09}. While this takes out any information related to the attenuation, it 
also allows the freedom from any uncertainties related to 
continuum calibration and can be thought of as fitting the equivalent widths of 
the absorption features only. An often under-appreciated caveat is, however, that this 
procedure throws away some information. In particular (nearly) featureless continua 
(such as from very young stars) become undetectable, yet still affect the equivalent widths. This 
``featureless'' continuum has to be appreciated in relation to the
noise in the observations and, in an intermediate age spectrum can 
even include a significant contribution of old stars. 
One expression of this caveat is described in Section \ref{s:methodcav} as the bias 
occurring when fitting single SSP models to extended SFHs. 

Also the effects of forward scattering on the optical SED are not included in 
Equation \ref{e:f_ssp} and cannot be linearized usefully. In principle the scattering 
can be modelled through radiation transfer modelling and could thus be included in 
non-linear minimization schemes. However, radiation transfer modelling is too 
time-consuming to make this a practical possibility.

\subsubsection{Validation}
\label{s:invervalid}

\begin{figure}
\includegraphics[angle=90,width=1\hsize]{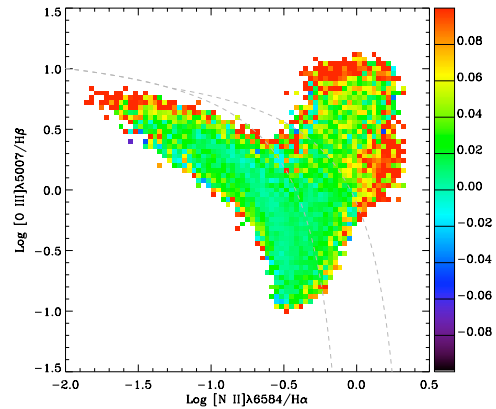}
\caption{Classic BPT diagram of [N{\sc II}]/H$\alpha$ versus[O{\sc
    III}]/H$\beta$, showing the distribution of SDSS emission line
  galaxies. The colours indicate the median $\log\chi^2$ of the
fit of the CB08 model library to the SDSS fibre spectra with those
emission line ratios. The upper and lower dashed lines indicate the
\citet{kewley01} and \citet{kauffmann03b} AGN dividing lines respectively.
[Courtesy J. Brinchmann]}
\label{f:chi2_bpt}
\end{figure}

In an unpublished contribution to the workshop, Jarle Brinchmann (and collaborators) 
showed that the $10^6$ spectra of galaxies in the Sloan Digital Sky Survey (SDSS)
offer a rich set of tests of the current models and techniques for fitting SEDs in 
the optical region. For the first series of tests 
they use a set of 274613 galaxies from the SDSS DR7 with reliable determinations 
of the emission line fluxes. They fit them using the latest version of the \citet{bruzual03} 
models (termed here CB08). In the optical the main improvement in these models 
is the switch from the Stelib library to the MILES library. The fitting procedure involves 
classical inversion using a non-negative least squares routine. Additionally they allow 
for $\sim 150$ {\AA} wide mismatches between models and spectrum using a smooth 
final correction. This is necessary because of uncertainties in the spectrophotometry 
of the data and the models. The result of the NNLS routine is, among others, the mean 
$\chi^2$ of the spectrum, which can in turn be plotted into the
classic \citet[][BPT]{baldwin81} diagram 
[N{\sc II}]$\lambda$6584\AA/H$\alpha$ versus
[O{\sc III}]$\lambda$5007\AA/H$\beta$, shown in Figure \ref{f:chi2_bpt}. 
Generally, the fit is good, with a reduced $\chi^2$ around 
$1\pm0.1$. Regions can be identified, where the fit is, while still in this range, 
somewhat worse. These are (A) galaxies with high star formation activity, 
high excitation and low metallicity, (B) strong, high excitation AGN, and (C) galaxies 
with high surface mass density (i.e.~high mass) and weak emission lines. 

For cases (A) and (B), i.e., the low metallicity starbursts and the high-ionization AGN, 
they relate the main increase in $\chi^2$ to the imperfect masking of some emission lines 
and to the lack of nebular or AGN continua in the spectra. In addition the BC03 models 
have some problems with hot stars \citep[e.g.][]{wild07} - the preliminary version CB08 
appears to work better, due to a better coverage of hot stars in the spectral library 
(G. Bruzual, priv. comm.). 

For case (C) one of the largest mismatches in the optical is located, as expected, around 
the Mg features at 5100 {\AA}. This is the region where the enhancement of the 
abundances of the $\alpha$ elements is known to play an important role. 
$\alpha$-enhancement is not covered by the CB08 models. 
Nevertheless, in the optical the mismatch due to stellar populations is modest, 
of the order of 0.02 magnitudes. 

\begin{figure}
\includegraphics[width=\hsize]{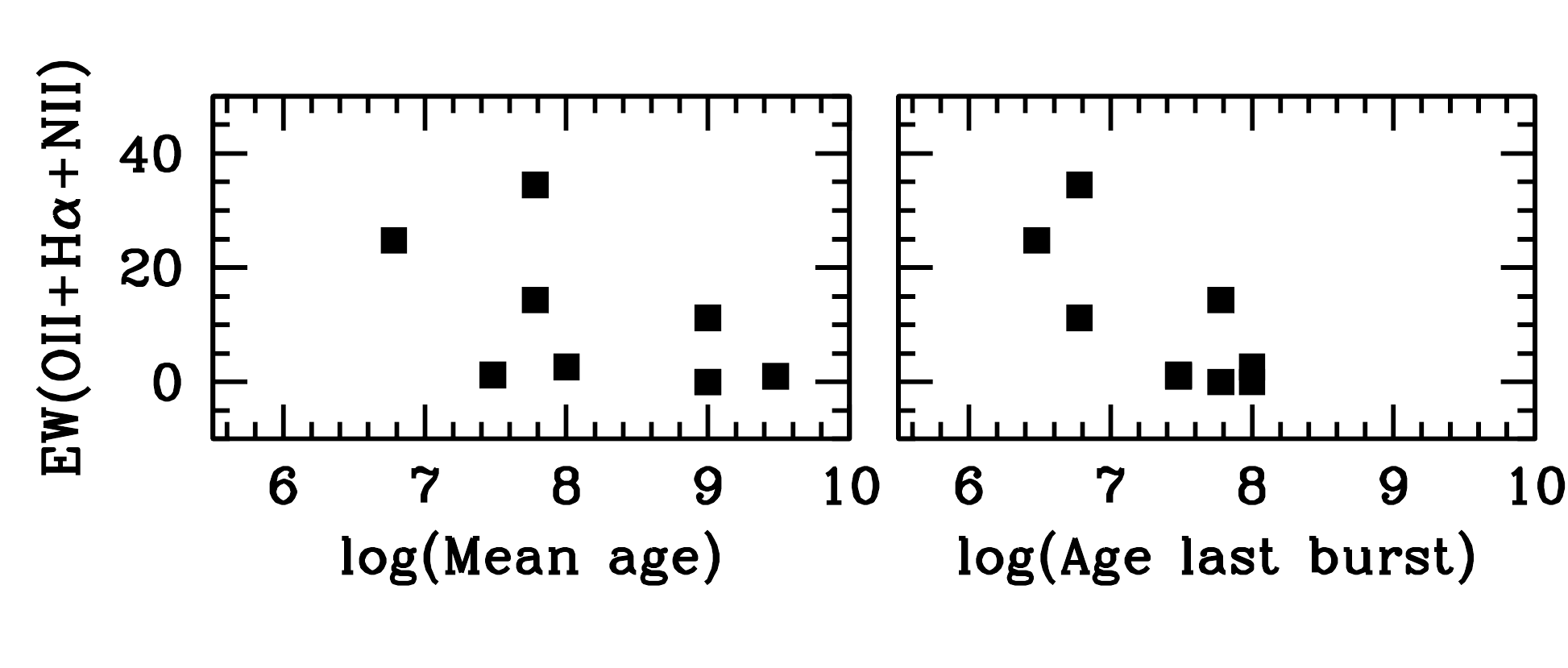}
\caption{Comparison between the equivalent widths of emission lines and 
the mean age (left panel) or the age of the last burst (right panel) as determined 
from spectral fitting for a sample of nine nuclear clusters from \citet{walcher06}. The clear 
correlation is evidence that the youngest age picked up by the spectral fit does 
bear significance. }
\label{f:ncageemlin}
\end{figure}

The best method to validate a fitting procedure is to compare the result to independent 
measurements of the same physical property. This can be done for
redshifts, as discussed in section \ref{s:photoz}, but also for
the spectra of nuclear clusters, as done by \citet{walcher06} for nine
nuclear clusters. The nuclear clusters have multi-aged 
stellar populations and are as such similar to entire galaxies, and 
their total stellar masses can be determined dynamically, 
independent from fitting the stellar spectra.. As shown in 
\citet{walcher06}, the stellar M/L ratios from inversion agree with those determined from 
independent dynamical modelling, albeit with a large scatter. This scatter reflects the general 
difficulty of constraining the oldest stellar population from SED
fitting. Additionally, \citet{walcher06} also found that the age of the
youngest population as determined from the spectral fitting  
is correlated to the measured emission line equivalent width, as shown
in figure \ref{f:ncageemlin} (\emph{right}). On the other hand, the
mean \emph{mass-weighted} age 
is not correlated, which is evidence that the fit result is not heavily biased towards the 
youngest population (in contrast to results from fitting single SSPs, see section \ref{s:validssp}).

Another careful work on validation of the spectral fitting procedures
was the work by \citet{koleva08}, who looked at Galactic clusters
which had individual stellar spectra and full colour-magnitude
diagrams to compare with the results of the fits to the integrated
spectra (also compare this with section \ref{s:validssp} for the validation of the 
predictions of SSP models for integrated spectra).

\subsection{Bayesian inference}
\label{s:bayesian}

We refer here in particular to using the ``Bayesian fitting'' method, in which 
multi-wavelength SEDs are fit by first pre-computing a discrete set or
library of 
model SEDs with varying degrees of complexity and afterwards determining the 
model SED and/or model parameters that best fit the data. This approach is favoured in particular for 
multi-wavelength data, as the problem of solving for the physical parameters 
is not linear if effects such as dust attenuation, line emission, and dust emission 
have to be taken into account. In addition, the influence of geometry and the physical conditions 
of the dust generally make the number of parameters very large, and with many possible degeneracies, 
thus making non-linear minimization schemes usually impractical.

\subsubsection{Method}

It is important to realise that the simple fact of \emph{pre-computing} a set of galaxy models 
and afterwards determining the one with the lowest $\chi^2$ is by itself a Bayesian approach. 
By choosing which models to compute one introduces a prior, which,
while possibly flat in terms of a given parameter, assumes that the
data can be represented by that model and that parameter space. By using $\chi^2$ as a 
maximum likelihood estimator one finds the most probable model, or in
bayesian terms, the probability of the data given that model. A mathematically much 
more rigorous description of the method is given in Appendix A of \citet{kauffmann03a} and currently 
builds the basis for Bayesian inference in SED fitting. A more practically oriented description 
can be found in e.g.~section 4 of \citet{salim07}. In a nutshell, the method uses the fact 
that --- assuming Gaussian uncertainties --- the probability of the data (D) given the 
model (M) is given by P(D$|$M) $ \propto e^{-\chi^2/2}$. 
The prior probability distribution of the model parameters is often taken to be flat within a given
parameter range (or flat in log space for larger parameter ranges).
To determine the  parameters for a specific galaxy dataset, the probabilities for all models are computed and 
then integrated over all parameters (model space) except the one to be derived, which yields a Probability 
Distribution Function (PDF). The median and width of this distribution yield the parameter estimate 
and associated uncertainties. The integration can be repeated and the
PDF can be built for all parameters of the models as well as the observables themselves, 
e.g.~with the aim to plot the model and observed
fluxes in a fitted band for comparison purposes.

An important step in Bayesian inference for SED fitting is the
construction of the library and the prior assumptions (Section \ref{s:lib}). 
It is equally important to synthesize the correct observables from the models, taking 
into account response functions and redshifts (Section \ref{s:response}).

Bayesian inference for SED fitting has several advantages that minimize potential sources 
of uncertainties: i) all available measurements 
contribute to the fit result, ii) the $K$-correction is integrated,
iii) non-linear effects such as from dust are accounted 
for as part of the models (see section \ref{s:isdust}), and thus in a
self-consistent way, and finally iv) uncertainties on the derived  
parameters include measurement uncertainties as well as intrinsic degeneracies. The big 
caveats concerning this method are the sensitivity to the prior
distribution of parameters (the ``library'')
and, connected to this, the dependence on realistic input physics in the modelling.

\subsubsection{Libraries and priors}
\label{s:lib}

The computation of the library is one of the essential steps when using Bayesian inference. 
It encodes our prior knowledge about the galaxies in the sample, but also our assumptions 
about which of the physical effects we can safely neglect. Concerning photometric SEDs, a 
prominent case of neglection is the internal chemical evolution of the model galaxies. This is 
justified because even the overall metallicity is not measurable from broad-band photometry 
alone \citep[e.g.~][]{walcher08}. Also the effects of forward scattering on the optical SED are 
often not included for computational reasons, but could potentially be important, in particular 
for starbursts \citep{jonsson06}.

One assumption that 
is generally used is that SFRs tend to decrease monotonically with cosmic time in a given 
object (with the exception of star formation bursts). This assumption is particularly appropriate 
for early-type galaxies, which has led to the wide-spread use of
so-called $\tau$-models, i.e.~exponentially falling functions
representing the SFH. However, it is not to be expected that  
galaxies in reality follow the falling exponential model. The SFHs of dwarf galaxies are 
expected to be dominated by intermittent bursts \citep[see e.g.][]{gerola80}. But even 
for large galaxies, the SFH is not expected to be as smooth as suggested by the simple 
$\tau$-models. Both semi-analytic \citep{quillen08} and hydrodynamic modelling
suggest some stochasticity, as discussed in section \ref{s:physprop}. However, 
due to the smooth variation of SED properties with age and 
due to the ensuing degeneracies, the $\tau$-models provide reasonable
fits to observed SEDs and cannot be easily refuted. Also, they 
provide a convenient, simple parametrization of SFHs which explains their widespread use. 

The danger of using only $\tau$-models lies in the neglect of degeneracies. Let us assume, 
the $\chi^2$ of a specific $\tau$-model is the same as that of a different model, which is the 
superposition of a smooth SFH and a burst. Both models will have different mean ages and 
M/L ratios. If they are both included in the library, they will thus lead to a larger error bar. 
However, if the burst model is omitted, two situations are possible: 1) If we indeed know that 
strong secondary bursts are unlikely (as e.g.~for a sample of morphologically selected 
early-type galaxies) one will obtain a
\emph{better} estimate of the actual error bar through use of a better prior. 2) If on the other 
hand one is fitting the entire population of galaxies in a given volume, it is unlikely that no 
galaxy shows a secondary burst and one will thus --- sometimes severely --- underestimate 
the true uncertainty of the recovered parameters. Often ``new'', supposedly 
more precise estimates of galaxy parameters are published that are based on too 
restrictive priors. In the following we showcase two examples of how to construct a library 
of star formation histories.

\paragraph{A library for Gaia}

One of the first cases where Bayesian inference is applied to spectra will be presented in 
Tsalmantza, et al., (in prep). This is also a good example of how to create a library with 
a suitable prior for a specific application. The Gaia mission will produce spectra of $~10^6$ 
galaxies at low resolution of R between 50 and 200. Automatic classification of a large number 
of low-resolution spectra is therefore crucial. Classification is done by comparison 
to a custom-created library of model galaxy spectra that simulate the properties of the 
Gaia spectra. This library is used for training and testing of the classification and 
parametrization algorithms. 

Two libraries have been created: The first library is an entirely synthetic one. The library 
creation starts from the synthetic spectra of 8 typical galaxies produced by PEGASE.2 
\citep{fioc99}. The SFHs of the library spectra are built from analytic prescriptions 
for either the dependence of SFR on the gas mass (SFR = $M_{gas}^{p1}/p2$, late types) 
or for the SFH directly (SFR=$p2.e^{-t/p1}/p1$, early types) as well as parameters 
for gas infall and galactic winds. In order to create a smooth grid,
the input parameters (i.e.~$p1, p2$, gas infall and galactic winds) were 
allowed to vary over a range of values\citep[see][]{tsalmantza07, tsalmantza09}. 
At this point, while of course guided 
by prior knowledge of the properties of galaxies, the parameter distribution is chosen 
nearly ad-hoc. 

A second library was created based on spectra from the SDSS \citep[DR5,][]{adelman07}. 
The SDSS spectra were shifted to z=0 and degraded to the resolution
of the PEGASE.2 output spectra. They were then classified by comparison to 
the pre-existing library of purely 
synthetic spectra through  simple $\chi^2$-minimization. Using the
best fit model spectrum the SDSS library spectra could be extended to cover the Gaia 
wavelength range and values for the most significant parameters could be assigned 
to them. Mainly, however, the results of the $\chi^2$ fitting were 
used to judge how the synthetic library compares to the empirical galaxy population in 
the SDSS. The distribution of input parameters governing the SFHs obtained from the 
observed SDSS sample can then be used to determine a suitable prior distribution 
for the synthetic Gaia library. This application is a good example of a case where the 
prior can be well established based on other observations.

\paragraph{Simplifying M/L determinations}

A good example of how the appropriate choice of prior can help with constraining the 
fit is provided by the work of \citet[][and R. de Jong's talk at workshop]{bell01}. 
This work centers on finding the simplest way to derive the mass-to-light ratio 
of a stellar population. Let us first remember that the stellar mass itself is 
the product of a galaxies luminosity and its $M/L$. $M/L$ in turn is a function of the 
star formation history of a galaxy. If the SFR at time t is written as $\Psi(t) = dM(t)/dt$ 
and is defined over the entire Hubble time $T_h$, $L_{\nu}(t)$ is the luminosity of an 
SSP at frequency $\nu$, at age $t$, and with mass $dM(t)$ and $T_{\nu}(t)$ is the mean 
transmission of the ISM at wavelength $\nu$ and for the SSP with 
age $t$, then we can write 
\begin{equation}
M/L_{\nu} = \frac{ \int_{0}^{T_h} dt \Psi(t) } { \int_{0}^{T_h} dt L_{\nu}(t) T_{\nu}(t)}
\end{equation}
This dependence of the $M/L$ on the SFH means that any prior assumption on the 
SFH is also a prior assumption on the $M/L$. 

The crucial simplification inherent in \citet{bell01} is thus the parametrization of the SFH. 
This simplification is provided by a prior derived from a 
hierarchichal formation scenario. Using their scenario, the SFHs of galaxies 
lead to a situation in which their intrinsic colours show a strong correlation with 
M/L. The dust attenuation vector is parallel to this relation, and thus the M/L 
ratio can be derived from a single optical colour with reasonable precision. 
Nevertheless there is a curvature in the difference between the M/L ratios determined 
from colours only and M/L ratios determined from an SED fit in the sense 
that for very blue (young) galaxies and for very old (red) galaxies, the colour-M/L 
will provide an overestimate, while for intermediate colour, intermediate age 
galaxies the colour-M/L will be an underestimate. 

De Jong \& Bell (talk at workshop) explore different possible effects. When comparing 
a smooth SFH with a single burst, the M/L values have large systematic offsets, due 
to the lack of faint, old stellar populations in the single burst model. Two burst models 
do not solve the problem, in particular if the fit is dominated by two young populations. 
In this case the two-burst model is still not complex enough, it again misses the 
faint, old population, which is crucial for the mass budget. The problem with the 
SFH template is further clarified by a comparison between M/L ratios as determined 
in \citet{bell03a} and those derived by k-correct \citep{blanton07}. While the scatter in 
the difference is relatively small and there is no systematic trend, there is a systematic 
offset of about 0.15dex. This can traced to the parametrization of the SFH. Indeed, 
while \citet{bell03a} use pure falling exponentials, the SFH assumed in in k-correct 
peaks 6 Gyr after the beginning of star formation. Thus, while the resulting SED is not 
affected, the number of low-mass stars is much higher in the \citet{bell03a} SFHs. 
This uncertainty of 0.15 dex is thus inherent to \emph{any} mass estimate from 
SED fitting.

\subsubsection{Validation}
\label{s:bayesvalid}

In order to validate the method and its intrinsic accuracy, a number of studies have 
been performed in the last years. What could be termed ``internal validation'' is the search 
for degeneracies and systematics in the models and the method itself 
\citep[see e.g.][]{walcher08, wuyts09,longhetti09}.

Extensive testing of stellar mass determinations through SED fitting was done with the 
COSMOS dataset \citep{ilbert10}. SEDs were generated with the stellar population 
synthesis package of \citet{bruzual03} assuming a \citet{chabrier03} initial mass 
function (IMF) and an exponentially declining star formation history
with a timescale, $\tau_{*}$, 
ranging from 0.1 to 30~Gyr.  The SEDs were generated on a grid of 51 timesteps 
between 0.1 and 13.5~Gyr.  Dust attenuation was simulated using the \citet{calzetti00} 
law with $E(B-V)$ in the range 0 to 0.5, with steps of 0.1.  The colors 
predicted by this library of SEDs in the COSMOS dataset for the photometric redshift 
of the galaxy were compared to the observed ones to select the best-fitting templates.

\begin{figure}
\includegraphics[width=\hsize,angle=0]{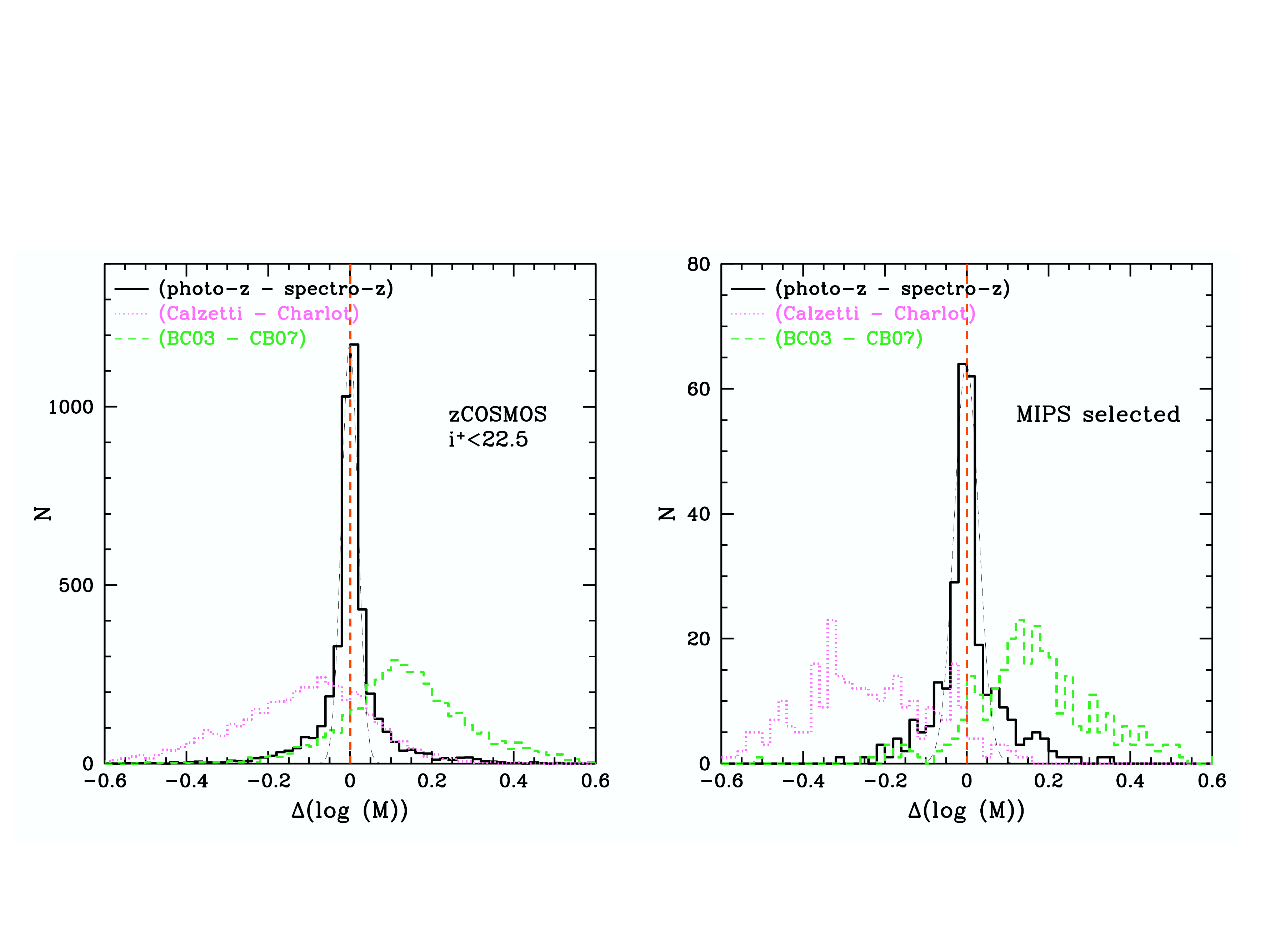}
\caption {Histograms of the difference between two stellar mass determinations, 
for the zCOSMOS bright sample selected at $i^{+}_{AB} < 22.5$ (right) and for a 
sample of sources selected at 24~$\mu$m in the S-COSMOS MIPS sample (right).  
The black histograms show the difference in stellar masses when using the 
photometric or the spectroscopic redshifts.  The thin black dashed lines are 
gaussian distributions with $\sigma = 0.02$ (left panel) and $\sigma = 0.03$ 
(right panel).  The green dashed lines show the differences between the stellar 
masses computed with \citet{bruzual03} and Charlot \& Bruzual (2007, private 
communication). The magenta dotted lines show the differences between the 
stellar masses computed using the \citet{calzetti00} and \citet{charlot00} 
attenuation laws.  The redshifts were set to the spectro-$z$ values in the two last 
cases.  Systematic uncertainties due to the models dominate the errors introduced 
by the photo-$z$, at least in the magnitude/redshift range explored with our 
spectroscopic samples. [Courtesy H.~Aussel]}
\label{f:HAussel1}
\end{figure}

How the potential inaccuracy of the photometric redshifts affects the stellar 
mass derivations was investigated, and results are displayed in 
Figure~\ref{f:HAussel1}.  Two spectroscopic samples were used: the 
zCOSMOS bright spectroscopic sample \citep{lilly07} selected with a 
magnitude cut only at $i^{+}_{AB} < 22.5$ and a spectroscopic follow-up 
of 24~$\mu$m sources selected from S-COSMOS \citep{sanders07}.  
These sources are typically dusty starbursts with $18 < i^{+}_{AB} < 25$.  
A median difference smaller than 0.002~dex is found between the photo-$z$ 
and the spectro-$z$ stellar masses, for both spectroscopic samples.

It is interesting to note that the stellar synthesis models have the same effect for 
both selections, as shown in Figure~\ref{f:HAussel1}.  This is not the case when 
studying the impact of the attenuation law.  The MIPS selection consisting preferentially 
of dusty galaxies, the difference due to the choice of attenuation (the \citealt{calzetti00} 
law in a case, the \citealt{charlot00} in the other) is amplified for this sample.  The 
median difference in stellar masses is $-$0.1~dex for the optical selection and 
$-$0.24~dex for the infrared selection.

The impact of the choice of stellar synthesis model on the stellar masses determinations 
was also examined. Figure~\ref{f:HAussel1} presents the difference between 
stellar masses computed using the \citet{bruzual03} models and the new models 
of Charlot and Bruzual (2007, private communication) including a better treatment 
of the pulsing asymptotic giant branch.  A median difference of 0.13--0.16~dex 
and a dispersion of 0.09~dex is observed. \citet{pozzetti07} found a similar 
difference (0.14~dex) when comparing \citet{bruzual03} and \citet{maraston05} models. 
\citet{walcher08} show that this offset changes with the measured sSFR of the galaxy, 
in the sense that the masses determined using BC03 for objects with an intermediate 
sSFR ($-12 < \log ({\rm sSFR}) < -9$) are, in the mean, more massive by 50\% than those 
determined using CB07. These are the objects where the TP-AGB stars are most likely 
to contribute significantly to the light in the NIR bands. Objects with lower or higher 
sSFRs do not show this offset in \citet{walcher08}. 

\citet{ilbert10} concluded from this study that the use of
photometric redshifts is only a weak source of uncertainty when
deriving stellar masses for a sample  of galaxies.  
The uncertainties are dominated by the uncertainties in the SED 
modeling itself, thus one has to be very cautious about the interpretations when 
selecting samples where a specific type of model is preferred.  

External validation of the method is possible for one parameter in particular, namely 
the SFR. Using the entire SED to derive the SFR  was considered less precise than 
other methods (based on specific tracers) in \citet{kennicutt98}. The availability of 
large samples with UV, optical and NIR photometry have lead to a reappraisal of 
the method in the last years. \citet{salim07} show a detailed comparison based 
on the data from the SDSS and the Galex satellite. They derive SFRs 
using the multi-wavelength SED. They then compare these to the SFRs derived by 
\citet{brinchmann04}, which used all emission lines that can be usefully measured in 
the SDSS spectra and detailed modelling of the emission line spectrum to derive 
dust-corrected, emission-line based SFRs. As shown in their figure 6, the 
SED-derived SFRs show in general a satisfying agreement. A similar test was performed 
in \citet{walcher08} on a  VVDS-SWIRE-GALEX-CFHTLS sample, 
where the SED SFR was compared to SFRs derived from the [OII]$\lambda3727$\AA\ 
emission line. While the latter are much less accurate than the emission line SFRs 
from \citet{brinchmann04}, the agreement between SED and emission line SFR was 
shown to be satisfying out to a redshift around one (their figure 8). 
An advantage of obtaining SFRs from broad-band SEDs compared to
emission-lines is the time needed to obtain sufficient S/N in the spectra
to get the lines. Nevertheless, 
an interesting residual correlation of SFR difference with stellar
mass is found in SED fitting, which 
remains yet to be fully understood. Another interesting problem exists in early-type 
galaxies, in which no H$\alpha$ emission can be measured, yet UV emission 
is commonly found \citep[see e.g.][Figure 3]{salim07}. This probably is the cause 
why some  objects with no
measurable emission lines have relatively high SED SFRs. 

\begin{figure}
\includegraphics[width=\hsize]{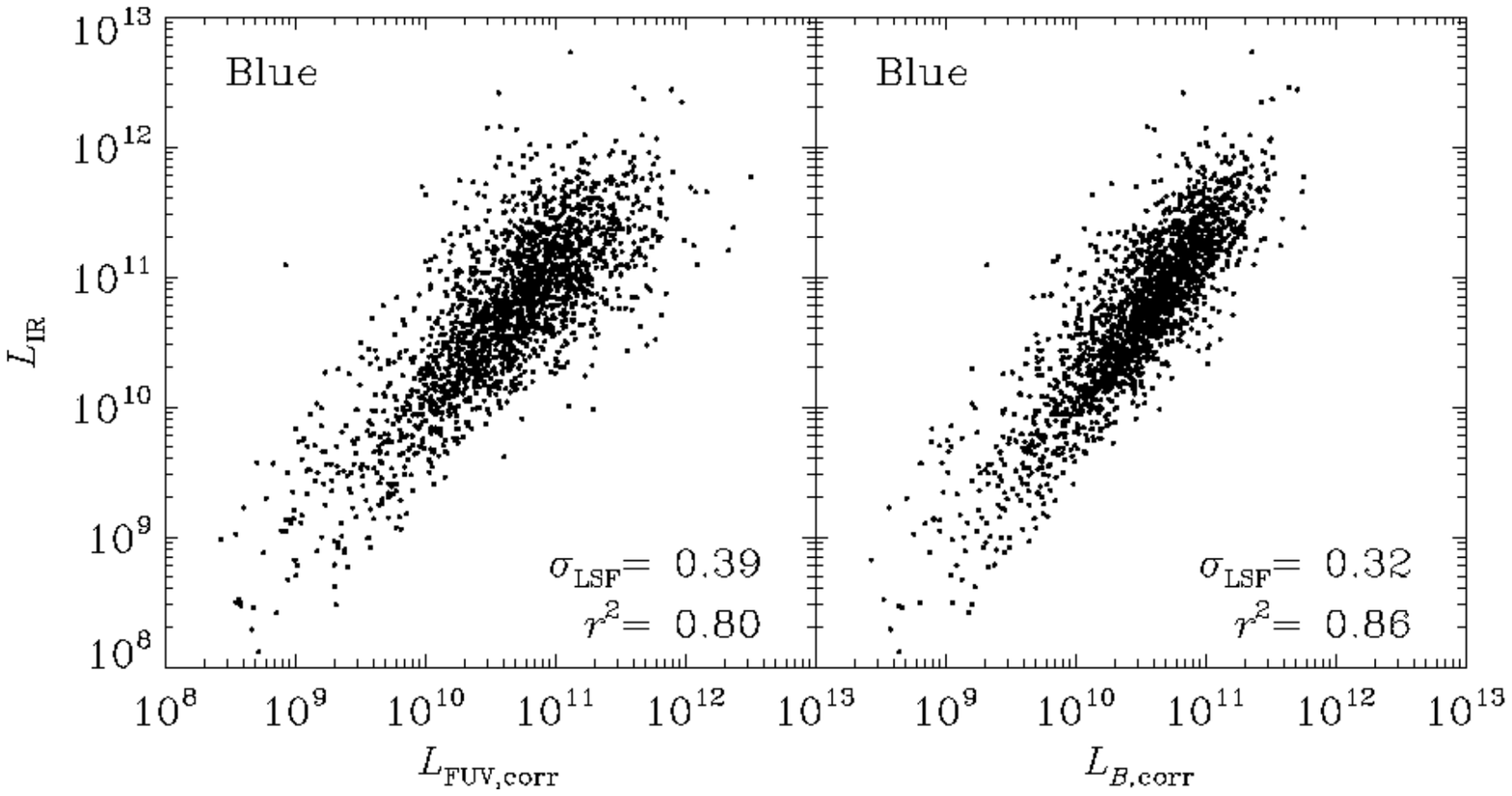}
\caption{Comparing UV and optical luminosities to the IR luminosity determined
  from the 24\mum\ flux. The offset ($\Delta$) and dispersion ($\sigma$) around
  the mean relation of the galaxy sample are marked in the lower right 
  corner. Figure from \citet{salim09} [Courtesy S. Salim].}
\label{f:salim09}
\end{figure}

In order to test the SFR determinations from SED fitting at z$\sim$1,
\citet{salim09} extended previous work to include the mid-IR dust emission. 
The consensus is that the mid-IR emission is heated by the UV, and
hence traces the emission of  young stars 
($\lapprox 100$ Myr) and it therefore provides a good means to measure the SFR. 
Deep MIPS 24 $\mu$m imaging exists for the Extended Groth Strip (EGS), as well 
as redshifts assembled in the context of the DEEP2 survey. The rest-frame 
12 $\mu$m emission was converted to a total dust luminosity using the \citet{dale02} 
prescriptions. Using the dust attenuation determined from fitting the UV-optical SED
one can calculate the dust-corrected UV and B-band luminosities. These correlate 
well with the 24\mum\ luminosity as seen in figure \ref{f:salim09}. Turning the IR 
luminosity into a SFR using the relations of \citet{kennicutt98}, the best correlation 
was found for an SED-SFR measured over the last 1--3 Gyr. That the mid-IR may 
be heated by older stars recently received support in a study by \citet{kelson10}, 
which predicts that C-type TP-AGB stars ($\sim$1.5 Gyr old) can produce a large fraction 
of the mid-IR flux. 

Nevertheless, for some very red objects in their sample, some SED SFRs seriously 
underpredict the SFR determined from the IR emission. These galaxies have a 
dominant old population, where the IR may have little to do with SF. Indeed, for the 
case of early-type galaxies Johnson et al.~(in prep.) find that 
it is important to consider the fraction of the stellar emission absorbed redwards of 4000 
{\AA} in order to obtain a good prediction for the measured FIR emission. Neglecting 
this emission leads to an underprediction of the FIR emission by factors of up to 5. 
 
Stellar mass is the second parameter that can in principle be calibrated empirically, 
with the hope of deriving the normalization of the M/L ratio, in other words 
the choice of IMF. However, uncertainties related to technical questions concerning 
the SED fitting and the dynamical mass measurements are still too high to derive 
very accurate overall IMF normalizations \citep[see e.g.][]{cappellari06, salucci08}. 
Careful attention should also be paid to issues such as the contribution of dark matter, 
recycled gas, and other dark components. 
In the formulation of \citet{bell01}, the slope of the colour-M/L relation is 
independent of the IMF, but the normalization depends on it. For this reason \citet{bell01} 
used the maximum disk approximation to provide an upper boundary of the total 
mass in stars allowed in spiral galaxies. They thus confirm that the Salpeter IMF 
leads to too high M/L ratios and they adapt the normalization of the colour-M/L 
relation to be 0.15 dex below Salpeter. Thus a Kroupa or Chabrier IMF seem to 
be good choices (see section \ref{s:ssp}). Variations between stellar population models and in dependence 
on the available data range also need to be considered carefully \citep[see e.g.][]{wel06}. 
As many more independent mass tracers exist, which require careful 
assessment, the interested reader is pointed to de Jong \& Bell (in prep.) for further 
discussion of this topic.

\subsection{Method-independent caveats}
\label{s:methodcav}

A good example of how the quality of the available models and data influence 
the use of tracers and the precision with which physical properties can be 
recovered is given by recent developments in the use of index fitting. 
Historically, early-type galaxies have been analyzed using simple stellar 
populations models as templates, effectively using the prior assumption 
that early-type galaxies were created in one single burst of star formation. 
However, detailed spectroscopic studies \citep{trager00} as well as 
near-ultraviolet photometry of early-type galaxies \citep{ferreras00,yi05,kaviraj07} 
have confirmed the presence of hot stars in early-type galaxies.

\begin{figure}
\begin{center}
\leavevmode
\plotone{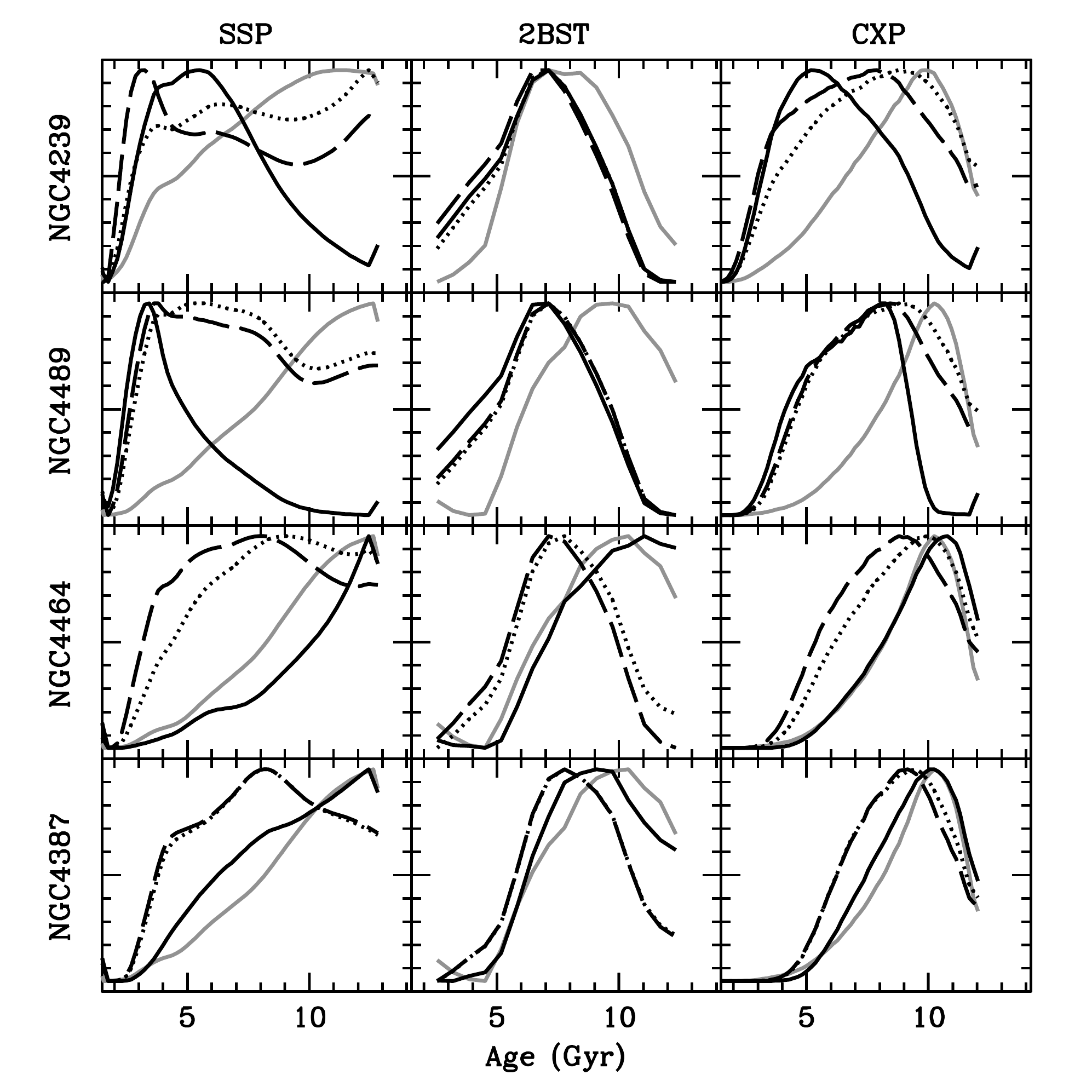}
\end{center}
\noindent
\caption {Marginalized age distribution in four Virgo cluster elliptical galaxies.  
Three models for the star formation history are considered as labeled (see text for details).  
The black lines correspond to the age extracted from [MgFe] plus either H$\beta$ 
(solid), H$\gamma$ (dashed) or H$\delta$ (dotted), respectively.  
The gray line is the result from the fit to the SED. 
[Courtesy I. Ferreras]
}
 \label{f:IFerreras2}
\end{figure}

Figure~\ref{f:IFerreras2} illustrates that it is now possible and necessary 
to go beyond SSPs for early-types \citep[see also e.g.][]{serra07}. 
The marginalized age distribution is shown for four galaxies corresponding 
to three different models (from left to right):  Simple Stellar Populations 
(SSP), a 2-burst model consisting of a superposition of an old and a 
young SSP (2BST), and a composite model assuming an exponentially 
decaying star formation rate, including a simple prescription for chemical 
enrichment (CXP).  All these models combine the population synthesis 
models of \citet{bruzual03}.  The figure shows that SSPs give mutually 
inconsistent age distributions, whereas composite models such as 
2BST or CXP give a more consistent picture.  Notice how lower mass galaxies 
such as NGC~4489 or NGC~4239 (top panels) give better fits for a 2 
burst scenario, whereas higher mass galaxies (NGC~4464, NGC~4387; 
bottom panels) are better fit by a smooth star formation history.

A similar analysis is presented in \citet{trager09}, who analyze mock data from 
semi-analytic models in standard ways, in particular computing the ``SSP-equivalent'' 
ages and metallicities. They quantify that the SSP-equivalent age is poorly 
correlated with the mass-weighted or light-weighted average ages. The 
SSP-ages tend to be younger, biased to the star formation in the last 0.1-2 Gyr. 
This has in particular the effect of exaggerating the signature of downsizing. 
On the other hand, the SSP-equivalent metallicity is mostly equivalent to the 
light-weighted metallicity. This serves as an important reminder, that the prior 
assumption that one puts into the analysis are very important for the resulting 
measurement. In the case of using SSP-equivalent ages, the assumption,
or prior, is 
that galaxies are single-age entities. This inevitably leads to important biases in 
the determined parameters. 

Among the important caveats of SED fitting, the issue of the age-metallicity degeneracy 
cannot remain unmentioned. This degeneracy is particularly important in the 
optical. In a nutshell, age effects on colors or line strengths can be mimicked by a 
change in metallicity.  This is especially important in the old populations found in 
early-type galaxies. Equally important and generally well-understood are the 
effects of using the optical SED only to determine the attenuation. Basically this 
is a very difficult undertaking and under almost any circumstance, measurements 
of the Balmer decrement, the FIR emission or the UV slope are essential to a 
reliable determination of the attenuation. 

In principle it should be possible and informative to combine information from 
the photometric SED and spectral information \citep[][]{gallazzi09,lamareille09}. 
While widespread use of this combination is hampered by possible systematic 
differences in the results \citep[e.g.][]{wolf07,schombert09}, general agreement 
between both types of information has been claimed at least 
for stellar masses \citep{drory04}. Brinchmann et al. 
(Talk at workshop) showed the results of a systematic comparison between 
photometric and spectral SED fitting. Comparing formal errors, they find no 
significant systematic offset. However, for actively star forming galaxies it might be 
better to use multiwavelength, broad-band photometry than the detailed spectrum 
information. The reason is, as mentioned in Section \ref{s:nonlinear}, that very young 
stars tend to ``hide'' in the normalized optical part of the spectrum, however they 
are readily visible as blue continuum in the multi-wavelength SED. On the other 
hand, spectra are better at picking up recent bursts through their Balmer lines 
\citep[e.g.][]{wild07,wild09}. Emission lines are a potential problem for fitting 
photometric SEDs. A quick back of the envelope calculation, however, shows 
that they produce a maximal offset in r-i colour of 0.1 at 
an equivalent width of 100 {\AA}. Thus, ELs are a minor issue in the
broad-band colours when a full SED 
is available and of limited concern for normal z$\sim$0 galaxies\footnote{But recall 
that equivalent widths are proportional to (1+z) and that the sSFR also tends 
to increase with redshift.}. 

Galaxies with significant redshifts can in principle be treated the same way as local 
ones, subject to two main caveats though: (1) we currently do not have the same 
kind of information available for large samples of galaxies at redshifts above 1 as 
for local galaxies. (2) Galaxies at high redshifts may have been significantly different 
from todays galaxies, e.g.~concerning their typical star formation histories, their 
metallicities or their gas content. As local 
analogs are rare or lacking, SED models are less well calibrated and may be subject 
to considerable systematic uncertainties. These need to be explored in detail, which 
is currently only possible through semi-analytic models \citep[see e.g.][]{schurer09}.
The design limits of SED fitting codes 
should thus be kept in mind when quoting results and -- in particular -- errorbars on 
high redshift properties.

\section{Results of SED Fitting: Photometric redshifts}
\label{s:photoz}

A special case of analyzing SEDs of extragalactic sources is the problem 
of redshift estimation, a topic that is usually refered to as {\em photometric 
redshifts} (hereafter photo-$z$). This problem is distinct from all other 
estimates of physical 
properties because independent and more precise measurements of 
the same property are available for large samples in the form of 
spectroscopic redshifts. The method can thus be tested extensively 
and even calibrated empirically. It is also one of the earliest forms of 
SED fitting, having been suggested as a manner to go beyond the limits 
of early spectroscopy \citep{baum57}.

For a working definition, \citet{koo99} suggests the following: ``photometric 
redshifts are those derived from {\em{only}} images or photometry with 
spectral resolution $\lambda/\Delta\lambda \la 20$. This choice of 20 is 
intended to exclude redshifts derived from slit and slitless spectra, narrow 
band images, ramped-filter imager, Fabry-Perot images, Fourier transform 
spectrometers, etc.'' This definition leaves room for a wide variety of 
approaches that are actively being explored by members of the community. 
While today most studies build on a set of magnitudes or colors, recently 
other observables have been utilized with good success, e.g., in the
work by \citet{wray08}. However, all methods depend on strong features 
in the SEDs of the objects, such as the Balmer break or even PAH features 
\citep{negrello09}.

Traditionally, photometric redshift estimation is broadly split into
two areas: empirical methods and the template-fitting approach.  
Empirical methods use a subsample of the photometric
survey with spectroscopically-measured redshifts as a `training set'
for the redshift estimators. This subsample describes the redshift distribution 
in magnitude and colour space empirically and is used then to
calibrate this relation. Template methods use libraries of either 
observed spectra of galaxies exterior to the survey or
model SEDs (as described in Section \ref{s:models}). As these are
full spectra, the templates can be shifted to any redshift and then 
convolved with the transmission curves of the filters used in the 
photometric survey to create the template set for the redshift estimators.  

Both methods then use these training sets as bases for the redshift 
estimating routines, which include $\chi^2$-fitting and various machine 
learning algorithms (e.g.~artificial neural networks, ANNs). The 
most popular combinations are $\chi^2$-fitting with templates 
and machine learning with empirical models. 
For a review of the ideas and history of both areas, see \citet{koo99}.
 
The preference of one over the other is driven by the limitations of 
our understanding of the sources and the available observations. 
Template models are preferred when exploring new regimes since 
their extrapolation is trivial, if potentially incorrect. Empirical models 
are preferred when large training sets are available and great 
statistical precision is required. Here we review these techniques 
and estimators, concentrating predominantly on the
template method which is closer to the idea of SED fitting as
discussed in the previous section.

\subsection{Methods}

\subsubsection{Empirical techniques}

Early on, the first empirical methods proved extremely powerful
despite their simplicity \citep[see e.g.][]{connolly95a,brunner97,wang98}.
This was partly due to their construction, which should provide both accurate
redshifts and realistic estimations of the redshift uncertainties. Even 
low-order polynomial and piecewise linear fitting functions do a reasonable 
job when tuned to reproduce the redshifts of galaxies \citep[see e.g.][]{connolly95a}.
These early methods provided superior redshift estimates 
in comparison to template-fitting for a number of reasons. By design
the training sets are real galaxies, and thus suffer no uncertainties
of having accurate templates. Similarly as the galaxies are a
subsample of the survey, the method intrinsically includes the effects
of the filter bandpasses and flux calibrations  

One of the main drawbacks of this method is that the redshift estimation 
is only accurate when the objects in the training set have the same 
observables as the sources in question. Thus this method becomes 
much more uncertain when used for objects at fainter magnitudes than the
training set, as this may extrapolate the empirical calibrations
in redshift or other properties.  This also means that, in practice,
every time a new catalogue is created, a corresponding training set needs to be
compiled.

The other, connected, limitation is that the training set must be large 
enough that the necessary space in colours, magnitudes, galaxy 
types and redshifts is well covered. This is so that the calibrations and 
corresponding uncertainties are well known and only limited 
extrapolations beyond the observed locus in colour-magnitude 
space are necessary. 

The simplest and earliest estimators were linear and polynomial
fitting, where simple fits of the empirical training set in terms of
colours and magnitudes with redshift were obtained \citep[see
e.g.][]{connolly95a}. These could then 
be matched to the full sample, giving directly the redshifts and their
uncertainties for the galaxies. 
Since then further, more computational intensive algorithms, have been
used, such as oblique decision tree classification, random forests,
support vector machines, neural networks, Gaussian process regression,
kernel regression and even many heuristic homebrew algorithms.

These algorithms all work on the idea of using the empirical training
set to build up a full relationship between magnitudes and/or colours
and the redshift. As each individual parameter (say the $B-V$ colour)
will have some spread with redshift, these give distributions or
probabilistic values for the redshift, narrowed with each additional
parameter. This process, in terms of artificial neural networks, is
nicely described by \citet{collister04}, who use this method in their
publicly available photo-$z$ code ANNz (described in the same
paper). They also discuss the limitations and uncertainties that arise
from this methodology.

\begin{figure}
\epsscale{1.1}
\plottwo{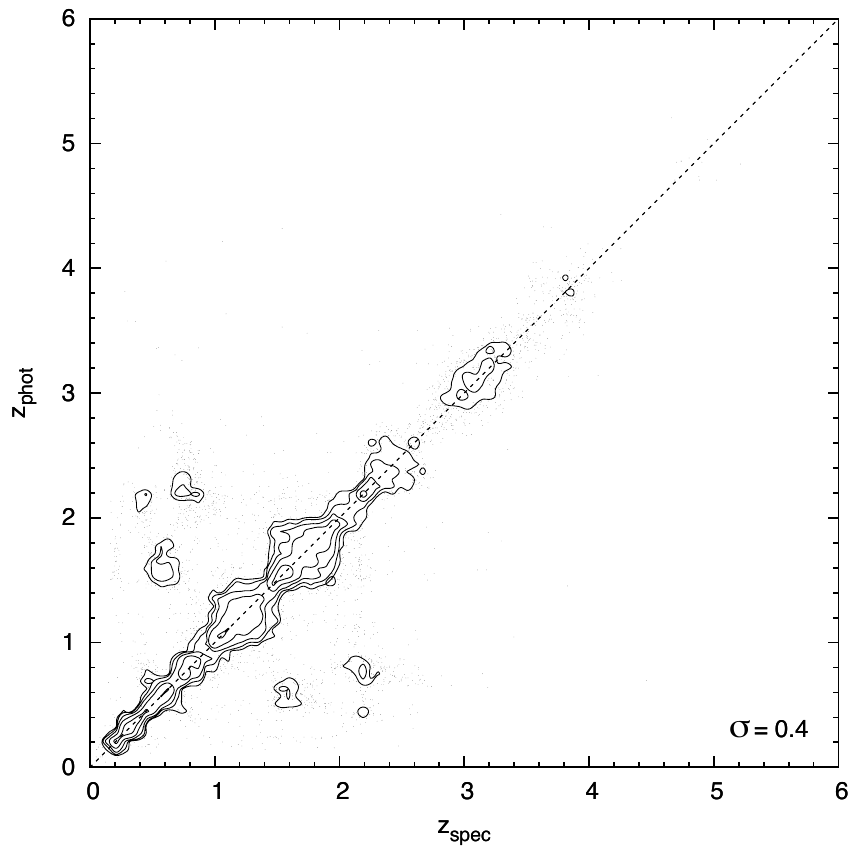}{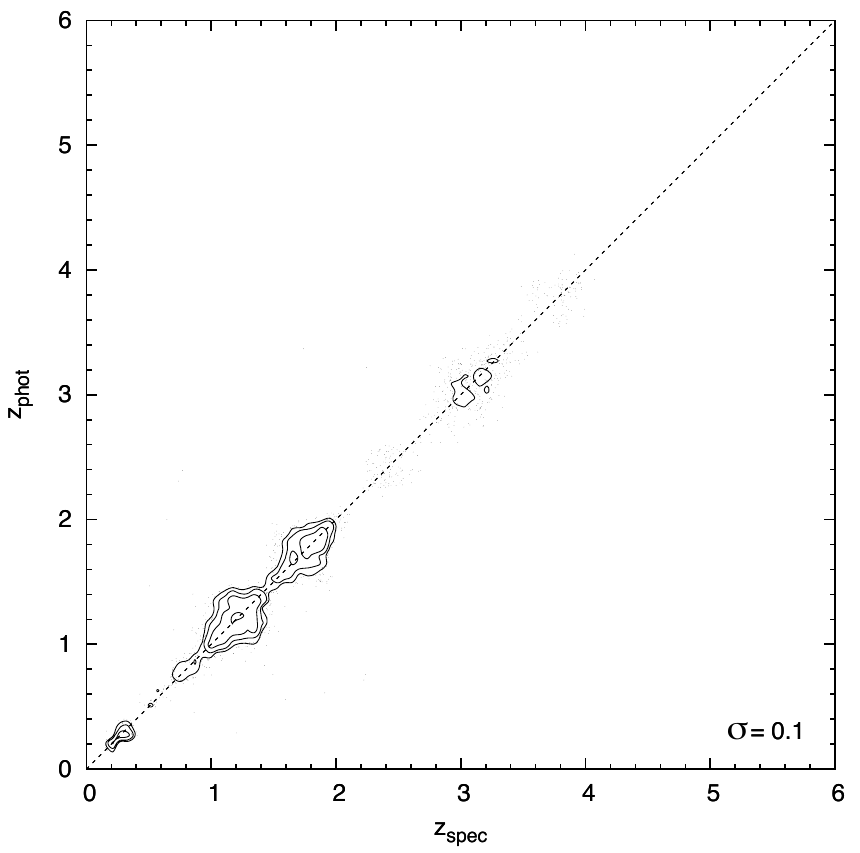}
\epsscale{1}
\caption{Improvement in quasar redshifts enabled by the \citet{ball08b}
data mining techniques,
shown as spectroscopic versus photometric redshift for the SDSS. {\bf Left:} A reproduction,
using their framework, a typical result prior to their work \citep[e.g.][]{weinstein04}
{\bf Right:} The result of using machine learning to assign probability density functions
then taking the subset with a single peak in probability. Contours indicate the areal
density of individual quasars (points) on the plot. Data from \citet{ball08b}. 
[Courtesy N.Ball]}
\label{f:photoz}
\end{figure}

Machine learning algorithms (of which neural networks is one) are one
of the strengths of the empirical method. These methods are able to determine
the magnitude/colour and redshift correlations to a surprising degree,
can handle the increasingly large training sets (i.e.~SDSS) and return
strong probabilistic estimates (i.e.~well constrained uncertainties, 
see figure \ref{f:photoz})
on the redshifts \citep[see][for a description of  machine learning
algorithms available and the strong photo-$z$ constraints
possible]{ball08a}. In addition, machine learning algorithms are also
able to handle the terascale datasets now available for photo-$z$
determination rapidly, limited only by processor speed and algorithm
efficiency \citep{ball08b}.

The additional benefit of the empirical method with machine learning, now 
increasingly being used, is that the constraining inputs for the photo-$z$s are 
not limited to the galaxies SED. Suggested first by 
\citet{koo99}, properties such as the the bulge-to-total flux ratio
\citep[e.g.][]{sarajedini99}, surface brightness
\citep[e.g.][]{kurtz07}, petrosian radii \citep[e.g.][]{firth03}, and
concentration index\citep[e.g.][]{collister04} have all been used in
association with the magnitudes and colours to constrain the redshift,
some codes even bringing many of these together \citep[e.g.][]{wray08}.

\subsubsection{Template Fitting}

Unlike the empirical method, the template-based method is actually a form of 
SED fitting in the sense of this review 
\citep[see e.g.][]{koo85,lanzetta96,gwyn96,pello96,sawicki97}. 
Put simply, this method involves building a library of observed
\citep[][is a commonly used set]{coleman80} and/or model 
templates \citep[such as][]{bruzual03} for many redshifts, and 
matching these to the observed SEDs to
estimate the redshift. As the templates are ``full'' SEDs or spectra, 
extrapolation with the template fitting method is trivial, if potentially incorrect.
Thus template models are preferred when exploring new regimes in a
survey, or with new surveys without a complementary large
spectroscopic calibration set. A major 
additional benefit of the template method, especially with the
theoretical templates, is that obtaining additional information, besides 
the redshift, on the physical properties of the objects is a built-in part 
of the process (as discussed in section \ref{s:physprop}). Note though, 
that even purely empirical methods can predict some of these properties 
if a suitable training set is available \citep[see e.g.][]{ball04}.

However, like empirical methods, template fitting suffers from several
problems, the most important being mismatches between the templates and the
galaxies of the survey. As discussed in section \ref{s:models}, model
templates, while good, are not 100\% accurate, and these
template-galaxy colour mismatches can cause systematic errors in the
redshift estimation. The model SEDs are also affected by modifiers
that are not directly connected with the templates such as the
contribution of emission lines, reddening due to dust, and
also AGN, which require very different templates \citep[see e.g.][]{polletta07}. 

It is also important to make sure that the template set is complete, 
i.e.~that the templates used represent all, or at least the majority, of
the galaxies found in the survey (compare also Section \ref{s:lib}). 
This is especially true when using 
empirical templates, as these are generally limited in
number. Empirical templates 
are also often derived from local objects and may thus be intrinsically 
different from distant galaxies, which may be at different evolutionary 
stage. A large template set is also important to gauge problems with 
degeneracies, i.e.~where the template library can give two
different redshifts for the same input colours.
Another potential disadvantage of template fitting methods comes from 
their sensitivity to many other measurements to about the percent level, 
e.g., the bandpass profiles and photometric calibrations of the survey. 

For implementations of the template fitting, the method of
maximum likelihood is predominant. This usually involves the
comparison of the observed magnitudes with the magnitudes derived from
the templates at different redshifts, finding the match that minimizes
the $\chi^2$ (compare section \ref{s:bayesian}). What is returned
is the best-fitting (minimum $\chi^2$) redshift and template (or
template+modifiers like dust attenuation). By itself this method does
not give uncertainties in redshift, returning only the best fit. For
estimations of the uncertainties in redshift, a typical process is to
propagate through the photometric uncertainties, to determine what is
the range of redshifts possible within these uncertainties. A good
description of the template-fitting, maximum likelihood method can be
found in the description of the publicly available photo-$z$ code,
\emph{hyperz} in \citet{bolzonella00}.

As mentioned above, one of the issues of the templates is the
possibility of template incompleteness, i.e.~ not having enough
templates to describe the galaxies in the sample. Having too many
galaxies in the template library on the other hand can lead to
colour-redshift degeneracies. One way to overcome these issues is
through Bayesian inference: the inclusion of our prior knowledge (see
Section \ref{s:bayesian}), such as the evolution of the upper age
limit with redshift, or the expected shape of the redshift distributions, 
or the expected evolution of the galaxy type fractions. As described
in Section \ref{s:bayesian}, this has the added benefit of returning a
probability distribution function, and hence an estimate of the
uncertainties and degeneracies. In some
respects, by expecting the template library to fit all observed
galaxies in a survey, the template method itself is already Bayesian.
Such methods are used in the BPZ code of \citet{benitez00}, who
describes in this work the methodology of Bayesian inference with
respect to photo-$z$, the use of priors and how this method is able to
estimate the uncertainty of the resulting redshift. 

It should be noted that, while public, prepackaged codes might provide reasonable
estimates for certain types of sources, no analyses should proceed without
cross-validation and diagnostic plots. There are common problems that appear
in data sets and issues that need to be understood first, and worked
around, if possible \citep[see e.g.][ for a comparison of some public
photo-$z$ codes]{mandelbaum05}.
Some further public photo-$z$ codes include \emph{kphotoz}
\citep{blanton03},  {\sc zebra}
\citep{feldmann06} and \emph{Le Phare} 
\citep{arnouts99,ilbert06,ilbert09a}. 

\subsection{Calibration and error budgets} 

Redshift errors are ultimately data-driven: they typically scale with $1+z$ given 
constant wavelength resolution of most filter sets; they also scale with photometric 
error in a transition regime between $\sim2$\% and $\sim20$\%. Smaller errors are 
often not exploited due to mismatches between data and model arising from data 
calibration and choice of templates, while large errors translate non-linearly into 
weak redshift constraints. If medium-band resolution is available, QSOs show strong 
emission lines and lead to deeper photo-$z$ completeness for QSOs than for galaxies.

Photometric redshifts have limitations they share with spectroscopic ones, and some 
that are unique to them: as in spectroscopy, catastrophic outliers can result from the 
confusion of features, and completeness depends on SED type and magnitude. Two 
characteristic photo-$z$ problems are mean biases in the redshift estimation and large 
and/or badly determined scatter in the redshift errors. Catastrophic outliers result from 
ambiguities in colour space: these are either apparent in the model and allow flagging 
objects as uncertain, or are not visible in the model but present in reality, in which case 
the large error is inevitable even for unflagged sources. Empirical models may be too 
small to show local ambiguities with large density ratios, and template models may lack 
some SEDs present in the real Universe.

Remedies to these issues include adding more discriminating data,
improving the match between data and models as well as the model
priors, and taking care with measuring  
the photometry and its errors correctly in the first place. Photo-$z$
errors in broad-band  surveys appear limited to a redshift resolution
near $0.02\times (1+z)$, a result of limited  spectral resolution and
intrinsic variety in spectral properties. Tracing features with higher  
resolution increases redshift accuracy all the way to actual spectroscopy. Future work 
among photo-$z$ developers will likely focus on two areas: (i) Understanding the diversity of 
codes and refining their performance; and (ii) Describing photo-$z$ issues quantitatively 
such that requirements on performance and scientific value can be translated into 
requirements for photometric data, for the properties of the models and for the output 
of the codes.

\subsubsection{Template accuracy}

\begin{figure}
\plotone{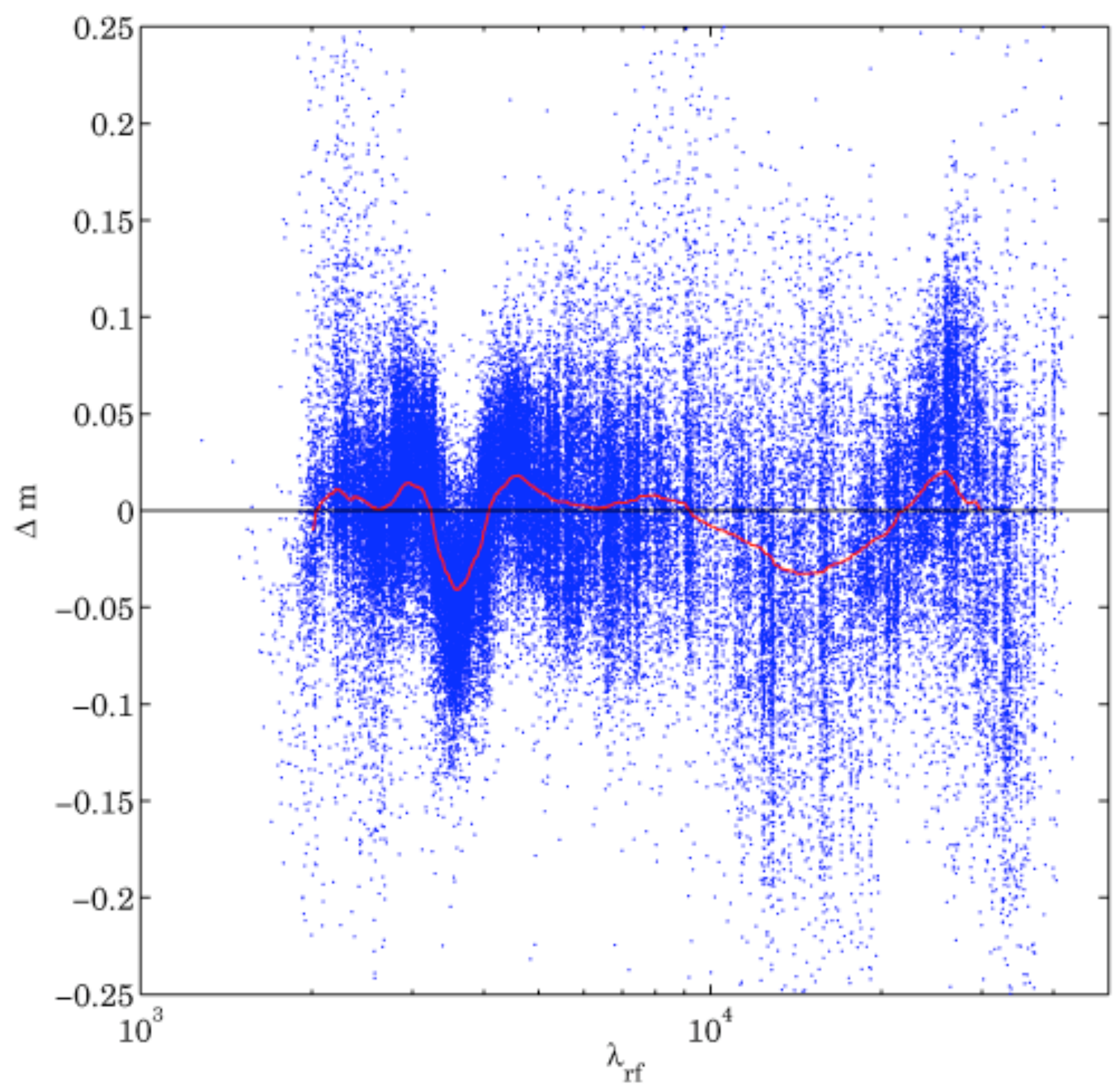}
\caption{Rest-frame residuals ($\Delta{\rm m}=(template-
galaxy)/template$) for all galaxies with reliable redshifts from the 
zCOSMOS sample. The feature with the strongest significance is at around 3500 
{\AA}, where the templates are too faint by 0.05 mag compared to the data. 
Additional nebular emission was added to the original SPS templates for this plot.
[Courtesy P. Oesch]}
\label{f:oesch}
\end{figure}

In general, template-based photo-$z$ estimates depend sensitively on the set 
of templates in use. In particular, it has been found that better photo-$z$ estimates 
can be achieved with an empirical set of templates \citep[e.g.][]{coleman80, kinney96} 
rather than using stellar population synthesis models 
\citep[SPSs; e.g.~][see Section \ref{s:models}]{bruzual03,maraston05}
directly. Yet the models are what are 
commonly used to compute stellar masses of galaxies. Since the use of these 
templates do not result in very good photometric redshifts, what is usually done, 
is to first derive photometric redshifts through empirical templates, and then estimate 
the stellar masses with the SSP templates. Obviously this is not self-consistent.

To investigate what causes the poorer photo-$z$ estimates of synthetic
templates, Oesch et al.~(in prep.) used the photometric data in 11 bands of the COSMOS survey 
\citep{scoville07}, together with redshifts of the zCOSMOS follow-up \citep{lilly07} 
and fit the data with SSP templates. In the resulting rest-frame residuals they identified 
a remarkable feature around 3500 \AA, where the templates are too faint with 
respect to the photometric data, which can be seen in figure \ref{f:oesch}. The feature 
does not seem to be caused by nebular continuum or line emission, which they 
subsequently added to the original SSP templates. Additionally, all types of 
galaxies suffer from the same problem, independent of their star-formation rate, 
mass, age, or dust content.

Similar discrepancies have been found previously by \citet{wild07,walcher08}, 
who found a $\sim 0.1$ mag offset in the D$_{\rm n}(4000)$ index. As this spectral 
break is one of the main features in the spectrum of any galaxy, it is likely that the 
poor photo-$z$ performance of synthetic templates is caused by this discrepancy. 
The cause of the discrepancy has been identified as a lack of coverage in the 
synthetic stellar libraries used for the models. It will thus be remedied in the next 
version of GALAXEV (G. Bruzual, priv. comm.).

\subsubsection{Spectroscopic Calibration of Photo-$z$s}

One of the strong benefits of the template method is that any
spectroscopic subsample of a survey can be used to check the
template-determined photo-$z$s. This can also be done for the
empirical methods, yet for this a very large spectroscopic sample is
necessary such that it can be divided into a large enough training set
and testing sets.

With the existence of a test spectroscopic sample, it is then possible
to \emph{calibrate} the template library, leading to a combined
empirical-template method. This means to correct for
errors in the photometric calibration or even the correction of the
templates themselves for example to allow for the
evolution of galaxies with a small library, or to account for
inaccurate models (see section \ref{s:models}).
Such calibration is typically an iterative
process, in which the photometry and/or template SEDs are modified to
minimize the dispersion in the resulting photometric redshifts.

\begin{figure}[bthp]
\plotone{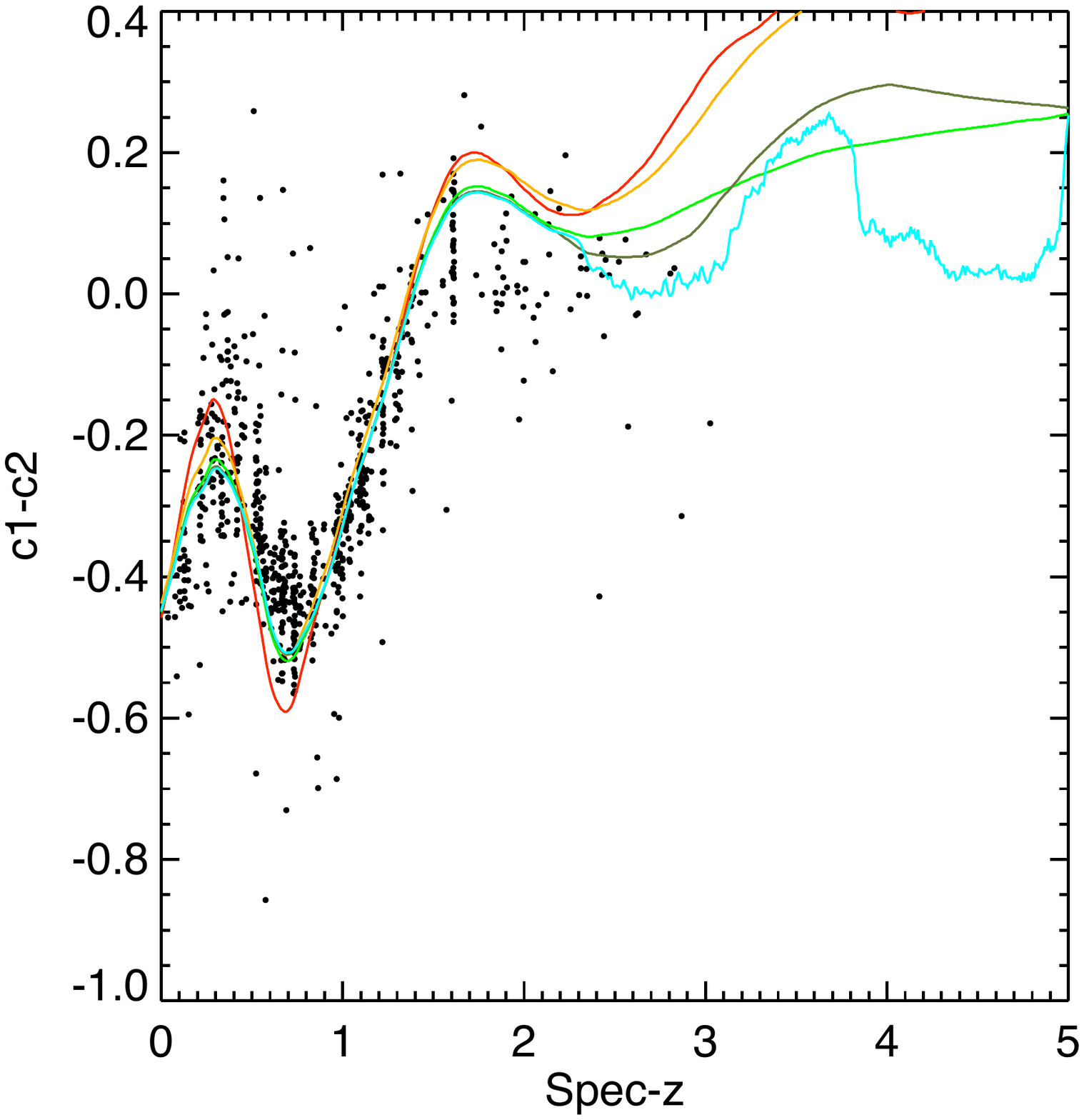}
\caption{IRAC [3.6]-[4.5] color-redshift plot for a sample of GOODS
  galaxies. The solid curves show the change in colours with redshift
  of five different empirical template SEDs. [Courtesy M.~Brodwin]}
\label{f:brodwin1}
\end{figure}

The simplest kind of calibration involves adding small zero-point
offsets to the photometry uniformly across the sample.  This does not
imply that the photometry is incorrectly calibrated (though in
practice the absolute calibration may well have small errors in the
zero-point), but rather that there is often a mismatch between the
real SEDs of galaxies and the templates used to fit them.  The
calibration is meant to minimize those differences. Plotting
color-color or color-redshift diagrams (figure~\ref{f:brodwin1}) with
the template SEDs overlaid will often indicate bulk offsets between
the two.

\begin{figure}[bthp]
\plotone{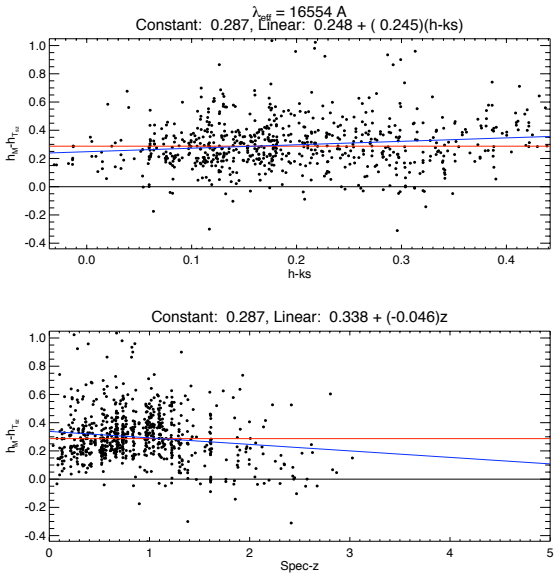}
\caption{$H$-band residuals vs.~color and redshift in a sample of
  GOODS galaxies.  An effective offset of $\approx 0.3$ mags is
  apparent. [Courtesy M.~Brodwin]}
\label{f:brodwin2}
\end{figure}

A more instructive approach, however, is to compute the residuals
between the predicted magnitude of the best-fit template at the
spectroscopic redshift and the observed magnitude \citep[for more details,
see][]{brodwin06a, brodwin06b}.  These residuals can be plotted
versus color or redshift for added diagnostic power.  In the example
in Figure~\ref{f:brodwin2}, there appears to be an effective magnitude
offset of $\approx 0.3$ mags in the $H$-band.

\begin{figure}[bthp]
\plotone{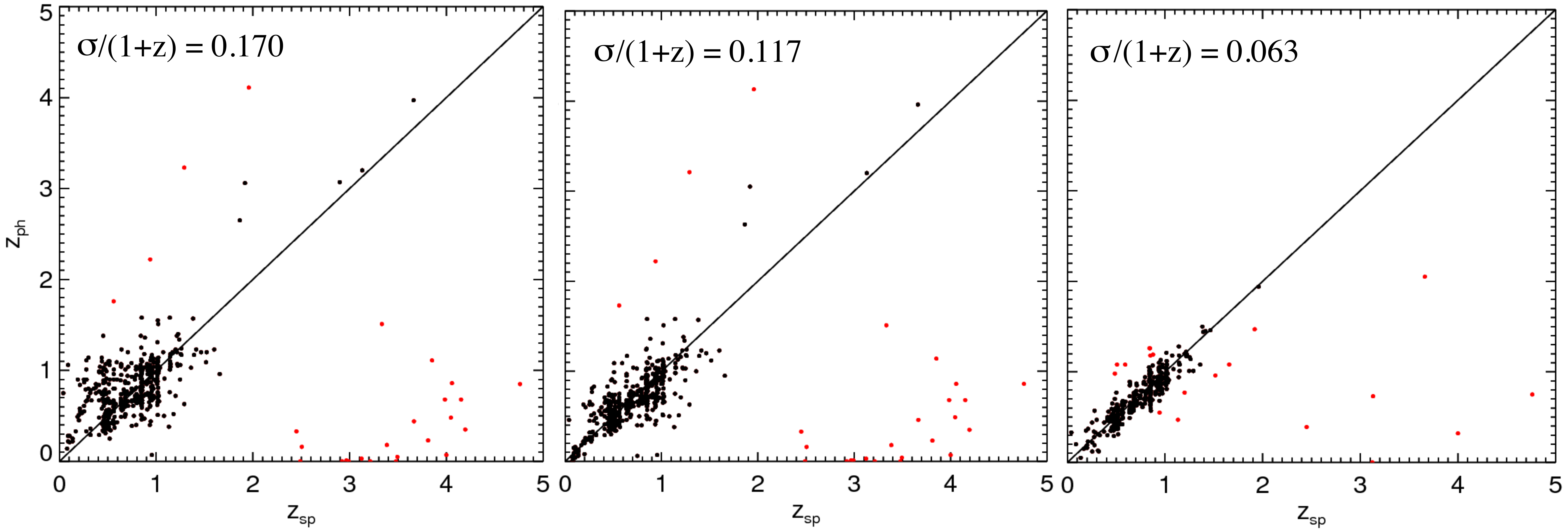}
\caption{Iterative improvement in photometric redshift estimation via
  this simple calibration technique. [Courtesy M.~Brodwin]}
\label{f:brodwin3}
\end{figure}

Applying such effective zero-point adjustments in all bands in an
iterative process minimizes the mismatch between the data and the
templates, and hence minimizes the resultant photometric redshift
dispersion, as shown in Fig.~\ref{f:brodwin3}.

Such calibration phases are used in the works of \citet{brodwin06a}
and as ``template-optimization'' in the codes {\sc zebra}
\citep{feldmann06} and \emph{Le Phare} \citep{ilbert06,ilbert09a} which
use template fitting with Bayesian inferences and this calibration
phase to give the most accurate photometric redshifts possible with
the template approach.  

With the most accurate photometric redshifts possible, the
template-fitting can then be used to estimate physical  properties
such as stellar masses, star-formation rates, etc. (see section
\ref{s:results2}). 

\subsubsection{Signal-to-noise Effects}

\begin{figure}
\plottwo{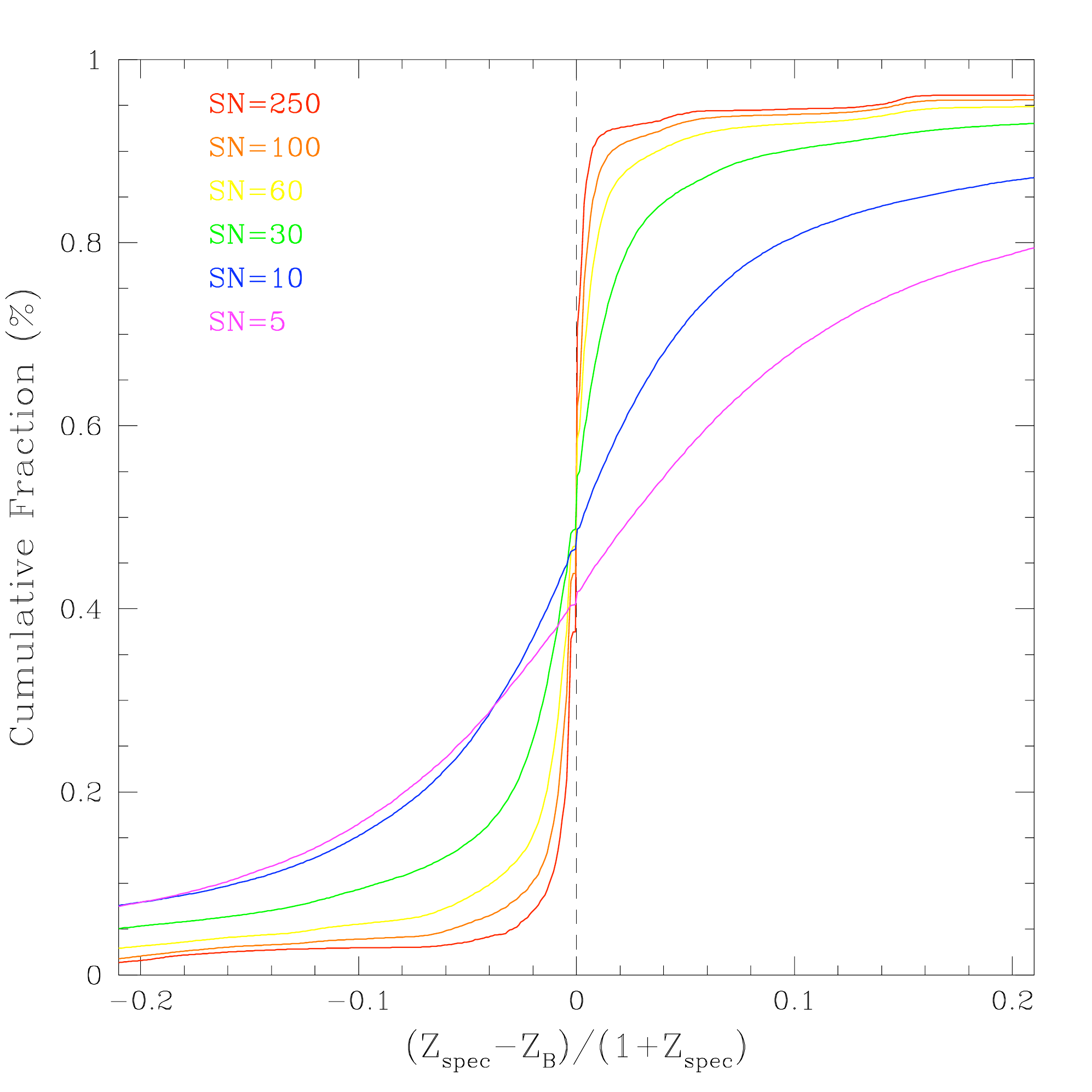}{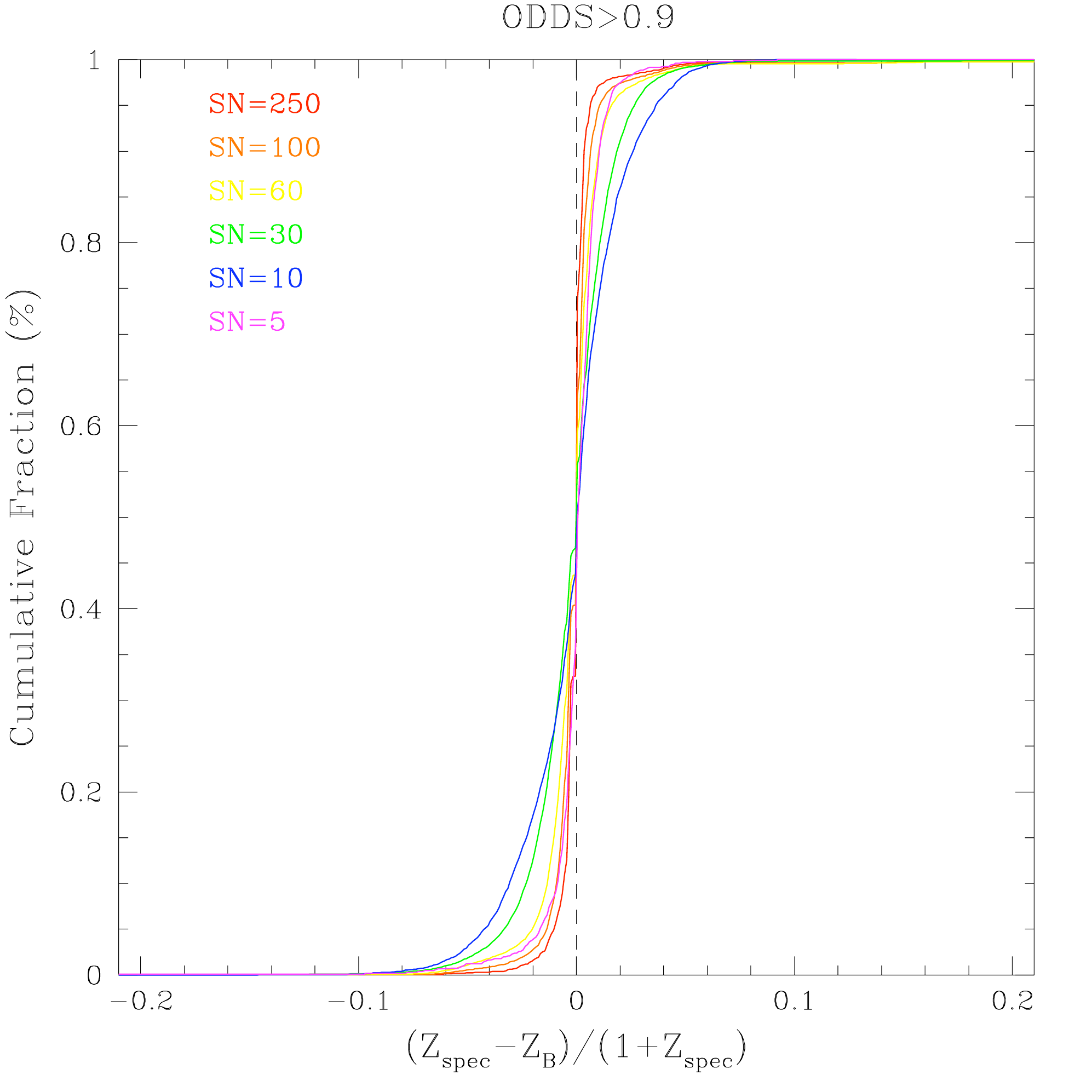}
\caption{Cumulative fraction of objects with $\delta_z$ smaller than
a given value. Red line indicates the simulation in which all galaxies
have been set to have SN=250 in all BVRz; orange indicates a simulation
ith SN=100; and so on. Right panel shows all galaxies, and left panel
shows galaxies with $ODDS>0.9$. Note that only 6.4\% and 1.2\% respectively
of objects with SN=10,5 have $ODDS>0.9$ \citep[figure from][]{margoniner08} [Courtesy V.~Margoniner]. }
\label{f:margnoniner}
\end{figure}

\citet{margoniner08} have specifically investigated the impact of
photometric signal-to-noise 
(SN) on the precision of photometric redshifts in multi-band imaging
surveys. Using simulations of galaxy surveys with redshift distributions (peaking at $z
\sim 0.6$) that mimics what is expected for a deep (10-sigma R band =
24.5 magnitudes) imaging
survey such as the Deep Lens Survey \citep{wittman02} they investigate
the effect of degrading the SN on the photometric redshifts determined
by several publicly available codes (ANNz, BPZ, hyperz) 

Figure \ref{f:margnoniner} shows the results of one set of their
simulations for which they degraded the initially perfect photometry
to successively lower SN. In these unrealistic simulations all
galaxies have the same SN in all bands. The 
figure shows the cumulative fraction of objects with $\delta_z$ smaller than
a given value as a function of $\delta_z$. The left panel shows the
cumulative fraction for all objects, while the right panel shows
galaxies for which the BPZ photo-$z$ quality parameter, $ODDS>0.9$.  
The number of galaxies in
the right panel becomes successively smaller than the number in the left
as the signal-to-noise decreases (64\% of SN=250, and only 6.4\% of SN=10
objects have $ODDS>0.9$), but the accuracy of photo-$z$s is clearly better.

The results of this work show (1) the need to include realistic
photometric errors when forecasting photo-$z$s performance; (2) that
estimating photo-$z$s performance from higher SN spectroscopic objects will
lead to overly optimistic results.

\subsection{A unified framework}

\begin{figure}
\plotone{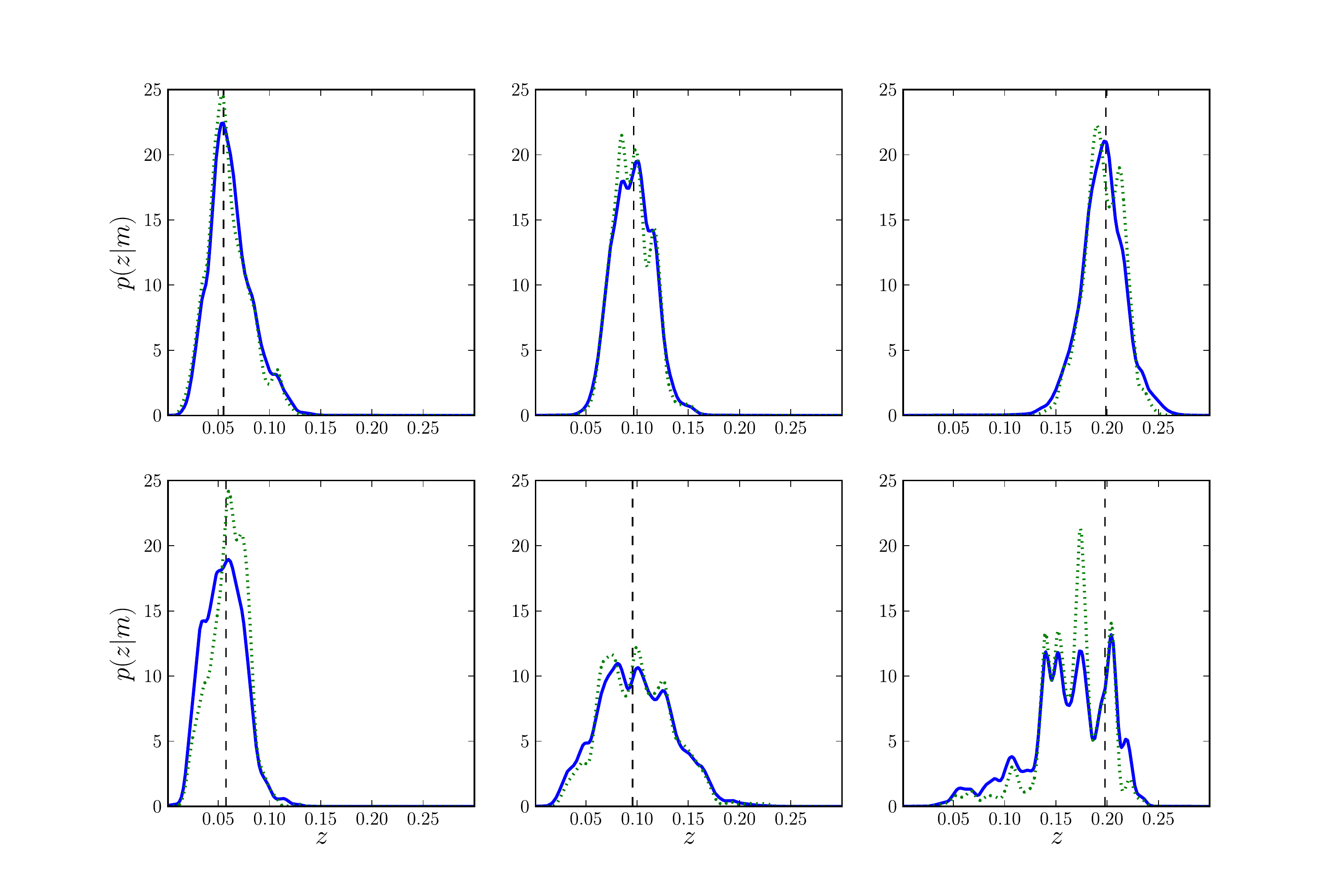}
\caption{The probability density as a function of the redshift for
early- and late-type galaxies ({\em{}upper} and {\em{}lower} panels, respectively)
at different distances marked by the vertical {\em{}dashed} lines. For
every object, the {\em{}dotted} line shows the empirical relation of 
$p(z|\vx=\apjvec{m}_q)$, and the
{\em{}solid} line illustrates the final result of $p(z|\vy=\apjvec{m}_q,M)$ after
properly folding in the photometric uncertainties via the mapping in the model
\citep[figure from]{budavari09a} [Courtesy T.~Budavari].}
\label{f:pdf}
\end{figure}

The field of photometric redshifts and the estimation of other physical properties
has been very pragmatic. Its development,
since the first attempts, has been incremental in the sense that
most studies focused on refining components but staying within the
concepts of the original ideas of the two classes.
Empirical and template-fitting approaches today follow very separate
routes, with these classes of methods even use different sets of
measurements. Only the semi-empirical approach of zero-point
calibration comes close to linking the two approaches.
However, a recent study by \citet{budavari09a}, tries to understand
this separation and possibly bring these methods together by
devising a unified framework for a rigorous solution
based on first principles and Bayesian statistics.

This work starts with a minimal set of requirements: a training set with some
photometric observables $\vx$ and spectroscopic measurements $\vk$, and
a query or test set with some potentially different set of observables $\vy$.
The link between these is a model $M$ that provides the mapping between 
$\vx$ and $\vy$, the probability density $p(\vx|\vy,M)$. This is more than 
just the usual conversion formula between photometric systems because 
it also incorporates the uncertainties. 

The empirical relation of $\vx\!\!-\!\!\vk$ is often assumed to be a function.
A better approach is to leave it general by measuring the conditional density function.
The simplest way is to estimate the relation by the densities on the training set as
$p(\vk|\vx)=p(\vk,\vx)/p(\vx)$. The final result is just a convolution of
the mapping and the measured relation:
$p(\vk|\vy,M) = \int\!\!d\vx\,p(\vk|\vx)p(\vx|\vy,M)$.
In figure~\ref{f:pdf}  we show the results from \citet{budavari09a},
where he plots the empirical
relation (dotted blue line) 
and the final probability density (solid red) for a handful of SDSS galaxies.
The top panels show intrinsically red galaxies, whose constraints are 
reasonably tight out to the highest redshifts. Blue galaxies in the bottom
panels however get worse with the distance as expected.

The aforementioned application follows a minimalist empirical
approach but already goes beyond traditional methods. Template-fitting is
in the other extreme of the framework where the training set is generated from
the model using some grid. Without errors on the templates, the equations
reduce to the usual maximum likelihood estimation that is currently used by
most codes. A straight forward extension \citet{budavari09a} suggests is
to include more realistic errors 
for the templates. Similarly one can develop more sophisticated predictors
that leverage existing training sets and spectrum models at the same time.

\subsection{The State of Photometric Redshifts} 

Generally there is an overall agreement in most aspects of photometric
redshift methodologies, and even technicalities. 
However there is a need for standardized
quality measures and testing procedures.
It is important to analyze to performance of each model spectrum as a
function of the redshift. This is best done by plotting the difference of
the observed magnitudes and template-based ones. These figure can pinpoint
problems with the spectra and even zero-points. Intrinsically these are
the quantities used inside the template-optimization procedures, e.g., in
\citet{budavari00} and \citet{feldmann06}.

In SED fitting, interpolation between templates is often used, which can be done
linearly or logarithmically. The latter has the advantage of being independent
of the normalization of the spectra. Yet, most codes appear to use linear
interpolation without a careful normalization. This might explain some of the
discrepancies among similar codes
found by the Photo-$Z$ Accuracy Testing (PHAT) project. 
\footnote{\url{http://www.astro.caltech.edu/twiki\_phat/bin/view/Main/WebHome}}

The determination of the quality of the estimates is also a crucial topic.
There is need for different measures that can describe the scatter of the points
without being dominated by outliers and that can estimate the fraction of
catastrophic failures.
It is also recommended to characterize the accuracy of the estimates by
a robust M-scale instead of the RMS; a measure that
is simple to calculate, yet, not sensitive to outliers.
Another aspect of this is the study of selection criteria that is often
neglected. Certain projects are not concerned with incomplete samples
as long as the precision of the ones provided is good \citep[e.g.,
weak lensing,][]{mandelbaum05}, 
while others, such as galaxy clustering, might require an unbiased selection.
Therefore, it is perceived that studies using methods with any quality flags
or quantities should provide details of their selection effects.

A common theme for future goals in most photometric redshift works
appears to be more detailed probabilistic analyses, with the need for
probability density functions.
Priors used in most bayesian analyses seem to be generally accepted in
the photo-$z$ community.
With such consensus amongst photometric redshifts obtained, the focus
of work now is shifting from the estimation of ``just'' the redshifts 
to simultaneously constraining physical parameters and the redshift 
in a consistent way.

\section{Results of SED Fitting: Physical Properties}
\label{s:results2}

SED fitting is a very versatile tool. From a rough estimation of the stellar masses of 
distant galaxies to the search for small subpopulations of stars in high S/N spectra, 
it can be applied to a large part of the problems in galaxy evolution. This is the strength 
and the weakness of the SED fitting technique: it does it all at once. 

We highlight here a few significant results. The intention is not to be complete or to 
mention the work that has been most in view, but rather to highlight the diversity of 
questions that can be adressed from fitting the integrated SEDs of stellar populations. 
Particular importance has been given to supply cautionary remarks, as it is easy to 
overinterpret the significance of the derived properties, in view of the complexity 
of the physical mechanisms and our frequent lack of detailed understanding.

\subsection{Stars}

\subsubsection{Stellar masses} 

Stellar masses are computed by multiplying a mass-to-light ratio $M/L$ with a 
luminosity $L$. While the uncertainties on $L$ depend on the quality of the data, 
the estimate over $M/L$ and its associated uncertainties depend mostly on the 
care taken with SED fitting. It is a good idea to search for a reference band that minimizes 
the effects of $M/L$ variations due to stellar populations (age, metallicity, chemical 
abundances) and due to dust absorption. While the common notion that the NIR (e.g.~the 
$H$ band at 1.65~$\mu$m) is close to ideal is correct in some cases, problems can 
arise because of thermally pulsing asymptotic giant branch stars (discussed
in section \ref{s:sspissues}) if young ages ($<$2~Gyr) are well represented 
in the SED. That reliable $M/L$ from SED fitting cannot be dispensed with is evident 
when looking at IRAC 3.6$\mu$m data of nearby galaxies, where star formation 
regions are evidently prominent. 

\paragraph{Stellar Mass Maps of Resolved Galaxies}

In the work of \citet{zibetti09} a ``data-cube'' approach is introduced to 
investigate the SEDs of nearby, resolved galaxies, aimed at preserving the maximum 
spatially-resolved information. One feature of the approach is that it allows to compare 
``global'' quantities, which are notoriously difficult to determine, with the integral over the 
local quantities, a useful test of how meaningful global quantities can be. 
A large part of the effort concentrated on developing a reliable method to obtain 
stellar surface mass density maps from a minimum set of broad-band observations.
This method relies on Bayesian inference, as discussed in Section \ref{s:bayesian}. 

The effective mass-to-light ratio correlates with optical / near-infrared colors 
\citep[e.g.,][]{bell01}, so $M/L$ can be expressed 
as a function of color(s). A better estimate is obtained if $M/L$ is mapped as a 
function of two colors, instead of one.  The colors adopted are $g-i$ and $i-H$.  
Their large wavelength separation allows to robustly describe the shape of the 
SED over the entire optical-near-infrared range, in a way that as insensitive 
as possible to photometric and modeling uncertainties.  

\begin{figure}
\plotone{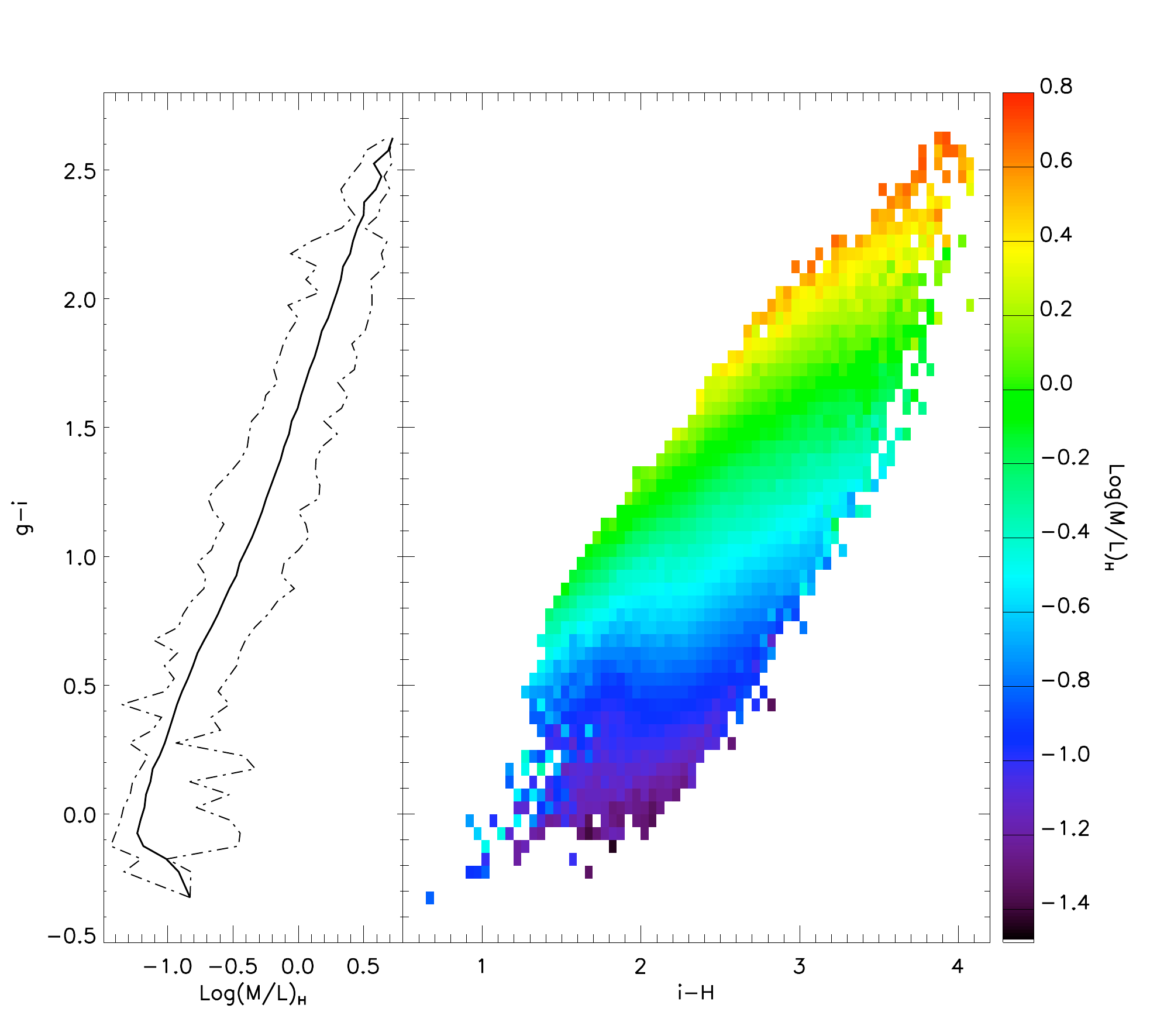}
\caption {Monte Carlo library of 50,000 models created from the Version 2007 of 
\citet{bruzual03}. The ($g$$-$$i$,$i$$-$$H$) space is binned in cells of 0.05~mag 
$\times$ 0.05~mag and marginalized over $M/L$ in each cell. The right panel 
shows $M/L_H$ as a function of ($g$$-$$i$,$i$$-$$H$).  
The scatter at each position is typically 0.11~dex, with peaks at $\sim$0.2 dex 
in the bluest corner.  The left panel shows $M/L$ as a function of $g$$-$$i$ 
(median value marginalized in a given $g$$-$$i$ bin).  The minimum and 
maximum $M/L$ values derived from the right panel at given $g$$-$$i$ are 
displayed by dashed lines. Figure from \citet{zibetti09} [Courtesy
S. Zibetti].}
\label{f:SZibetti1}
\end{figure}

To study the dependence of $M/L$ on ($g$$-$$i$,$i$$-$$H$) the authors use 
a Monte Carlo library of 50,000 models created from the 2007 version of BC03, 
which include also dust in different amounts and spatial distributions.. 
The ($g$$-$$i$,$i$$-$$H$) space is binned in cells of 0.05~mag $\times$ 
0.05~mag and marginalized over $M/L$ in each cell. The median is chosen as 
the fiducial $M/L$ at each position of the color-color space. A look-up table is 
created to derive $M/L_H$ as a function of ($g$$-$$i$,$i$$-$$H$).
Figure~\ref{f:SZibetti1} illustrates that the information in the second color 
improves the M/L determinations systematically by $\pm$0.3--0.4~dex. 

\begin{figure}
\plotone{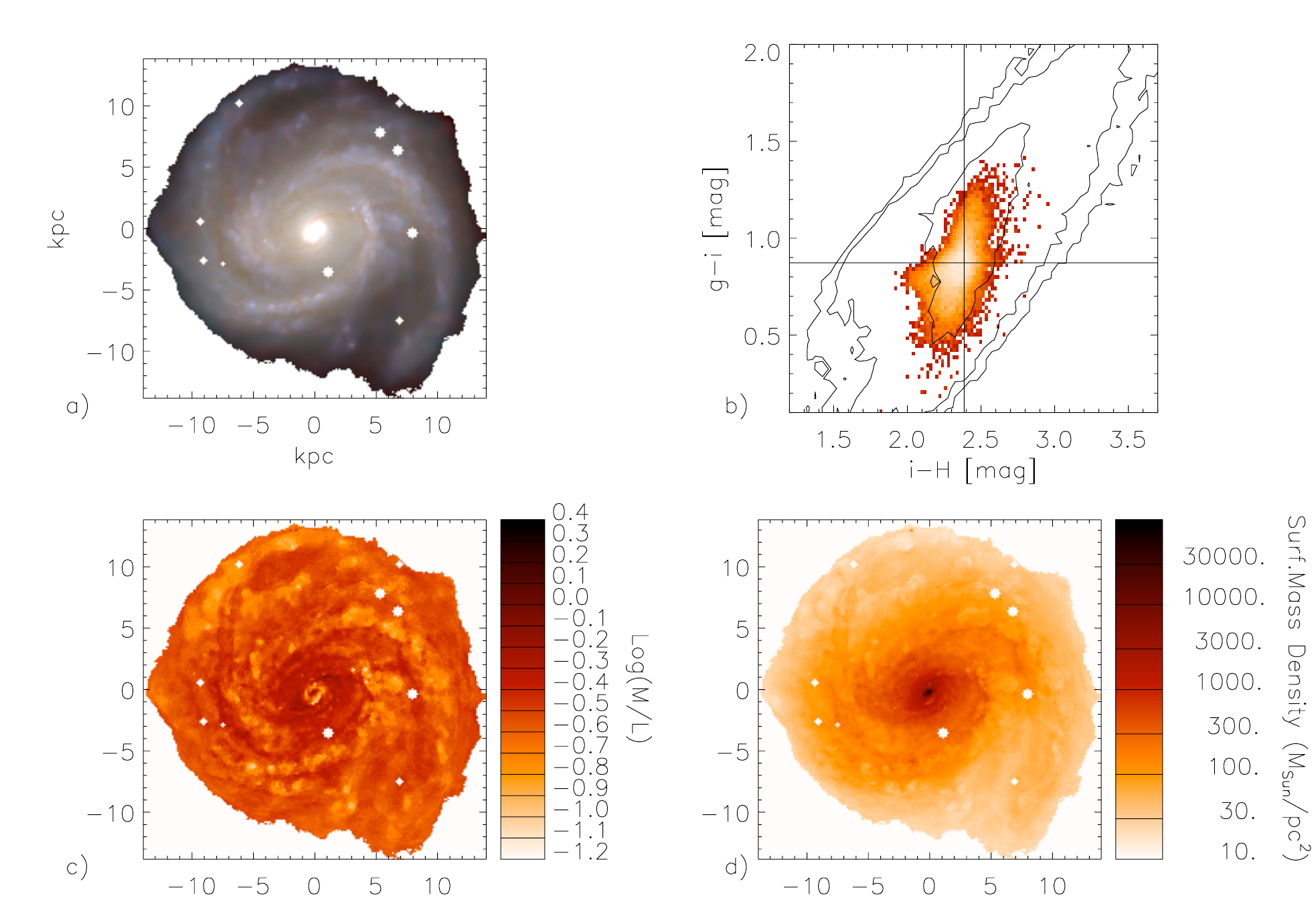}
\caption {Panel a) shows the color-composite 
RGB image ($H$ in red channel, $i$ in green and $g$ in blue) after adaptive 
smoothing. Panel b) plots the distribution of pixels in color-color space (log 
intensity scale).  The cross marks the position of the global colors of the galaxy. 
The over-plotted contours show the number density distribution of models from 
the library described above. Interestingly, observed colors are reproduced by the 
models over the full range.  Panel c) shows the mass-to-light ratio map derived 
with the method described above.  Finally, Panel d) displays the 
stellar mass surface density map, obtained by multiplying the $H$ band surface 
brightness times $M/L$.  Images a) and d) show that the galactic structure is much 
smoother in the mass image, where spiral arms have much less contrast with 
respect to what the light shows. Figure from
\citet{zibetti09} [Courtesy S. Zibetti].}
\label{f:SZibetti2}
\end{figure}

The method described above is applied to each pixel of the image of a galaxy, where 
``pixel'' implies the pixel that results after matching the images in the three bands to a 
common resolution.  In order to provide sensible results, it is crucial that color 
measurements do not exceed 0.1~mag errors, which requires S/N$\sim$20.
The results for M~100 (NGC 4321) are displayed in Figure~\ref{f:SZibetti2}.

An important result appears from the comparison between total stellar mass as 
obtained by integrating the stellar mass surface density maps (Figure~\ref{f:SZibetti2}d) 
and the one obtained using global colors to estimate the ``average'' $M/L$ ratio to 
be multiplied by $L(H)$.  This second method, the one commonly adopted in 
extragalactic studies, agrees within $\sim$10\% of the mass map integral only 
when the color distribution is quite homogeneous, i.e., for early type galaxies.  
When substantial color inhomogeneities and especially heavily obscured regions 
are present within a galaxy, using global colors and fluxes can lead to underestimates 
of the actual stellar mass of a galaxy by up to 60\%.  This can be understood if 
dust-obscured regions can contribute a significant amount of mass but are heavily 
under-represented in the global color, which is flux weighted and hence biased 
toward the brightest (and bluest) parts of the galaxy.  While the pixel-by-pixel 
$M/L$ gives the correct mass weight to these obscured regions, the globally 
computed $M/L$ severely underestimates their mass contribution.  

\paragraph{Stellar mass functions}

One of the holy grails of current observational efforts in galaxy evolution studies 
is a consistent picture of the build-up of stellar mass over the age of the universe. 
An important constraint on this is the stellar mass function, or its integral, the 
stellar mass density. A comprehensive discussion of these results would warrant 
a review of its own. Suffice it here to point out that not only the local mass function  
has been measured with great precision \citep[e.g.][]{bell03a}, but these results 
have also been extended to redshifts of 1 \citep{pozzetti07, bundy06}. At  redshifts above 1.2 
an observed-frame optical selection corresponds to a rest-frame UV selection, subject 
to large biases. These therefore have to be circumvented by a K-band selection 
\citep[e.g.][]{cirasuolo07}  or, better, by a selection at 3.6$\mu$m \citep[e.g.]{arnouts07, ilbert10}. 
For observational and conceptual reasons (detailed in Sections \ref{s:lib} and \ref{s:methodcav}), 
determining stellar masses and therefore mass function at redshifts higher than z=2 is very difficult. 
Most authors thus prefer to restrict themselves to luminosity functions instead 
\citep[see e.g.][for just one very recent example]{cirasuolo10}, thus leaving the burden 
of transforming luminosities to stellar masses to the interpreting model 
\citep[but see e.g.][]{kajisawa09}. 

\paragraph{Stellar masses of high redshift galaxies}

Many authors have used some kind of Bayesian inference-based method 
(Section \ref{s:bayesian}) to determine stellar masses for high redshift galaxies 
\citep[e.g.][]{sawicki98, papovich01, forster04}. There is good hope that these mass 
estimates are reasonably good \citep{drory04}, despite important caveats on the 
methodology that become more important with improving data quality (see e.g.~ 
Section \ref{s:physprop}). A recent result has been 
the discovery and study of high redshift galaxies with high stellar masses and 
low star formation rates (early type galaxies, ETGs). In the following we describe 
only one "family" of papers, as presented at the workshop, but see \citet{cimatti09} 
for a review. Massive ETGs are the first objects to populate the red sequence 
\citep[see e.g.][]{kriek08b}. Objects in the redshift range 2-3 can be identified 
by multi-band photometry \citep[e.g.][]{dokkum06}. For determining the physical 
properties however, significant uncertainties are due to photometric redshift 
determinations. For example \citet{kriek08} conclude that  while stellar masses 
are reasonably robust to small errors arising from photometric redshifts, the 
actual star formation history is generally very poorly constrained with broad 
band data alone. 

The obvious next step is thus to analyze these galaxies with spectroscopy, 
despite this being an expensive undertaking in terms of telescope time. 
When doing this, \citet{kriek08} also go further in blurring the limits between 
spectroscopy and photometry by binning their ``low" S/N spectra into bins of 400 
{\AA}. While the information content remains unchanged, this certainly leads 
to improvements in presentability and fitting speed. \citet[][]{kriek09a} then 
show explicitly  that provided enough photons can be assembled through 
either exposure time or telescope size, the spectra of galaxies at redshifts 
2-3 are amenable to the same kind of analysis as in the local universe. 
The upshot of these studies is that massive, compact ETGs with very little 
residual star formation are in place already at redshifts between 2 and 3. 

Despite these successes, the study of \citet{muzzin09a,muzzin09b} confirms 
that even using spectroscopic data, model uncertainties mean that 
SED-derived stellar masses are affected with uncertainties of factors 2-3 
at these redshifts. For further discussion on stellar mass determinations 
the reader is also referred to the review by de Jong \& Bell (in prep.).

\subsubsection{Deriving SFHs from spectroscopy}

\paragraph{Comparing observations to semi-analytic models} 

\citet{trager09} extend the semi-analytic model of \citet{somerville08} to predict the line 
absorption strengths (Section \ref{s:indices}) of the resulting galaxies. 
This allows them to use the same analysis tools that 
would be used in the analysis of the measured line strengths of an observed sample of 
galaxies on objects with known properties, in particular star formation histories. They 
select in particular early type galaxies from the mock catalogues they produce and 
compute the index strengths of the resulting spectra. These index strengths can then 
be plotted on the same plots as real data. They come to the sobering conclusion that 
while the sample of \citet{trager08} is of sufficient quality do do a meaningful comparison, 
it remains too small. On the other hand large samples of galaxies, as the one of 
\citet{moore02}, still lack the required precision. 

\paragraph{The archeology of the universe}

The database of the SDSS spectra has been used to derive the SFH of galaxies from their 
current spectra \citep[e.g.~][ see also Section \ref{s:inversion}]{heavens04}, 
a procedure sometimes called ``unlocking the 
fossil record'' or simply ``astro-archeology''. A recent update on this has been 
presented in \citet{tojeiro07}, who applied VESPA (see Section \ref{s:inversion}) to the 
SDSS sample of spectra and derived a catalogue that was made available to the 
community at http://www-wfau.roe.ac.uk/vespa/. It provides detailed star formation 
histories and other parameters for SDSS's latest data release (DR7) of the Main 
Galaxy Sample and the Luminous Red Galaxies sample. Details of the catalogue, 
including description, basic properties and example queries can be found in \citep{tojeiro09}. 

\paragraph{Combining spectroscopic and dynamic ages} 

The use of spectroscopy in combination with dynamical arguments to understand the 
evolution of a single object was presented in \citet{pappalardo10}. 
The galaxy NGC4388 is a member of the Virgo cluster 
and sports a huge trail of HI gas \citep{oosterloo05}. It represent an ideal study case 
for the effects of ram stripping on a galaxy moving in the intra-cluster medium \citep{vollmer03}. 
The stripping age has been estimated to be of order 200 Myr from dynamical arguments. 
Using VLT/ FORS spectroscopy of the outer and inner regions of the galaxy and the 
STECKMAP program \citep{ocvirk06}, \citet{pappalardo10} were able to show that, while 
the inner region of the galaxy is of solar metallicity and has continued forming stars to 
the present day, the outer regions of the galaxy have sub-solar metallicity and have 
stopped forming stars roughly 200 Myrs ago, in accordance with the dynamical estimate. 
Single cases like this can thus help identify the processes and timescale associated 
with shutting down the star formation in galaxies, one of the most profound changes 
a galaxy can experience. 

\paragraph{Star formation rates}

Star formation rates from SED fitting have been little used, with most authors preferring 
to rely on single tracers (see Section \ref{s:bayesvalid} for a comparison). \citet{walcher08} 
have used stellar masses and star formation rates consistently derived from the 
same SED fit to compare the predicted evolution of the stellar mass function to the 
observed one. The main result is that while stars form in blue cloud galaxies, most 
of the growth of the stellar mass function occurs in quiescent galaxies, in agreement with
studies based on different tracers of star formation \citep[e.g.][]{bell07}. From comparison 
with merger studies in the same field, \citet{walcher08} conclude that about half of the 
mass growth on the red sequence comes from major mergers and half from minor mergers.

\citet{salim07} have compared their SED-fitting SFRs to SFRs determined from emission lines. 
They find that some galaxies with no detected emission lines nevertheless have substantial 
SED-based SFRs. They attribute this result to ``recent" star formation, i.e.~stars that formed 
long enough ago that their emission lines already vanished, but still recently enough to 
be revealed in the galaxy SED. Recent HST imaging in the UV which clearly shows SF 
structures seems to confirm this (Salim \& Rich 2010, ApJ submitted).

\subsubsection{Identifying and studying outliers}

This is an underexplored use of SED fitting, in the opinion of the authors. One example, objects 
with differing SFR measurements from emission lines and SED fitting has been covered in 
the last section. 

\paragraph{Finding Wolf-Rayet stars}
\label{s:wolfrayet}

The availability of large databases of spectra, such as from the SDSS, and of accurate stellar 
population model predictions enables the search for rare objects or systematic deviations 
that are not predicted by the model itself. An example of this are Wolf-Rayet stars, 
evolved, massive stars with characteristic features. These have ages between 
$2\times10^6$ and $5\times10^6$ years,  and are thus a transient feature of galaxy 
spectra. They are useful as a tracer of recent star formation history as well as the 
metallicity of their host galaxies. As shown in \citet{brinchmann08}, a 
systematic search in the SDSS database yields a sample several times larger than 
previous serendipitous searches. The essential ingredient of such a search is the 
accuracy of the stellar population model that allows an inversion technique (Section 
\ref{s:inversion}) to be applied on a large sample. Indeed, either a smooth correction or 
residual features from inaccurate models would severely impair the identification 
of the specific features. As an example \citet{brinchmann08} show that the use of the 
\citet{bruzual03} models produced a large number of false positives, while an updated 
version of the same model using different stellar spectra (CB09) provides much better fits.

\subsection{Dust}

Dust cannot be ignored when fitting a galaxy's SED, as shown by the
cosmic infrared background, which has comparable power to the distinct
peak of the cosmic UV-optical background
\citep{hauser01}. The relative strength of the cosmic background in
the infrared suggests a significant processing of the galactic stellar
light over the age of the universe. This processing must have also
been more significant with increasing redshift as the percentage of
stellar light re-radiated by dust is only $\sim30$\% locally
\citep{popescu02}, as supported by the increasing number density of
luminous IR galaxies up to $z\sim 1.3$ \citep{magnelli09}.

As discussed in section \ref{s:models},  the absorption and emission
of light by dust are generally treated as separate processes in
modelling, and this is similarly true in SED fitting.

\subsubsection{Attenuation by Dust}

Dust between the observer and the individual stars of a galaxy
acts to extinguish and redden the light from those stars. When the
stars in our own Galaxy were examined it was found that a simple
relation with wavelength was able to describe the extinction
and reddening by dust for a wide range of galactic environments, with
the only strong feature occurring at $\sim$2175\AA\
\citep{cardelli89}. A similar but steeper extinction law was found for
the Magellanic Clouds, with weaker or non-existent feature at 2175\AA\
\citep{gordon98,misselt99}. It is these extinction laws that have
given rise to the contemporary model of dust in the ISM
\citep[i.e.][]{mathis77}, and the understanding that the dust
composition between the Milky Way and Magellanic clouds is different.

Yet when integrated over the whole of a galaxy the situation becomes complex,
with the geometry of the stars and dust strongly affecting the
resulting spectrum. The effects of varying amounts of extinction of
the different stellar populations due to the spatial distribution of
stars and clumpy dust, and the scattering of blue stellar
light into our line of sight act to flatten the effects of dust on the
spectrum, creating an attenuation law, where the amount of reddening
with extinction is less (or `greyer') than we observe locally
\citep{witt92}. This was exactly what was found in starburst galaxies
by \citet{calzetti94}, and \citet{charlot00} found that a simple
screen effective attenuation (i.e.~a screen of dust between the galaxy and
observer) with a power-law relation, $\tau_{\rm
  ISM}\propto\lambda^{-0.7}$, was able to account reasonably well for
the diffuse ISM attenuation in galaxy observations. 
It is this complexity that makes disentangling the effects of geometry
and differing dust difficult, and thus the extraction of physical dust
properties from galaxy SEDs problematic. There are two areas where some progress has
been made.
 
\paragraph{The 2175\AA\ feature}
The 2175\AA\ feature has been associated with small carbonaceous
grains in the ISM \citep{mathis77}, and is observed in both the Milky
way, M33 \citep{gordon99}, and (weakly) in the LMC, but is almost
non-existent in the SMC. This feature is not observed in the
attenuation law of starburst galaxies \citep{calzetti94}. Whether this
lack is due to the clumpy geometry of dust and stars
\citep{fischera03} or is actually indicative of SMC-like dust in
starburst galaxies \citep{gordon99} is still under debate, yet this
feature is generally not needed to fit the attenuation of galaxies.
In QSOs, which, being dominated by a nuclear source, are closer to
the galactic extinction situation, an average attenuation curve does not
show this feature, suggesting processing of the ISM in these active
objects \citep{czerny04}. However, in a few non-local galaxies where direct
extinction lines of sight are available, this feature has been
observed, suggesting it may be more common than the attenuation curves
of local galaxies suggest \citep{wang04,eliasdottir09}. At higher redshifts, 
where UV spectra are more commonly observed, recent studies find 
evidence for the existence of the 2175{\AA} bump \citep[e.g.~][]{noll07, noterdaeme09}. 

\paragraph{Young versus old attenuation}

One important progress made in the treatment of galaxy attenuation is
the realisation that the effective attenuation of a galaxy is
dependent upon its star formation history. \citet{calzetti97} found
that in starburst galaxies the effective attenuation of the stellar continuum
was less than that suffered by the nebula emission, as measured
through emission lines. This clear indication of the clumpiness of the
dust in galaxies has been interpreted as an indication of differential
attenuation of different stellar populations, with young stars, and
their associated ionized nebula, strongly attenuated by the clouds
from which the stars formed, while older stars have evolved out of
their `birth clouds' either through cloud or stellar dispersion, and
are only attenuated by the diffuse ISM dust, which acts on both the
young and old stars \citep[see e.g.][]{charlot00}. Exactly what is the
clearing time of these clouds and the differential attenuation is
still uncertain, and may be galaxy specific, but this forms a basis
for current galaxy SED models as discussed in section \ref{s:models}.

\subsubsection{Dust Emission}

Extracting physical properties from dust emission in the IR is difficult for
both theoretical and observational reasons: excluding the mid-IR, there
are no observed dust features, most being washed out due to the broad
shape of the blackbody emission; the IR suffers from strong
observational constraints, with most data coming from space- and
balloon-based observations; associated with this is the, until
recently, limited sensitivity and spatial resolution and in the
far-IR, at wavelengths $>100$\mum, the sparsity of data. 

With \emph{ISO} and, especially, \emph{Spitzer} space telescopes this
situation has recently improved, and will improve more so in the near
future with the recent launch of \emph{Herschel} and \emph{ALMA}
beginning to take form. So we touch upon here some of the galaxy physical
properties that have been determined from the dust IR emission.

\paragraph{PAH emission in the Mid-IR}
\label{s:pahs}

As mentioned in section \ref{s:models}, the 5-20\mum\ mid-infrared
spectrum of galaxies is generally dominated by broad emission features
arising from large molecules, polycyclic aromatic
hydrocarbons \citep[see e.g.][]{smith07}. Underlying these features is the stellar continuum at
short wavelengths and hot dust emission.  Confusing the interpretation
of the emission features are strong ionic emission lines arising from
species such as Ne$^+$ and  Ne$^{++}$ and strong, broad absorption features
from silicate grains at 9.8\mum\ and 18\mum.

\begin{figure}
\plotone{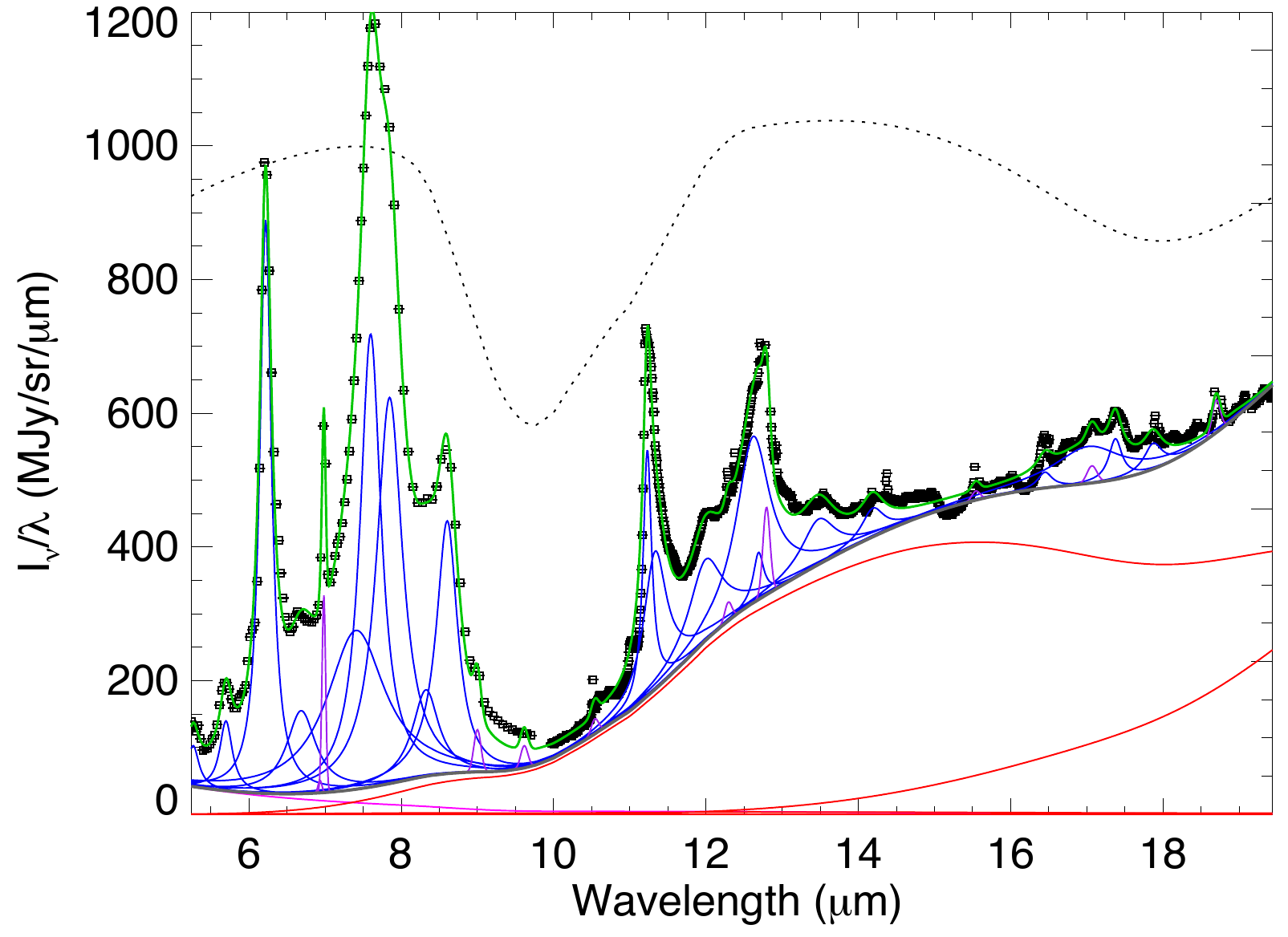}
\caption {An example of combining \emph{Spitzer}-IRS data with a
PAHFIT mid-infrared model. The IRS data arise from the central
region of M82 and are marked by the boxes. The PAHFIT final
spectrum passes through these points (green line) and can be
seperated into PAH features (broad blue curves), fine-structure
emission lines (purple narrow cuves), hot dust continuum
(underlying red cuves) all experiencing broad dust extinction,
especially the pronounced silicate 10 and 18 \mum\ features
(represented by the dotted curve). Data are from \citep{beirao08}.}
 \label{fig:PAHFIT}
\end{figure}

A recent tool, PAHFIT, has been developed to decompose the
mid-infrared spectra into its stellar, PAH, dust continuum, and line
emission constituents, using functional forms and templates for the
features in this wavelength range \citep{smith07}. An example of this
can be seen in figure \ref{fig:PAHFIT}. 
The PAH feature luminosity has been used as star formation rate tracers (see
section \ref{s:SFRIR}), and the relative strength of these features to
the continuum have been found to be strongly linked to the presence of
AGN \citep[see e.g.][and below]{spoon07}, and to the gas phase
metallicity \citep[see e.g.][]{smith07}. The relative strengths of
these features can also be used to diagnose the mean size and
ionization state of the PAHs, which is related to the average
radiation field and dust size distribution \citep{draine07a}.

\paragraph{Diagnosing the energy source in ULIRGs}
Due to the high
obscuration by dust in IR bright galaxies, especially ultra
luminous IR galaxies (ULIRGs), diagnosing the dominant heating
source is problematic. The diagram put forward by
\citet{spoon07} helps resolve this issue by using the
strength of the strong silicate absorption feature that is
determined from fitting the mid-IR SED (as discussed
above in Section \ref{s:pahs}) in association with the
equivalent width of the PAH features. This diagram cleanly separates
different classes of ULIRGs, from obvious Seyfert galaxies, strongly
starbursting galaxies, and to deeply
buried AGN ULIRGs and represents one of the strengths of IR SED
fitting, extracting information from objects which are heavily obscured at
shorter wavelengths.

\paragraph{Dust masses} 

One of the more important properties obtained by fitting the IR SED is
the dust mass. Through fitting of the far-IR SED the temperature(s)
and the relative contributions of the different dust
components that make up the SED can be constrained. Then, using
knowledge of the emissivity per unit mass of dust,  the total dust
mass $(M_{\rm d})$ can be determined, 
using an equation such as \citep[based on][]{dunne01};
\begin{equation}
M_{\rm d}=\frac{L_{850}}{\kappa_{\rm d}(850)}\left[\sum_k \frac{N_k}{B(850,T_k)}\right],
\end{equation}
with $L_{850}$ the 850\mum\ luminosity, and $N_k$ and $T_k$ the
relative contribution and temperature of dust component
$k$. The sum of dust components is usually limited ($\le 3$) by the
sparse observational points at long wavelengths, but can also be
represented by an integral of temperatures, parametrized by the
strength of the heating radiation field \citep[such as used by, e.g.][]{dale02,draine07a}. $\kappa_{\rm d}(850)$ is the
dust mass opacity coefficient, taken to be 0.077 m$^2$kg$^{-1}$ by
\citet{dunne00,dunne01}, an intermediate value between graphite and
silicate. It is generally with this parameter that most of the
uncertainties in determining dust masses remain.

Longer wavelength fluxes ($>300$\mum, such as 850\mum) are preferable to shorter wavelengths when
determining dust masses as these sample the Rayleigh-Jeans part of the
Planck curve, where the flux is least sensitive to temperature. Longer
wavelengths are also more sensitive to the mass of the emitting 
material, as they are sensitive to cold dust as well as warm.

Clear examples of fitting the far IR SED using simple,
emissivity-modified black-bodies and determining the total
dust masses can be found in \citet{dunne00},\citet{dunne01} and more
recently in \citet{clements09} \citep[see also][Section \ref{s:fullsedfit} below]{dacunha10}. 
These works detail nicely the pertinent issues with
both the data and fitting the far-IR SEDs. One of the best examples
of determining the total dust mass, as well as other parameters, using
the full IR SED was done by \citet{draine07b}. Their physically based
SED models \citep[described in detail in][]{draine07b} were fitted to
the far-IR SEDs of galaxies from the SINGS sample, and gave
determinations of the total dust mass, PAH fraction and information on
the interstellar radiation field heating the dust. They found that
dust in spiral galaxies resembled that found in the local Milky Way
ISM, with similar dust-to-gas ratios, and that generally it is the
diffuse ISM that dominates the total IR power, excluding strong
starbursting systems. These results thus confirmed the earlier 
ISO discoveries \citep[see the review by][]{sauvage05}. Note also 
that even earlier detailed radiative transfer modelling of individual 
galaxies had pointed to the dominance of the diffuse component 
\citep{popescu00}.

\paragraph{Sub-mm excess emission}
SED fitting can not only return physical properties, but can also
indicate where our current knowledge is failing. 
As mentioned above, the long wavelength dust emission is a good 
handle for the total dust mass. However, when fitting the IR SED of
several dwarf galaxies it has been found that the sub-mm flux is in
excess to a standard cool dust-body emission, requiring additional dust
components at a unreasonably low temperatures ($\lapprox 7$ K) to fit
the SED \citep[e.g.]{lisenfeld02,israel10}. While very cold large
grains could be one possible cause, other suggestions have included
small stochastic grains that spend most of their time at cold
temperatures \citep{lisenfeld02}, rotating dust grains
\citep{israel10}, or some modification of the dust emissivity at these
wavelengths or temperatures \citep{draine84, weingartner01}. Either
way until this issue is resolved on the cause of this excess, the dust
mass of these dwarf galaxies such as NGC 1569 will have large
uncertainties. 
It is hoped that telescopes such as Herschel and ALMA may find more of these
objects in the near future and help find the cause of this  excess
emission. 

\subsubsection{Dust in the UV to IR}
\label{p:IRXbeta}

The infrared-to-ultraviolet ratio is a coarse measure of dust extinction in the ultraviolet, 
and thus should be related to the amount of reddening in ultraviolet spectra.  Indeed, 
starbursting galaxies follow a tight correlation between the ratio of infrared-to-ultraviolet 
emission and the ultraviolet spectral slope  \citep[e.g.][]{calzetti97,meurer99}.  Compared 
to the relation defined by starbursts, normal star-forming galaxies are offset to redder ultraviolet 
spectral slopes, exhibit lower infrared-to-ultraviolet ratios, and show significantly larger scatter
\citep{kong04,buat05,burgarella05,seibert05,cortese06,boissier07,gildepaz07,dale07}.  
Offsets from the locus formed by starbursting and normal star-forming galaxies can be 
particularly pronounced for systems lacking significant current star formation, such as 
elliptical galaxies, systems for which the luminosity is more dominated by a passively 
evolving older, redder stellar population. 

Using a sample of 1000 galaxies with spectroscopy from the SDSS and 
homogeneous photometric coverage from the UV to 24$\mu$m from SDSS and the Galex and 
Spitzer satellites, \citet{johnson07a} found that the sample galaxies span a plane in the three-dimensional space 
of NUV-3.6$\mu$m colour, D$_n$(4000) index \citep[as defined
by][]{balogh99}, and infrared excess, IRX (=L$_{IR}$/L$_{FUV}$).   
The three-dimensional relation can be expressed in terms of empirical functions, where
IRX is a function of NUV-3.6$\mu$m (or more weakly with other colours)
and D$_n$(4000). They suggest that this relation 
can be explained primarily through SFH and dust attenuation, with both
acting to steepen the optical-UV slope (as measured by the
NUV-3.6$\mu$m color), but only attenuation increasing the IR flux and
hence IRX \citep{johnson06}. 

\begin{figure}
\plotone{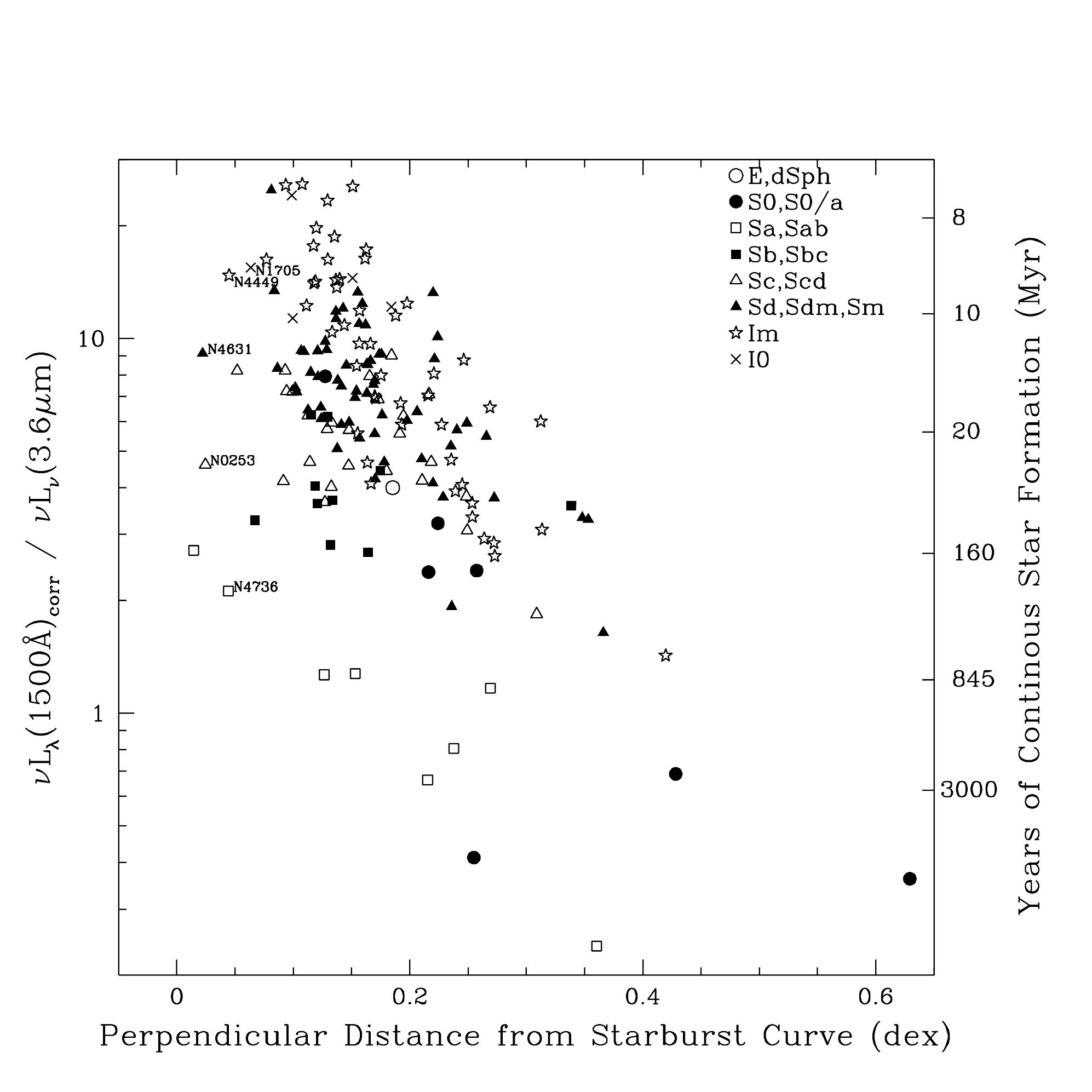}
\caption {The 
dependence of galaxy star formation history as a function of distance from the 
starburst relation of \citet[][]{calzetti97,meurer99}.  The lefthand axis is an observable 
diagnostic of the birthrate parameter, the current star formation rate normalized 
to the average star formation rate.  The righthand axis shows the number of years 
(continuous) star formation has been occurring, as measured from theoretical 
spectra. }
 \label{f:DDale1_2}
\end{figure}

A similar analysis was presented at the workshop by D. Dale using the LVL survey 
(see Section \ref{s:LVL}), which consists of a statistically complete set of star-forming 
galaxies, nearly two-thirds of which are dwarf/irregular systems.  
Figure~\ref{f:DDale1_2} shows the ratio of far-ultraviolet-to-near-infrared 
luminosity as a function of the (perpendicular) distance from the starburst
curve \citep[e.g.][]{calzetti97,meurer99} for the LVL galaxies, with
the far-ultraviolet emission is corrected for attenuation using the
infrared-to-ultraviolet-based recipe formulated in \citet{buat05}.  
By correcting for dust, the FUV/3.6 \mum\ ratio measures only the ratio of past-to-present star formation, 
sometimes referred to as the birthrate parameter 
\citep[see also, for example,][]{boselli01,cortese06}.  This ratio represents the 
birthrate parameter since the far-ultraviolet traces star formation over 100~Myr 
timescales whereas the near-infrared probes the total stellar mass built up over 
much longer timescales.  This plot is as such a compression of the
plane discussed by \citet{johnson07a}, and shows a clear trend, with lower 
birthrate systems exhibiting larger distances from the starburst trend, consistent 
with the study of \citet{kong04}. To further quantify this, theoretical
models with solar metallicity, 
1~$M_\odot~{\rm yr}^{-1}$ continuous star formation curves assuming a double 
power law initial mass function, with $\alpha_{\rm 1,IMF}=1.3$ for
 $0.1<m/M_\odot<0.5$ and $\alpha_{\rm 2,IMF}=2.3$ for $0.5<m/M_\odot<100$ 
 were run \citep{vazquez05} and were matched with their determined
FUV/3.6 \mum\ ratio on the right axis, demonstrating that those with
the oldest SFH (i.e.~lowest birthrate parameters) lie furthest from
the theoretical starburst curve.

\subsubsection{Star Formation Rate from the IR}
\label{s:SFRIR}

One of the  most commonly extracted galaxy properties from the IR is
the star formation rate. As discussed above, recently formed stellar
populations tend to be more obscured than older stellar
populations. They are also more luminous and emit more in the
ultraviolet where dust opacity peaks, and thus dust emission is in
principle a good tracer of recent star formation, assuming a simple
calorimetric situation. It is these assumptions that lead to the
widely-used \citet{kennicutt98} relation between total IR luminosity
(8--1000\mum) and star formation rate. 

Of course the situation is more complex than this, as discussed in the
same work. A range of ages contribute to dust heating \citep{kennicutt09}, 
and star forming regions in galaxies suffer a range of obscurations, from
totally obscured (ULIRGs) to unobscured (blue compact dwarfs). It is for this 
reason that this relation has been re-examined and empirically calibrated 
with new data from
\emph{Spitzer}. In general, all recent studies have found that the IR can
be used as a SFR indicator, albeit without a direct one-to-one connection. 
Complications arise due to the correlation of SFR, luminosity, and galaxy gas
and dust masses, and possible non-linearities due to metallicity
effects \citep{wu05}.

More specifically, \citet{calzetti07} using spatially resolved
observations, found strong correlations between the 24\mum\ and
P$\alpha$ luminosity densities (a proxy for SFR, assuming little attenuation in
the near-IR), and a correlation between the 8\mum\ and P$\alpha$
luminosity densities, though this failed at low metallicities. 
Using these, they were able to create new calibrations for SFR versus
24 \mum\ luminosity, and  SFR versus 24 \mum\ and observed H$\alpha$
luminosities, with the latter relation accounting better for the
escaping radiation not accounted for by the dust emission.
\citet{rieke09} took this further, showing that for higher IR
luminosity objects, P$\alpha$ was no longer a good tracer for SFR as
even it was obscured, and gave their own calibrations for SFR with the
IR luminosities. 

On galactic scales, \citet{zhu08} showed with a larger galaxy sample
from SWIRE  that the \citet{calzetti07} relations between 24\mum\
luminosity and extinction corrected H$\alpha$ luminosity hold, and
thus L(24\mum) is a good SFR indicator. In addition they also showed
that 24\mum\ is well correlated with 70\mum\ and total IR
luminosities, indicating that these too can be used as SFR indicators,
albeit with larger scatter. 

The 8\mum\ (and other PAH bands) and longer wavelength observations,
such as the \emph{Spitzer} 160 \mum\ band, are observed to be
correlated with each other \citep{bendo08}, and are thought to be more
associated with the cooler diffuse ISM. While the diffuse ISM is
heated by the radiation from star forming regions, it is also heated
by the diffuse radiation field from older stars, meaning that these
bands are not as strong SFR traces, especially at low SFRs.

\subsection{Fitting the full UV to FIR SED}
\label{s:fullsedfit}

Fitting a self-consistent model over the divide between stellar and dust 
emission in the SED is of course one of the ultimate goals of SED fitting. 
However, while modelling efforts are well-underway (see Section \ref{s:fullsed}), 
unfortunately, few authors have attempted to apply these to large samples of galaxies. 
Inversion techniques are not applicable here, as the problem is highly non-linear 
with many free parameters and therefore time consuming. Even for calculation 
of a library of model galaxies, it is challenging to provide models that are sufficiently simple, 
complete and fast to make this a practical possibility. 

We here need to bypass instances where modellers test their codes on single galaxies 
\citep[e.g.~][]{silva98, popescu00, groves08}. This is of course a most valuable and necessary 
step to make sure that the model does bear on our understanding of reality and to further our 
knowledge of the underlying physics. Most of the results from these studies have been 
presented above in Section \ref{s:models}. For the potential novice reader of this manuscript we 
nevertheless emphasize at this point that proper filter convolution and 
$\chi^2$ fitting cannot be replaced by by-eye passing of spectra through 
photometric data points. 

Using the model by \citet{dacunha08}, in a follow-up paper 
\citet{dacunha10} have demonstrated the
strength of fitting the full SED from UV to IR wavelengths. By fitting
the full SEDs of $\sim3000$ galaxies with GALEX, SDSS, 2MASS, and IRAS
data, they were able to determine the star formation rate, the star formation
history as measured by the specific star formation ($\psi_{\rm s} =$current
$0-10^8$Myr star formation rate divided by the past average star
formation rate), dust and stellar masses and other parameters. They found a
strong correlation of dust mass ($M_{\rm d}$) to star formation rate
($\psi$, in M$_{\odot}$ yr$^{-1}$), shown below in
equation \ref{eq:dacunha10}), as well as relations between the dust to
stellar mass ratio and $\psi_{\rm s}$, and the fraction of IR emission
arising from the diffuse ISM and $\psi_{\rm s}$.

\begin{equation}\label{eq:dacunha10}
M_{\rm d}=(1.28\pm0.02)\times10^7\psi^{1.11\pm0.01}{\rm M_{\odot}}
\end{equation}

This work demonstrates clearly the connection between dust mass, star
formation history and stellar evolution.

\citet{iglesias07} have gone to the length of using the GRASIL code \citep{silva98} to compute 
a library of ~5000 model galaxies and then use Bayesian inference to derive the properties of their 
sample. Their general results agreed well with independent studies by other authors, thus lending 
support to the notion that full SED fitting is a reliable tool to derive galaxy properties. More 
importantly in the present context, they show that their reduced $\chi^2$ distribution has a median 
value of 2.6, albeit with a long tail extending well above 10. Thus the GRASIL library is found 
to reproduce their sample fairly well. Nevertheless, from the point of view of reliable SED 
fitting tools, a more thorough analysis of the outliers (model uncertainties, incomplete libraries, 
AGN, etc.) would be valuable, not only in this but in many other works.

\citet{noll09} present a new code, which they call "CIGALE", which effectively computes a 
library of model galaxies and then uses a modified version of the Bayesian inference 
described in Section \ref{s:bayesian} to determine the galaxy properties. Diagnostic plots like their 
Figure 14, which shows the residuals between best fit model and data for their full sample, are a very 
useful tool to understand model systematics. In their case for example, they conclude that 
"For MIPS 160 $\mu$m the significant deviations can partly be explained by the lacking flexibility 
of the one-parameter models of \citet{dale02}."

As shown by the last two examples, most UV-FIR SED fitting codes are still in their 
testing phase and have mostly been used to confirm results already obtained from more traditional 
single-tracer analyses. The large number of derived parameters 
and our still limited knowledge of their respective degeneracies and systematic uncertainties 
make it difficult to go a step further and fully use the full power of SED fitting. Indeed, for the moment 
it is still questionable whether it is not more fruitful to use a combination of single tracers to 
derive one property well \citep[e.g.~][for SFR]{kennicutt09}. On the long run, however, SED fitting 
holds the promise to provide a large set of galaxy properties for large samples. Self-consistent 
inter-comparison of 
sub-samples with different properties, such as masses and SFRs, and the exploitation of constraints 
on hitherto unconstrainable parameters, such as the relative weights of young, intermediate age and old 
populations, are an exciting avenue to explore further in the future.

\section{Conclusions}

We have presented an overview of some of the achievements and challenges 
related to fitting the Spectral Energy Distributions of galaxies. 
SED fitting can be used effectively to derive a range of physical 
properties of galaxies, such as redshift, stellar masses, star formation rates, dust masses, 
and metallicities, with care taken not to over-interpret the available data. To allow for 
more progress in galaxy evolution studies from SED fitting, we suggest two main areas. 
On the one hand there still exist many specific issues such as estimating the age of the 
oldest stars in a galaxy, finer details of dust properties and dust-star geometry, and the 
influences of poorly understood, luminous stellar types and phases. The challenge for 
the coming years will be to improve both the models and the observational data sets to 
resolve these uncertainties. On the other hand, the robustness and accuracy of 
SED-fitting-derived properties still need to be assessed more completely. The challenge 
here is to develop and understand the interplay between the fitting routines and the 
available data and models. 

In the hope of accompanying these challenges, the present review will be made 
available on a webpage (sedfitting.org) together with links to relevant models, fitting codes 
and datasets. We would like to encourage the 
community to send in suggestions for additions and changes to the 
text\footnote{Please note that reference to the online text should always include 
a citation to the present work.} through this webpage. The 
intention is twofold: 1) We hope to bolster the information currently available 
in this review and keep it up to date over the coming years. 2) Due to our bias 
to the workshop participants we did mention many important works in passing, 
or indeed missed them. We therefore hope that particularly 
those members of the community whom we missed will take to opportunity to 
add their part of the story, thus expanding the current text beyond its original scope.

\acknowledgments

The authors of this review and organizers of the workshop would like 
to thank the Lorentz center for making this workshop possible and 
for providing a first-class meeting environment. We also thank NOVA for 
additional support. 

We would like to thank the participants in the workshop for their 
motivation and for sharing their expertise, without which this review 
could not have been written. The talks and discussions of the 
attendees formed a large part of this review. Furthermore, we 
are deeply indebted to the following participants for supplying 
many pages worth of text: H. Aussel, N. Ball, M. Brodwin, S. Charlot, 
L. Dunne, I. Ferreras, V. Margoniner, M. Polletta, A. Sajina, J.D.~ Smith, 
P. Oesch, V. Wild, C. Wolf, S. Zibetti. We thank N. Ball, J. Brinchmann, 
E. da Cunha, S. Charlot, M. Dopita, L. Dunne, I. Ferreras, P. Oesch, and 
V. Wild for custom making their figures. 

We thank an anonymous referee for a critical and thorough reading of 
the manuscript, which led us to clarify several important points in the 
text. We also thank L. Spezzi for a careful reading of an early version of 
the manuscript. 

This research has made use of NASA's Astrophysics Data System Bibliographic Services.

\end{document}